%% file: main.tex
\documentclass[10pt]{article}

\usepackage[a4paper,margin=0.9in]{geometry}
\usepackage{amsmath}
\usepackage{amssymb}
\usepackage{array}
\usepackage{longtable}
\usepackage{graphicx}
\usepackage{dcolumn}
\usepackage{bm}
\usepackage{iftex}
\ifPDFTeX
  \usepackage[T1]{fontenc}
  \usepackage[utf8]{inputenc}
\else
  \usepackage{fontspec}
\fi
\usepackage{subcaption}
\usepackage{physics}
\usepackage{booktabs}
\usepackage{tabularx}
\usepackage{makecell}
\usepackage{multirow}
\usepackage{epstopdf}
\usepackage{pifont}
\usepackage{adjustbox}
\usepackage{dsfont}
\usepackage{quantikz}
\usepackage{tikz}
\usetikzlibrary{arrows.meta,positioning,shapes.geometric,fit}
\usepackage{pdflscape}
\usepackage{fvextra}
\usepackage{authblk}

\newcolumntype{Y}{>{\raggedright\arraybackslash}X}

\usepackage[bottom]{footmisc}           
\usepackage{tablefootnote}              
\newcolumntype{L}[1]{>{\raggedright\arraybackslash}p{#1}}
\newcolumntype{C}[1]{>{\centering\arraybackslash}p{#1}}
\newcommand{\public}{Public}
\newcommand{\gated}{Gated}
\newcommand{\NA}{\textemdash}
\newcommand{\cellcite}[1]{\,{\scriptsize\cite{#1}}}

\usepackage{array}

\newcolumntype{L}[1]{>{\raggedright\arraybackslash}p{#1}}
\newcolumntype{C}[1]{>{\centering\arraybackslash}p{#1}}

\usepackage{abstract}

\usepackage[hypertexnames=false]{hyperref}

\usepackage[
  backend=biber,
  style=numeric,          
  citestyle=numeric-comp, 
  sorting=none,
  url=true,
  doi=false,
  isbn=false,
  maxbibnames=5,
  minbibnames=5,
  maxcitenames=5,
  mincitenames=5
]{biblatex}

\AtEveryBibitem{%
  \clearfield{note}
  \clearfield{urldate}
  \clearlist{language}
}

\renewbibmacro{in:}{}

\renewbibmacro*{url+urldate}{%
  \printfield{url}}

\title{Quantum computing for transport research: \\
an introduction, systematic review, and perspective}

\author{Lachlan Oberg$^{1,*}$}
\author{Paul Corry$^{2}$}
\author{Moji Ghadimi$^{3}$}
\author{Ashish Bhaskar$^{1}$}

\affil{${}^{1}$School of Civil and Environmental Engineering, Queensland University of Technology, Queensland 4000, Australia}
\affil{${}^{2}$School of Mathematical Sciences, Queensland University of Technology, Queensland 4000, Australia}
\affil{${}^{3}$QCIF Ltd, Brisbane, Queensland 4072, Australia}

\date{September 7, 2026}

\usepackage{ragged2e}
\usepackage{colortbl}
\usepackage{helvet}
\usepackage{pdfpages}
\definecolor{EvidenceN}{HTML}{EBEEED}
\definecolor{EvidenceP}{HTML}{DADADA}
\definecolor{EvidenceM}{HTML}{B9D3C0}

\begin{document}

\maketitle

\begingroup
\renewcommand{\thefootnote}{}
\footnotetext{$^{*}$Corresponding author: \href{mailto:l.oberg@qut.edu.au}{l.oberg@qut.edu.au}}.
\endgroup

\begin{center}
\begin{minipage}{0.9\textwidth} 
\begin{abstract}
Transport research has significant potential to benefit from quantum computing. The rise of intelligent transport systems, autonomous vehicles, and the Internet of Things has created an unprecedented demand for efficient information processing and computational optimisation. Accordingly, transport engineers and scientists have explored the ever-improving capabilities of quantum computers in an effort to meet this demand. Motivated by this growing interest, this paper sets out three aims: (1) to introduce the fundamental aspects of quantum computing relevant to the transport domain; (2) to develop a pipeline for identifying transport-related problems which are suitable for quantum advantage, solving them, and assessing their performance fairly; (3) to provide a systematic review of the existing literature and a perspective. For the latter, a systematic search of the Scopus and IEEE Xplore databases, supplemented by additional citation sources, identified 91~studies for inclusion following PRISMA 2020 guidelines. While a diverse set of use cases have been proposed, we conclude that future research should prioritise problems where quantum computation offers a clearer practical benefit. To this end, we suggest promising directions to guide further work in this burgeoning subfield.
\end{abstract}
\end{minipage}
\end{center}

\setcounter{tocdepth}{2}
\tableofcontents
\bigskip

\newpage

\section{Introduction}

The demand for information processing in transport engineering has never been greater. Large urban networks can facilitate tens of millions of daily users across many transport modes, millions of links, and hundreds of thousands of nodes. Analysing the vast amount of data produced by these systems, let alone optimising them, is a daunting task with immense practical impact. For example, in 2024, U.S. drivers lost an average of 43 hours to traffic congestion, representing \$74 billion in nationwide productivity losses \cite{pishue_inrix_2025}. These delays also carry significant health and environmental consequences. Hence, computational techniques that deliver even slight improvements can generate substantial value at scale.

Quantum computing is a nascent paradigm heralded as a panacea for this ever-growing computational burden. This promise derives from a known set of quantum algorithms with provable computational advantage (in terms of time complexity, memory, or energy usage) over their classical counterparts. Moreover, there exists a secondary class of variational quantum algorithms (combining both quantum and classical computing) designed for combinatorial optimisation problems, whose computational advantage has proved more difficult to ascertain and whose ultimate prospects are more controversial. Nonetheless, any theoretical advantage has yet to be fulfilled in practice. Realising the touted `quantum revolution' has proved exceptionally difficult from a physics and engineering standpoint due to the challenges of manipulating quantum information (`qubits') while maintaining its useful quantum properties. Notably, no problem of practical interest to the transport engineer has yet benefited from an existing quantum computer relative to a classical device.

Despite the challenges facing hardware development, it is undeniable that the capabilities of quantum computers are improving rapidly. Industry and government have taken notice, with McKinsey \& Company predicting that the incremental value impact of quantum computing on travel, transport, and logistics could range from \$200--\$500 billion by 2035\cite{noauthor_quantum_2025}. The basis for this prediction is well-founded if the advertised potential for quantum computing is fully realised in a timely manner. Many proposed applications of quantum computing target NP-hard combinatorial optimisation problems where marginal gains of 1--2\% can translate into substantial real-world impact at scale. For example, in 2013, UPS found that eliminating one mile per driver per day is worth \$50 million annually, a figure which has likely increased significantly since\cite{business_for_social_responsibility_bsr_looking_2016}. 

There are also potential applications for autonomous vehicles, intelligent transport systems, computer vision, machine learning, and the emergent Internet of Things. Firstly, modern automobiles often possess dozens of on-board cores for local processing of their sensor data. This is essential for real-time tasks such as braking or obstacle avoidance where even a 100~ms delay can have catastrophic safety consequences. On-board quantum technology could potentially reduce this signal processing time with lower power consumption. Secondly, the U.S. operates more than 330,000 traffic signals which are typically re-timed every three to five years at a cost of around \$4,500 per intersection\cite{federal_highway_administration_automated_2020}. Because most intersections lack continuous performance monitoring, highway agencies rely on simulated models built from periodic manual data collection, adding significant time and expense. Quantum technology could address both fronts, enabling rapid on-demand processing of high-volume sensor data at individual intersections and optimising signal coordination across entire urban networks. In summary, the potential applications of quantum computing to transport engineering are diverse and impactful.

The promises of quantum computing have spawned a burgeoning academic field dedicated to exploring transport-related applications. Beginning with the availability of cloud-accessible quantum hardware around 2017, transport engineers have now explored a myriad of use cases within both an academic context and through partnerships with industry and government. However, there remains a pressing need for targeted applications which offer tangible computational advantage in the intermediate and long term. Recall that the functional utility of quantum computers relies on obtaining `quantum advantage': solving a given problem more efficiently or accurately than the best classical counterpart. As achieving this advantage is ultimately dependent on future improvements to quantum hardware, the role of transport researchers in the near term should be identifying suitable problems for quantum enhancement and formulating compelling use cases. While some works have seriously considered the efficiency of different algorithms for their particular problem instance, we contend that too much of the literature has focused on use cases \textit{applying} quantum computing rather than use cases which could conceivably \textit{benefit} from quantum computing. In particular, there has not been a sufficient emphasis on whether the chosen quantum algorithm can theoretically ever provide a quantum advantage, irrespective of the eventual state of quantum hardware.

The reasons for this state of affairs are likely multifaceted. Firstly, researchers may overestimate the effectiveness of general quantum algorithms for optimisation problems relative to instance-specific and highly efficient classical solvers, many of which have benefited from decades of research. Secondly, the field of quantum computing is rapidly evolving in both theory and experiment. Many assumptions made about the effectiveness of certain quantum algorithms may therefore be outdated, leading to inaccurate assessments of good applications for quantum computing in transport. As industry leaders expect quantum computers to obtain some degree of practical computational benefit over the next decade, the goal of this work is to establish a framework for transport researchers to effectively leverage this quantum advantage.

Specifically, we aim to: (i) provide the foundational knowledge required for applying quantum computation to transport engineering problems; (ii) develop a structured method for identifying transport problems suitable for quantum advantage, solving them, and assessing the results fairly; (iii) broadly review the academic literature for quantum computing in transport and provide a perspective on future directions for the field. This work sits amongst several recent reviews and perspectives on the intersection between transport studies and quantum computing\cite{niu_quantum_2026,liu_quantum_2025,cooper_exploring_2022,du_overcoming_2026,smith-miles_travelling_2025,bentley_quantum_2022}. Our review consolidates and extends these works in two ways. To our knowledge, this is the first transport-focused review to combine a PRISMA-guided systematic search with an explicit framework for assessing evidence of quantum advantage.

The structure of this work is as follows. In Section~\ref{sec:quantum_foundations} we introduce the fundamentals of quantum computing including the current state of hardware, algorithms, and their performance. This introduces a significant amount of quantum vocabulary, and we therefore provide a reference dictionary in Table~S1 of the Supplementary Material. In Section~\ref{sec:pipeline} we develop a pipeline for identifying transport problems suitable for quantum advantage and how to implement quantum solutions. We also introduce three critical criteria for assessing quantum advantage fairly. In Section~\ref{sec:transporTpplications} we provide a detailed literature review for the application of quantum computing in transport studies. Finally, in Section~\ref{sec:future} we review the state of the field and provide a perspective on future directions.

\section{Quantum computing foundations}
\label{sec:quantum_foundations}

\subsection{Quantum information}

Quantum computing is not mystical. It is a method of information processing which obeys the well-defined rules of quantum mechanics instead of classical mechanics. These rules can be exploited to run specialised quantum algorithms which, in principle, can be significantly faster than their classical counterparts. However, quantum computers are not a universal panacea and in many contexts display comparable or inferior performance to classical devices. Understanding the difference between genuine quantum advantage and mere quantum hype is therefore critical. Hence, transport researchers must understand the foundations of quantum computing before considering use cases. The following is intended as an introduction. For a comprehensive treatment, the interested reader is directed to the seminal textbook by Nielsen and Chuang\cite{nielsen_quantum_2010}.

Quantum information is encoded in qubits. A qubit is a quantum system with two distinguishable configurations (termed ``quantum states'') which represent the binary values ``0'' and ``1''. For example, these states could be energy levels of an electron or different polarisation of a photon. A key difference between classical bits and qubits is that the latter can exist in a ``quantum superposition'' of both 0 and 1. At an abstract level, this superposition is not so mysterious. Consider representing the two quantum states mathematically through a complex-valued vector called a ``ket''. We write $|0 \rangle = \begin{bmatrix} 1 & 0 \end{bmatrix}^\top$ and $|1 \rangle = \begin{bmatrix} 0 & 1 \end{bmatrix}^\top$. Then the state of the qubit can be described in ket notation as the two-dimensional vector
$$|q\rangle = \alpha |0\rangle + \beta|1\rangle = \begin{bmatrix} \alpha  & \beta \end{bmatrix}^\top,$$
where $\alpha$ and $\beta$ are complex numbers satisfying $|\alpha|^2 + |\beta|^2=1$ (this ensures probability normalisation as explained below). These coefficients are termed quantum amplitudes of the quantum state. So while classical bits are restricted to either $0$ or $1$, a qubit can occupy any normalised linear combination of $|0\rangle$ and $|1\rangle$.

A quantum computer works by interacting many qubits within a single device. This produces a collective quantum state which can encode all possible bit strings of the qubit register. For example, two qubits have four possible binary configurations $00$, $01$, $10$, and $11$. These are represented in ket form through the so-called ``computational basis states'' of $|00\rangle$, $|01\rangle$, $|10\rangle$, and $|11\rangle$. An arbitrary two-qubit state can therefore be described as the vector
\begin{equation}\label{2-qubit-TP}
|q\rangle = c_{00}|00\rangle
+c_{01}|01\rangle
+c_{10}|10\rangle
+c_{11}|11\rangle = [c_{00} \ c_{01} \ c_{10} \ c_{11}]^\top,
\end{equation}
where the $c$ are complex numbers and $|c_{00}|^2+|c_{01}|^2+|c_{10}|^2+|c_{11}|^2=1$. Similarly, an $n$-qubit system has $2^n$ possible binary configurations, each represented by a computational basis state labelled by an $n$-bit string. The state of the system is therefore represented by a vector in a $2^n$-dimensional complex vector space.

This reveals a critical difference between quantum and classical computing. Whereas a classical register of $n$ bits occupies a single $n$-bit configuration at any given time, an $n$-qubit register can exist in a superposition of all $2^n$ computational basis states simultaneously. This is not to be misconstrued as the quantum computer storing all $2^n$ bitstrings in analogy to classical memory. Explicitly we mean that up to $2^n$ complex numbers are required to fully describe a quantum superposition of $n$ qubits. In some quantum algorithms, this compact representation can be exploited to avoid storing all $2^n$ configurations in classical memory\cite{wilkening_quantum_2025}. 

Unfortunately, the benefits of superposition come with a huge caveat: readout. A classical bit can be measured repeatedly without changing its value (up to measurement error). Although the computational basis states coexist in superposition during a quantum computation, the superposition itself is never directly observed. Instead, readout is performed through ``quantum measurement''\footnote{This is a physical process which is poorly understood. Termed the ``measurement problem'', it is one of the most long-standing and controversial enigmas in modern physics.}, which collapses the state and returns a single classical outcome. For a single qubit, $\lvert q\rangle=\alpha\lvert0\rangle+\beta\lvert1\rangle$, measurement returns 0 with probability $\lvert\alpha\rvert^2$ and 1 with probability $\lvert\beta\rvert^2$. This is termed the Born rule, and it requires that $
\lvert\alpha\rvert^2+\lvert\beta\rvert^2=1$ since the qubit must be observed as either $0$ or $1$. For an $n$-qubit register, measurement returns a single $n$-bit string, with probability given by the squared modulus of its corresponding complex amplitude. Thus, for the state in equation~\eqref{2-qubit-TP}, the probability of observing the bitstring $\lvert01\rangle$ is $\lvert c_{01}\rvert^2$. The quantum state may therefore contain amplitudes for exponentially many configurations, but a single readout reveals only one of them.

Quantum computation proceeds by encoding information in a quantum superposition, manipulating that superposition according to a quantum algorithm, and finally reading out the result through measurement. The algorithm controls the amplitudes associated with the computational basis states so that the desired answer is assigned a high probability of being observed. Because measurement is probabilistic and returns only a single bitstring on each run, the computation may need to be repeated many times to estimate the resulting output probabilities.

\begin{landscape}
\begin{figure}
    \centering
    \includegraphics[width=1\linewidth]{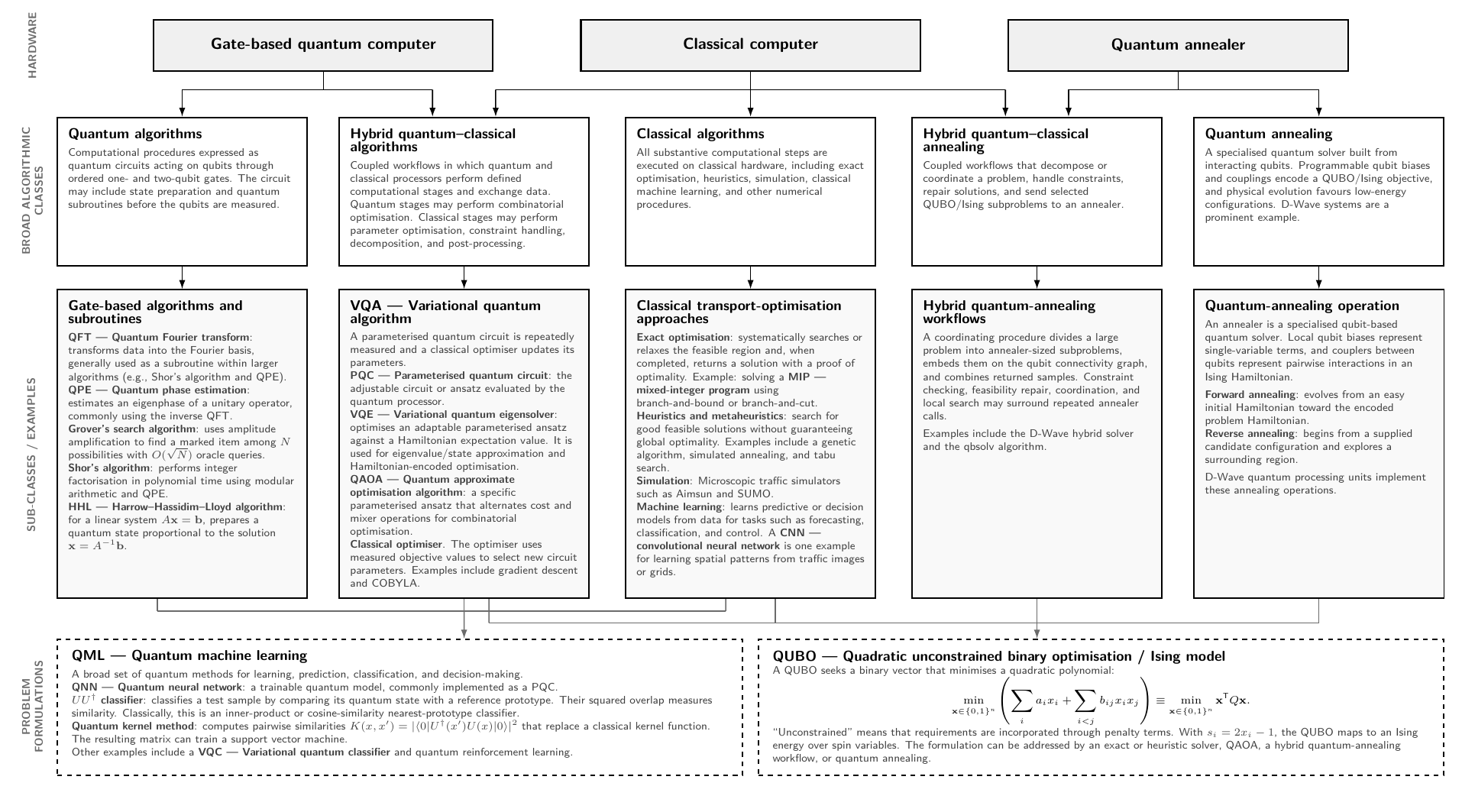}
    \caption{Flow diagram displaying the relationship between types of computing hardware and their associated algorithms. Curly brackets indicate that algorithms below are a subset of algorithms above.}
    \label{fig:algo_space}
\end{figure}
\end{landscape}

There are two types of quantum computing hardware which are currently used to perform quantum algorithms for transport engineering. The first type comprises gate-based quantum computers, which use quantum logic gates to manipulate qubits. These devices are universal in the sense that they can approximate any quantum operations on a finite number of qubits to arbitrary precision. The second type comprises quantum annealers, which are special-purpose devices that perform analogue control of qubits. They are not universal; their capabilities are largely restricted to solving quadratic unconstrained binary optimisation (QUBO) problems which occur frequently in transport research. We now introduce the operating principles of both gate-based computers and annealers, discuss key algorithms, and demonstrate the current state-of-the-art in hardware and algorithmic performance. As an overview, a flow diagram explaining each type of computing hardware and its associated algorithms is presented in Figure~\ref{fig:algo_space}.

\clearpage

\subsection{Gate-based quantum computers}

Gate-based quantum computers manipulate qubits using quantum logic gates, which are the quantum analogues of classical logic gates. A quantum algorithm is implemented as a sequence of these gates, each modifying the quantum state in a controlled way. Physically, a gate may be realised by a process (such as a laser or microwave pulse) targeting one or more qubits. At the abstract level, gates are represented by reversible unitary operations that preserve the total probability of the quantum state. Quantum algorithms commonly employ one-qubit gates, which act on individual qubits, and two-qubit gates, which couple pairs of qubits and can apply operations conditionally on the state of another qubit. Although such gates act directly on only one or two qubits, they necessarily modify the state of the entire register. Consequently, when the register is in a superposition, a single gate can change the amplitudes associated with many (or even all $2^n$) computational basis states simultaneously. This differs from a classical gate, which acts only on the single bit configuration currently stored by the computer. Some quantum gates are fixed operations, whereas others are parameterised by continuous variables. For example, the two-qubit gate $R_{ZZ}(\theta)$ depends on a continuously variable parameter $\theta$, commonly termed a rotation angle, which controls the transformation applied to the quantum state.

Gate-based quantum computing can be performed using a variety of different hardware (commonly termed architectures). The different hardware realise qubits and quantum gates in different ways. Table~\ref{tab:hardware} presents some of the leading quantum computing companies, their qubit architecture, their best-performing model, and device specifications. These realise qubits through current loops in superconductors, the internal electronic levels of trapped ions, or neutral atoms. Other architectures not listed realise qubits through photons, defects in semiconductors\cite{oberg_bottom-up_2025}, and a myriad of other ways. Each qubit architecture also realises gate operations through a different mechanism. For example, trapped-ion and neutral atom qubits are controlled through laser pulses whereas photonic qubits may be controlled by passive optical elements.

While there are many architectures currently vying for dominance in the nascent quantum ecosystem, there is not yet any clear ``winner'' in the race for quantum supremacy (indeed, it is unclear whether there will be a single winner at all). Nonetheless, benchmarking device performance is critical for assessing current and future capabilities. One obvious metric is qubit capacity: the total number of qubits in a device which can be initialised, controlled, and read out. A cursory glance at Table~\ref{tab:hardware} indicates that leading devices may possess a few dozen, hundreds, or even low thousands of qubits. Presumably, then, these devices could perform algorithms using this many qubits. Unfortunately, this is not the case. Qubit capacity alone does not determine the size of the algorithm which can be performed, primarily because of two fundamental challenges: control errors and decoherence.

Executing error-free quantum gates remains an immense engineering challenge. While classical gate operations fail roughly $10^{-23}$ of the time~\cite{proctor_benchmarking_2025}, two-qubit quantum gates currently possess infidelities of approximately $0.1\text{--}1$\%. The number of gate operations generally increases with the number of qubits involved, and complex algorithms may require thousands or millions of gates. These errors concatenate through successive operations until the final quantum state becomes effectively random. Decoherence is a quantum phenomenon which also degrades device performance. Unwanted interactions between qubits and their environment (e.g., atomic-scale vibrations, stray photons, magnetic-field fluctuations) cause information stored in the superposition to spread into surrounding quantum systems. Because this vast environment cannot be read out or controlled, the useful information contained in the qubit state is effectively lost. Each qubit therefore has a characteristic coherence time which limits how many gates can be applied before decoherence occurs. This is why many quantum architectures must be housed in dilution refrigerators at milli-Kelvin temperatures: higher temperatures produce more environmental interactions and degrade device performance.

Decoherence is ultimately unavoidable because useful qubits can never be completely isolated from their environment. The future success of gate-based quantum computing therefore depends upon fault tolerance through quantum error correction. The aim is to combine many error-prone physical qubits into reliable logical qubits using carefully designed sequences of gates and measurements. While this field has made significant progress, fully fault-tolerant quantum computers have not yet been realised at scale. Current hardware therefore remains in the noisy intermediate-scale quantum (NISQ) era, where qubit capacity is a poor indicator of practical performance\cite{Preskill2018quantumcomputingin}. Metrics such as ``quantum volume'' attempt to account for gate fidelity, connectivity, and circuit depth, although no single benchmark perfectly captures performance for a given algorithm. For example, Quantinuum's 56-qubit System Model H2 reported a quantum volume equivalent to 25 qubits and two-qubit gate fidelities of 99.895\%\cite{quantinuum_quantinuum_2025}. In a large demonstration of the quantum approximate optimisation algorithm (termed QAOA and discussed below) using up to 18 qubits and approximately $10^3$ two-qubit gates, the authors obtained results comparable to noise-free simulations for a 10-qubit instance. The output became progressively noisier as the problem size approached 18 qubits~\cite{shaydulin_evidence_2024}. Hence, the device possessed 56 physical qubits, a quantum volume equivalent to 25 qubits, and produced an accurate answer for approximately 10 qubits on that particular algorithm.

\subsubsection{Algorithms}
\label{ssec:algos}

There are two broad classes of quantum algorithms which have been developed for gate-based quantum computers. The first category comprises ``pure'' quantum algorithms, in which all substantive computational steps are performed on a quantum processor (although classical control and readout may still be required). We will refer to them simply as quantum algorithms. A relatively small number of these algorithms offer up to exponential speedups for particular problems, but their practical realisation is severely limited on current NISQ-era devices by limited qubit capacities and the large number of gate operations they require. The second class comprises ``hybrid algorithms'', which seek to obtain advantage through the combination of classical and quantum computation. This may involve integrating a quantum subroutine into a predominantly classical framework or using an iterative loop of quantum and classical subroutines. The latter are broadly termed variational quantum algorithms (VQAs) and will be discussed in depth below. These have received significant attention in the transport literature because their quantum circuits are relatively shallow and hence more compatible with NISQ-era hardware. Nonetheless, their practical implementation is also limited and evidence for quantum speedup remains scant.

\begin{figure}
    \centering
    \includegraphics[width=0.9\linewidth]{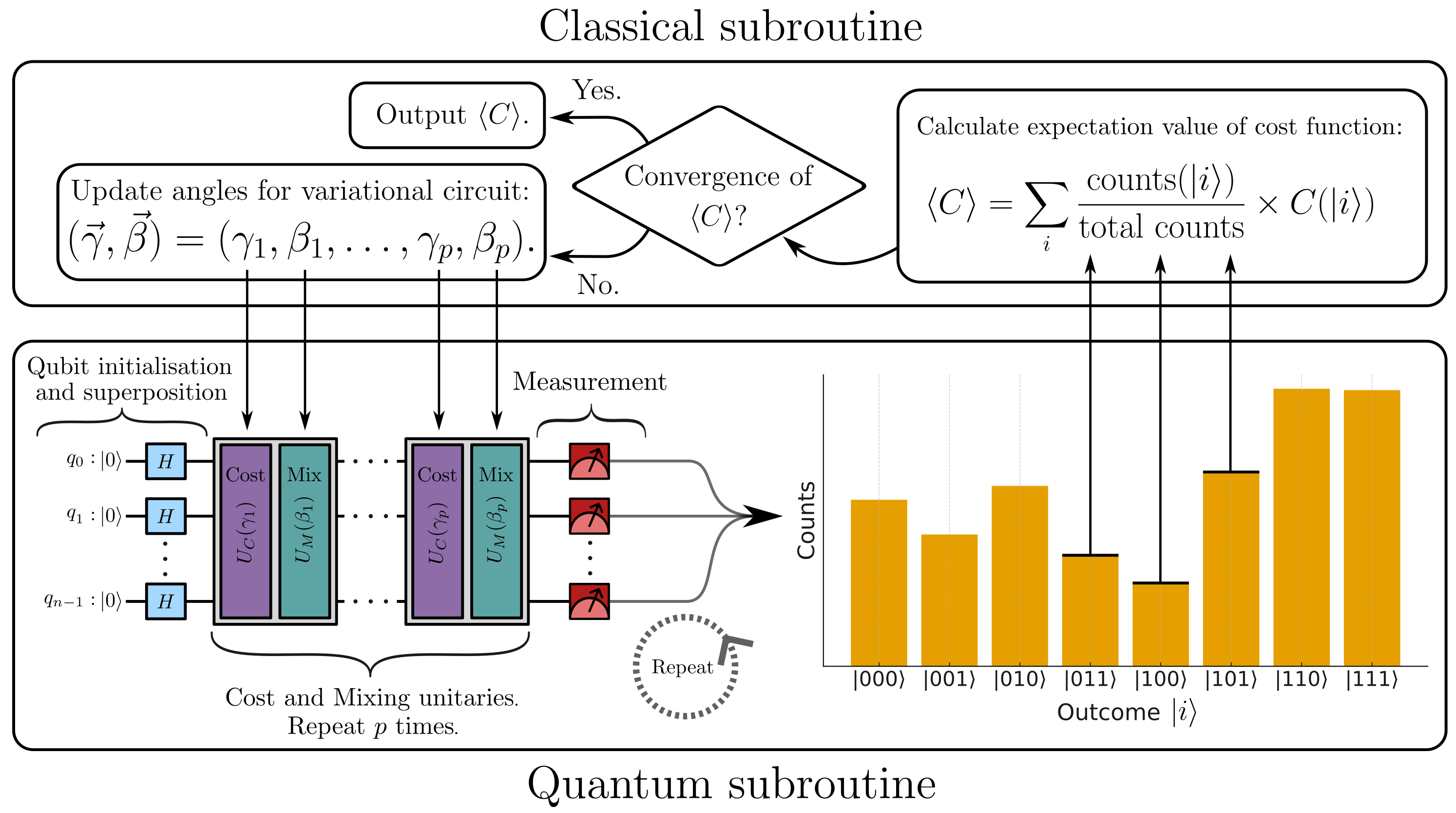}
    \caption{Basic structure of QAOA, a VQA designed for combinatorial optimisation problems. A classical optimiser selects the variational angles which parameterise the quantum circuit. The circuit is executed and measured repeatedly to obtain a distribution of bitstrings and an expectation value for the cost function. This value is returned to the classical optimiser, which updates the angles until convergence is reached.}
    \label{fig:VQE}
\end{figure}

Circuit diagrams are a standard way to visualise a quantum algorithm applied to a collection of qubits. Each line represents a different qubit, while the boxes represent quantum gates. Like sheet music, a circuit diagram is read from left to right, with the horizontal axis indicating the ordering of operations, although spacing does not generally represent physical execution time. An example for the QAOA is provided in Figure~\ref{fig:VQE}. Transport researchers who employ quantum algorithms need not construct such circuits gate by gate. As discussed in Section~\ref{sec:pipeline}, quantum software frameworks can generate abstract circuits from higher-level algorithmic descriptions and then compile those circuits into hardware-compatible forms for execution on a target device. In the following, we provide a high-level description of several quantum algorithms, including hybrid variational quantum algorithms (VQAs), that recur in the transport literature.

\paragraph{Quantum algorithms}

The original and ongoing impetus underlying the pursuit of quantum computing is the practical realisation of quantum algorithms with provably lower computational complexity than their best-known classical counterparts. A summary of several key algorithms, their time complexity, and the largest demonstration to date is provided in Table~\ref{tab:algorithms}. Three important examples are the quantum Fourier transform (QFT), quantum phase estimation (QPE), and Grover's algorithm. The QFT performs the discrete Fourier transform using exponentially fewer elementary operations than classical algorithms; QPE estimates eigenvalues with a quadratic improvement in precision scaling; and Grover's algorithm searches an unstructured space using $O(\sqrt{N})$ rather than $O(N)$ function calls. These routines can be adapted and combined within more specialised algorithms. For example, Shor's algorithm uses the QFT and QPE to factorise integers exponentially faster than any known classical algorithm, while the Harrow--Hassidim--Lloyd algorithm uses related subroutines to solve certain sparse, well-conditioned linear systems. For optimisation, the Dürr and Høyer method uses Grover's algorithm to find the index $x$ that gives the smallest value $f(x)$ in an unsorted database of size $N$ in $O(\sqrt{N})$ time.

It is important to emphasise the caveats required to obtain these advantages in practice. Algorithms such as Grover's require a quantum \textit{oracle}: a ``black-box'' function which must be encoded entirely within the quantum circuit. For example, Grover's algorithm could in principle search over traffic-signal parameters using fewer evaluations of a microscopic transport simulator than a classical unstructured search. In practice, however, the simulator would itself need to be implemented coherently as a quantum oracle within the quantum computer. This is an immense challenge, and the costs of encoding the oracle may overwhelm the advantage obtained from reducing the number of function calls.

A further caveat is the absence of scalable and fault-tolerant quantum random access memory (QRAM). In the worst case, $O(N)$ gate operations are required to encode an arbitrary superposition over $N$ classical data values, while measurement returns only a sample from the output state. These input and output costs can erase the apparent advantage of an isolated quantum subroutine. For example, consider using the QFT to extract the frequency spectrum of a classically stored time series with $N$ samples. Loading the data into amplitudes requires $O(N)$ gates and the circuit must then be repeated and measured many times to reconstruct the output distribution. With these overheads, the overall procedure can scale as $O(N^2)$ up to polylogarithmic factors, compared with $O(N\log N)$ for the classical fast Fourier transform. Similar state-preparation difficulties arise for the HHL algorithm and some forms of quantum machine learning\cite{jaques_qram_2025,aaronson_read_2015}.

\paragraph{Hybrid algorithms}
\label{par:HA}

As summarised in Figure~\ref{fig:algo_space}, hybrid quantum-classical algorithms couple gate-based quantum computation and classical computation within a single workflow. We divide these hybrid algorithms into two different categories. The first embeds a quantum algorithm or subroutine within a classical method to accelerate a specific computational bottleneck, while other tasks such as decomposition, constraint handling, coordination, and post-processing may remain classical. The second approach comprises VQAs. These use a variational loop in which quantum and classical processors repeatedly exchange information. VQAs are principally used for heuristic optimisation and constitute the dominant gate-based approach in the quantum transport literature. This is principally because their relatively shallow quantum circuits are more compatible with NISQ hardware.

The core idea underlying VQAs is an iterative loop between a classical optimiser and a quantum processor. The classical optimiser selects a set of continuous parameters \(\boldsymbol{\theta}\) (termed angles) which are passed to a so-called parameterised quantum circuit (PQC). The PQC consists of one- and two-qubit gates which are controlled by \(\boldsymbol{\theta}\). Although the circuit contains many individual gates, their combined action can be represented by a single parameter-dependent unitary operator, denoted \(U(\boldsymbol{\theta})\). If an \(n\)-qubit quantum register is initialised in the all-zero state \(\lvert 0\rangle=\lvert 0\ldots 0\rangle\), applying the circuit prepares the state
\[
    \lvert\psi(\boldsymbol{\theta})\rangle
    = U(\boldsymbol{\theta})\lvert 0\rangle.
\]
Reading this equation from right to left, \(U(\boldsymbol{\theta})\) acts on the initial state \(\lvert 0\rangle\) to produce the parameterised state \(\lvert\psi(\boldsymbol{\theta})\rangle\). The fixed structure of the PQC is termed the \textit{ansatz}. It determines the family of quantum states that the VQA can explore, while \(\boldsymbol{\theta}\) specifies a particular state within that family. The quantum processor executes and measures the circuit to estimate an objective value. This value is returned to the classical optimiser, which then updates the parameters \(\boldsymbol{\theta}\) accordingly. The loop continues until a termination criterion is reached. This is usually when the objective function reaches a local minimum.

Two important types of VQAs are the variational quantum eigensolver (VQE) and the QAOA. The VQE uses a problem-inspired or hardware-efficient ansatz to minimise the expectation value of a Hamiltonian and thereby approximate its eigenvalues and eigenstates. It was originally designed for quantum chemistry, but can also be applied to optimisation problems. In contrast, the QAOA is designed specifically for combinatorial optimisation problems and employs a prescribed ansatz that alternates between cost and mixing operations\cite{farhi_quantum_2014}. Its objective is to concentrate measurement probability on high-quality bitstrings. QAOA is the most widely studied VQA in the transport literature and is most commonly applied to quadratic unconstrained binary optimisation (QUBO) formulations. Given their importance to the literature, the QUBO representation and the detailed operation of QAOA are introduced below.

\subparagraph{QUBO}

Quadratic unconstrained binary optimisation represents a class of problems which are well-suited for both gate-based quantum computers and quantum annealers. Due to their versatility in describing many transport-related phenomena, the majority of transport studies applying quantum technology focus on solving QUBO problems. These problems seek to find the global minimum (or maximum) of a cost function expressed in terms of binary variables up to second order. Namely,
\begin{equation}
\label{QUBO}
C(x)=h^\top x+x^\top Qx,
\qquad x_i\in\{0,1\},
\end{equation}
where $h$ is the linear cost vector and $Q$ is the quadratic cost matrix. A constant may also be added without affecting the location of the global minimum. The formulation is termed \emph{unconstrained} because it imposes no explicit feasibility conditions beyond binarity. However, linear equality constraints $A_k^\top x=b_k$ can be incorporated through quadratic penalties,
\begin{equation}
\label{QUBO_penalty}
C(x)=h^\top x+x^\top Qx+
\sum_k\lambda_k\left(A_k^\top x-b_k\right)^2,
\end{equation}
where sufficiently large $\lambda_k>0$ make constraint violations more expensive than improvements to the original objective. Inequality constraints may similarly be represented using slack variables. In general, QUBO problems are NP-hard, with no known classical algorithm able to find the exact solution in polynomial time.

To apply QAOA or quantum annealing, the binary variables are replaced by spin variables through $\sigma_i=1-2x_i$. The QUBO can then be rewritten, up to a constant offset, through the following operator
\begin{equation}
\label{H}
H(\sigma)=C_0+\sum_i w_i\sigma_i+
\sum_{i,j}J_{ij}\sigma_i\sigma_j.
\end{equation}
On a quantum computer, each $\sigma_i$ is represented by a Pauli-$Z$ operator acting on qubit $i$, whose eigenvalues $+1$ and $-1$ encode the two possible spin values. Equation~\eqref{H} is a matrix termed a ``Hamiltonian'' in the quantum computing literature. The computational basis states are eigenstates of $H$, and solving the original QUBO is equivalent to finding the state with the lowest eigenvalue, termed the ground state.

\subparagraph{QAOA}

QAOA is a heuristic search algorithm for QUBO problems which iterates between classical and quantum subroutines
\cite{farhi_quantum_2014}. When solved using classical computing, researchers often adopt general heuristic solvers such as simulated annealing to approximate their global minima. QAOA may be conceptualised similarly, as a general-purpose heuristic solver which instead leverages quantum computation. The iterative structure for the QAOA algorithm is presented in Figure~\ref{fig:VQE}. Given a set of angles, the quantum subroutine prepares a superposition over candidate bitstrings. Repeatedly measuring the circuit therefore produces a distribution of bitstrings derived from the quantum amplitudes. From this distribution one calculates an expectation value for the QUBO cost function. The classical subroutine uses this value to update the angles, and the process repeats until convergence. In essence, the quantum subroutine defines a mapping from the angles to a probability distribution over bitstrings, while the classical subroutine searches this parameterised cost landscape for a local optimum.

The standard QAOA ansatz begins by applying a Hadamard gate to each qubit to produce an equal superposition over all bitstrings. It then applies $p$ alternating cost and mixing unitaries, each controlled by a different angle. The cost unitary assigns each bitstring a phase determined by its QUBO objective value, while the mixing unitary redistributes amplitude between nearby bitstrings. Implementing the cost unitary can become expensive for dense QUBO formulations because each non-zero quadratic term $Q_{ij}$ generally requires a two-qubit $R_{ZZ}$ interaction. Increasing the layer depth $p$ makes the ansatz more expressive, but also increases the number of gates, the number of parameters, and the circuit's sensitivity to noise. For the shallow depths practical on NISQ hardware, QAOA is therefore best viewed as a parameterised heuristic rather than an exact solver.

The choice of classical optimiser also has critical importance. QAOA landscapes may contain numerous local minima, and different gradient-based or derivative-free optimisers can produce substantially different results. Methods such as COBYLA and Nelder--Mead are commonly adopted, although their relative performance is problem dependent and is considered later in the literature review\cite{palackal_quantum-assisted_2023}. Optimisation can become increasingly difficult as the number of qubits and layers grows, particularly when the landscape develops broad regions with vanishing gradients known as barren plateaus\cite{mcclean_barren_2018,larocca_barren_2025}. There are multiple variations of QAOA designed to improve performance. One notable example replaces penalty terms with a constraint-preserving mixer\cite{hadfield_quantum_2019}. The register is initialised in the feasible subspace and the mixer is designed to prevent transitions into infeasible states. This avoids tuning penalty weights and may simplify the cost landscape, but typically requires more complicated state preparation and gate sequences.

\subparagraph{Quantum machine learning}

Quantum machine learning (QML) encompasses a broad collection of techniques which synthesise machine learning and gate-based quantum computing\cite{wang_comprehensive_2024, peral-garcia_systematic_2024}. Several of these are summarised in Figure~\ref{fig:algo_space}, but here we focus on two methods which occur repeatedly in the transport literature. These are so-called $U U^\dagger$ classifiers and quantum neural networks (QNNs). Both are compatible with small-scale and near-term implementations, although their overall resource requirements depend strongly on how one encodes the classical data.

Firstly, the $U U^\dagger$ approach is a quantum similarity method that compares a test input with one or more reference patterns encoded as quantum states. In its basic form, prototype feature vectors are first computed classically and then mapped to quantum states through a sequence of unitary gates $U_R$. That is, $\lvert\psi_R\rangle=U_R\lvert 0\rangle$. A separate set of gates, $U_T$, are then used to prepare the test input vectors as $\lvert\psi_T\rangle=U_T\lvert 0\rangle$. These two unitaries are combined to prepare the circuit $U_T^\dagger U_R$. The whole $U U^\dagger$ approach then amounts to calculating the probabilities
$ \left|\langle 0|U_T^\dagger U_R|0\rangle\right|^2 = \left|\langle\psi_T|\psi_R\rangle\right|^2.$
This quantity is the squared overlap between the test and reference states and provides a measure of their similarity. In effect, the quantum device estimates the squared magnitude of a classical dot product. Because only shallow circuits are required and no variational training loop is involved, $U U^\dagger$ classifiers have low quantum resource requirements.

Throughout the transport literature, QNNs are conceptualised as a type of VQA with a specially designed ansatz for their target application. Classical data are encoded into quantum states (e.g., using angle or amplitude encoding) and then processed by a sequence of trainable quantum gates in a PQC. Measurement produces classical outputs and a classical optimiser updates the gate parameters to minimise an empirical loss. In practice, QNNs function as hybrid layers that can be integrated into classical models. They provide a non-classical feature transformation while training and evaluation remain predominantly classical. These methods are attractive because they allow researchers to experiment with quantum layers without redesigning entire modelling pipelines or relying on large-scale hardware. The QNN layers themselves are appealing because they can generate rich, nonlinear feature maps with relatively few trainable parameters. This is because quantum circuits acting on encoded inputs effectively perform high-order trigonometric transformations of the data. However, QNNs also face drawbacks, including the overhead of data encoding in the absence of practical QRAM and well-documented trainability challenges such as barren plateaus.

\subsubsection{Performance}
\label{sssec:gate_perf}

\paragraph{Quantum algorithms}

The performance of NISQ-era quantum computers in executing quantum algorithms is extremely limited. Simply put, there are not enough reliable qubits, quantum gates are too noisy, and there is too much error. Current implementations are therefore largely restricted to small-scale demonstrations for benchmarking purposes, and there has not yet been a demonstration of practical quantum advantage over the best classical devices. Table~\ref{tab:hardware} compares leading hardware platforms, while Table~\ref{tab:algorithms} demonstrates the small problem sizes reached by many canonical algorithms. Note that relatively simpler quantum algorithms such as the QFT have been performed for up to 52 qubits, whereas more challenging algorithms such as Grover's algorithm are limited to searches of toy instances up to 8 elements.

Current capabilities are rapidly improving as hardware developers scale device size, lower gate errors, and pursue robust logical qubits through error correction. As demonstrated in Table~\ref{tab:algorithms}, their published road maps are extremely optimistic, with many predicting hundreds or more logical qubits by 2030. Quantum capabilities have undoubtedly increased enormously over the past decade, and significant further improvements are reasonable to expect. However, it remains unclear whether near-term devices will be capable of executing quantum algorithms for industrially relevant transport problems with any practical advantage.

\paragraph{Hybrid algorithms}

VQAs present one possible route towards an early demonstration of practical quantum advantage, but no such algorithm has yet achieved a lower time-to-solution than the best classical methods for an optimisation problem. A more relevant question in the NISQ era is therefore whether their performance scales more favourably with problem size. Simulation and analytical studies provide some evidence of limited scaling advantage for particular problems, although VQAs are generally not expected to provide more than a polynomial advantage over the best classical algorithms\cite{aaronson_bqp_2010}.

The strongest evidence is confined to specialised problem classes and strong assumptions. Studies of the max-cut problem have produced analytical performance bounds and tentative numerical evidence for improved scaling, but these results are sensitive to simplified noise models and extrapolation from small instances\cite{guerreschi_qaoa_2019,blekos_review_2024}. More recently, Boulebnane and Montanaro~\cite{boulebnane_solving_2024} demonstrated a polynomial QAOA speedup over the best-known classical algorithms for random instances of 8-SAT at sufficiently large depth. Critically, this result assumes noiseless execution and optimal angles, thereby removing the classical parameter-optimisation problem encountered in a typical QUBO application. Claims of advantage must therefore be assessed by carefully reading the fine print\cite{aaronson_read_2015}.

Caution is also required before generalising these results to transport engineering. Gate errors, measurement noise, decoherence, and barren plateaus can all destroy a prospective scaling advantage\cite{stilck_franca_limitations_2021,de_palma_limitations_2023,scriva_challenges_2024,wang_noise-induced_2021}. Moreover, even a fault-tolerant implementation of QAOA may be outperformed by classical algorithms if the underlying QUBO formulation is inefficient. This concern is particularly relevant for routing problems, and a recent review argues that QUBO formulations of the travelling salesman problem may be fundamentally uncompetitive with state-of-the-art classical methods\cite{smith-miles_travelling_2025}. In summary, VQAs such as QAOA may provide useful advantages for particular structures, but further work is needed to identify transport problems for which a clear and defensible quantum advantage is plausible. Similar considerations arise for quantum annealing.

\begin{landscape}
\begingroup
\small
\setlength{\tabcolsep}{1.8pt}
\setlength{\arrayrulewidth}{0.45pt}
\renewcommand{\arraystretch}{1.20}
\begin{longtable}{@{}|L{2.45cm}|L{2.65cm}|L{2.15cm}|L{2.05cm}|L{3.15cm}|L{4.10cm}|L{6.25cm}|C{1.15cm}|@{}}
\caption{Current gate-model systems from seven quantum-computing providers. Qubit counts use the provider's label for programmable,
computational, or physical qubits in the named system, but definitions may differ.
Gate fidelities are representative reported 1Q and 2Q values, but measurement methods may differ.  Quality benchmarks are provider-specific and are
not directly comparable.  Roadmaps are company targets.
``Public'' means ordinary-user paid access to the named system; ``Gated'' means
application-, sales- or private-beta access.}\label{tab:hardware}\\
\hline
\textbf{Provider} & \textbf{Architecture} & \shortstack[l]{\textbf{Leading}\\\textbf{hardware}\\\textbf{system}} &
\textbf{Qubit count} & \textbf{Gate fidelity} &
\shortstack[l]{\textbf{Qubit quality}\\\textbf{benchmark}} & \textbf{Roadmap to 2030} &
\textbf{Public cloud access}\\
\hline
\endfirsthead
\caption[]{Continued.}\\
\hline
\textbf{Provider} & \textbf{Architecture} & \shortstack[l]{\textbf{Leading}\\\textbf{hardware}\\\textbf{system}} &
\textbf{Qubit count} & \textbf{Gate fidelity} &
\shortstack[l]{\textbf{Qubit quality}\\\textbf{benchmark}} & \textbf{Roadmap to 2030} &
\textbf{Public cloud access}\\
\hline
\endhead
\multicolumn{8}{|r|}{\emph{Continued on next page}}\\
\hline
\endfoot
\hline
\endlastfoot

IBM\cellcite{IBMQuantumHardware2026,IBMBostonCompute2026,GambettaIBMBostonQV2026} &
Superconducting transmons &
Heron r3 &
156 &
1Q 99.98\%; 2Q 99.83\% &
Quantum Volume: 12 &
2029: 200 logical qubits; $10^8$ gates &
\public\\
\hline

Google\cellcite{GoogleWillowNature2025,ChenDevoret2025WillowHardware,SivakEtAl2026RLQEC,Neven2026QuantumModalities} &
Superconducting transmons &
Willow &
105 &
1Q 99.97\%; 2Q 99.88\% &
Logical error per cycle: 0.077\% (distance 7) &
By 2030: commercially relevant quantum computer &
\gated\\
\hline

Rigetti\cellcite{AWSCepheus2026,Rigetti10Q2026Q2,PelofskeEtAl2022QV,RigettiRoadmapAug2026} &
Superconducting transmons &
Cepheus-1 &
108 &
1Q 99.9\%; CZ 99.1\% &
Quantum Volume 8 (Aspen-M-1, 2022) &
2029+: 1,000+ physical qubits; 99.9\% median 2Q fidelity &
\public\\
\hline

IonQ\cellcite{ionqCompare2026,ionqRoadmap2026} &
Trapped ions &
Forte &
36 &
1Q 99.98\%; 2Q 99.6\% &
36 algorithmic qubits (\#AQ) &
2030: 2 million physical; 80,000 logical qubits &
\public\\
\hline

Quantinuum\newline{\scriptsize\cite{Ransford2026Helios,Quantinuum424B4_2026,quantinuum_helios_2025}} &
Trapped ions &
Helios &
98 &
1Q 99.9975\%; 2Q 99.921\% &
48 error-corrected logical qubits  &
2029: hundreds of logical qubits &
\gated\\
\hline

QuEra\cellcite{quera_computing_gemini-class_2025,BluvsteinEtAl2026FaultTolerant,QuEraRoadmap2026} &
Neutral atoms &
Gemini &
260 &
1Q 99.9\%; 2Q 99.2\% &
Up to 96 logical qubits active simultaneously  &
2028--29: $>20{,}000$ physical; 1,000+ logical qubits &
\gated\\
\hline

Infleqtion\cellcite{infleqtion2026product,infleqtion2026nqcc,rines2026sqale} &
Neutral atoms &
Sqale &
100 &
1Q 99.98\%; 2Q 98.8\% &
12 error-detecting logical qubits  &
2030: 100,000 physical; 1,000+ logical qubits &
\gated\\
\hline
\end{longtable}
\endgroup

\clearpage
\begingroup
\small
\setlength{\tabcolsep}{1.5pt}
\setlength{\arrayrulewidth}{0.45pt}
\renewcommand{\arraystretch}{1.20}
\begin{longtable}{@{}|L{2.65cm}|L{4.95cm}|L{3.55cm}|L{4.40cm}|L{3.50cm}|L{5.30cm}|@{}}
\caption{Algorithmic comparisons and selected experimental demonstrations. Scalings use standard theoretical models and omit implementation overheads such as state preparation, oracle construction, error correction and full readout.  Symbols follow the notation of the cited results: \(N\) denotes input size in the QFT, Grover, HHL, and LABS rows. \(n\) denotes the number of Boolean variables (and hence qubits) in the Max-Cut and 8-SAT rows.
$M$ is the integer factored, $d$ is dimension, $p$ is QAOA depth, $\epsilon$ is precision, and $\kappa$ is condition number.  $\widetilde O$ suppresses logarithmic factors.}
\label{tab:algorithms}\\
\hline
\shortstack[l]{\textbf{Quantum}\\\textbf{algorithm}} & \textbf{Description} & \textbf{Quantum scaling} &
\shortstack[l]{\textbf{Task-matched}\\\textbf{classical comparator}} & \textbf{Quantum speedup} &
\shortstack[l]{\textbf{Selected experimental}\\\textbf{demonstration}}\\
\hline
\endfirsthead

\caption[]{Continued.}\\
\hline
\shortstack[l]{\textbf{Quantum}\\\textbf{algorithm}} & \textbf{Description} & \textbf{Quantum scaling} &
\shortstack[l]{\textbf{Task-matched}\\\textbf{classical comparator}} & \textbf{Quantum speedup} &
\shortstack[l]{\textbf{Selected experimental}\\\textbf{demonstration}}\\
\hline
\endhead

\multicolumn{6}{|r|}{\emph{Continued on next page}}\\
\hline
\endfoot

\endlastfoot

Quantum Fourier transform (QFT)\cellcite{CleveWatrous2000} &
Fourier transform of amplitudes stored in a quantum state &
$O[(\log N)^2]$ gates &
FFT: $O(N\log N)$ &
Conditional exponential &
Approximate QFT on 100 qubits\cellcite{CooteBiercukBaum2026}\\
\hline

Quantum phase estimation (QPE)\cellcite{kitaev_quantum_1995} &
Estimate an eigenvalue (phase) of a unitary operation &
$O(1/\epsilon)$ total evolution time &
$O(1/\epsilon^2)$ independent measurements &
Quadratic over independent measurements &
Error-corrected H$_2$-energy QPE on Quantinuum H2-2\cellcite{YamamotoEtAl2026QECQPE}\\
\hline

Grover search\cellcite{grover_fast_1996} &
Search an unstructured set of $N$ items &
$\Theta(\sqrt N)$ oracle calls &
$\Theta(N)$ oracle calls &
Quadratic &
$N=32$ search with better-than-classical success probability on IBM hardware\cellcite{pokharel_better-than-classical_2024}\\
\hline

Shor factoring\cellcite{shor_polynomial-time_1997} &
Prime factorisation &
$\widetilde O[(\log M)^2]$ logical gates &
Subexponential number-field sieve &
Superpolynomial over the best-known classical method (theoretical) &
Compiled five-qubit factorisation of 15 ($3\!\times\!5$) on IBM  device\cellcite{amico_experimental_2019}\\
\hline

Jordan gradient estimation\cellcite{jordan_fast_2005} &
Estimate a multivariable gradient &
One function query for all $d$ components &
$d+1=O(d)$ function queries &
Linear query reduction &
One-dimensional four-spin NMR demonstration\cellcite{MaheshSuter2006Gradient}\\
\hline

Harrow--Hassidim--Lloyd (HHL)\cellcite{harrow_quantum_2009} &
Solve a sparse, well-conditioned linear system and return the solution as a quantum state &
$\widetilde O(\kappa^2\log N/\epsilon)$ &
$O[N\kappa\log(1/\epsilon)]$ for a comparable sparse solver &
Exponential in $N$ under stated conditions &
Fault-tolerant HHL for a $4\!\times\!4$ Poisson system on five logical qubits (48 physical qubits used)\cellcite{PerlinEtAl2026FTHHL}\\
\hline

QAOA for Max-Cut\cellcite{guerreschi_qaoa_2019} &
Split a graph into two sets while maximising edges between them &
$p=4$: $\sim\!\exp(0.0141n)$ (empirical runtime fit) &
AKMAXSAT: $\sim\!\exp(0.0409n)$ (empirical runtime fit) &
Projected crossover only, not demonstrated &
Modified $p=1$ QAOA pipeline for 156-node Max-Cut\cellcite{sachdeva_quantum_2024}\\
\hline

QAOA for random 8-SAT\cellcite{boulebnane_solving_2024} &
Find a satisfying assignment &
$\approx2^{0.30n}$ at $p=60$ &
WalkSATlm: $\approx2^{0.33n}$ &
Projected scaling improvement &
\NA\\
\hline

QAOA for LABS\cellcite{shaydulin_evidence_2024} &
Find binary sequences with minimal self-correlation &
QAOA $1.46^N$; with quantum minimum finding $1.21^N$ ($p=12$) &
Memetic tabu search: $1.34^N$ &
Projected with quantum minimum finding; QAOA alone is slower &
18 qubits at $p=1$\\
\hline

\end{longtable}

\vspace{-0.7em}
\noindent\begin{minipage}{0.98\linewidth}
\end{minipage}
\endgroup
\clearpage
\end{landscape}

\subsection{Annealers}

\subsubsection{Operating principle}

Quantum annealing is a specialised quantum computing paradigm designed for solving QUBO problems and related Ising Hamiltonians\cite{albash_adiabatic_2018}. Unlike gate-based quantum computers, current quantum annealers are not generally considered universal and are therefore not suited to implementing arbitrary algorithms such as Shor's, Grover's, or the QAOA. Rather than manipulating qubits through a sequence of discrete gates, quantum annealers operate as analogue devices which physically realise a problem Hamiltonian whose ground state encodes the QUBO objective. That is, the linear and quadratic terms of the cost function are programmed as physical parameters in the hardware, and the aim is to obtain the bitstring which minimises this function.

The annealing process begins from a simple Hamiltonian with a known ground state which is easy to prepare. The system is then gradually transformed into the final problem Hamiltonian representing the desired QUBO. If this transformation is sufficiently slow, the qubit register has a high probability of remaining near the ground state and hence returning a low-cost solution upon measurement. If it is too rapid, the system is more likely to finish in a higher-energy state corresponding to a non-optimal solution. The annealing time may therefore require tuning for each problem instance. As with gate-based quantum computers, readout is probabilistic and the procedure is repeated many times to produce a distribution of candidate bitstrings. Similar to QAOA and other VQAs, quantum annealing does not provide a guarantee of optimality and should be conceptualised as a heuristic method.

D-Wave Systems currently provides the only widely available cloud-based quantum annealing service, and its devices feature prominently throughout the transport literature. Their current state-of-the-art Advantage2 system contains approximately 4,400 physical qubits and more than 40,000 couplers \cite{d-wave_quantum_inc_performance_2025}. However, the physical qubit count does not directly indicate the number of logical variables which can be represented. Each qubit can couple directly to only a limited number of others, whereas a dense QUBO matrix may require interactions between many pairs of variables. If the connectivity required by the QUBO matrix exceeds that of the hardware, a procedure termed ``minor embedding'' is used. Multiple connected physical qubits are grouped into a chain which collectively represents a single logical variable. This embedding overhead can dramatically reduce the number of logical variables which can be encoded directly on the annealer, particularly for dense QUBO matrices. Embedding may also require significant computational time to perform optimally.

D-Wave also offers a hybrid annealing service which partially overcomes these connectivity and resource limitations. Classical algorithms divide a larger optimisation problem into smaller subproblems which can be embedded and solved on the quantum processing unit. The resulting solutions are then combined and improved through further classical processing. This enables much larger QUBO problems to be addressed than could be directly embedded on the annealer alone, with D-Wave previously reporting support for up to $10^6$ sparse or $2\times10^4$ fully connected variables \cite{mcgeoch_d-wave_2020}. These values describe the capacity of the overall hybrid solver and should not be interpreted as the number of variables solved simultaneously through quantum annealing. Constraints may also be included and are handled through additional classical processing. Accordingly, the hybrid service should be regarded as a proprietary heuristic optimisation pipeline which combines classical and quantum processors. Its runtime and solution quality therefore reflect the complete classical-quantum workflow rather than the quantum annealing subroutine alone.

\subsubsection{Performance}

Solution quality and compute speed are the two critical factors which define the practical performance of quantum annealers. For the transport engineer, annealers provide utility only if they obtain solutions of sufficient quality within a shorter time-frame than competing classical algorithms. A common benchmarking metric is the time-to-solution (TTS), defined as the expected time required to obtain the optimal solution at least once with 99\% probability:
\begin{equation}
\label{TTS}
\text{TTS} =T\frac{\ln(1-0.99)}{\ln(1-p)},
\end{equation}
where $T$ is the time required for each annealing run and $p$ is the probability of finding the optimal solution in a single run \cite{ronnow_defining_2014}. In benchmarking studies, $p$ is estimated by performing many runs and calculating the proportion which return the known optimum. Because the annealing time affects both $T$ and $p$, it should be optimised when comparing quantum and classical methods.

To be clear, no quantum annealer has demonstrated a lower TTS than the best-known classical algorithm for a general QUBO problem. Several studies have shown that quantum annealers can outperform simulated annealing for selected benchmark problems \cite{albash_demonstration_2018,king_scaling_2021,king_quantum_2023}. However, simulated annealing is only one classical heuristic and is not necessarily representative of the best available classical method. No work has yet demonstrated an unambiguous general scaling advantage against state-of-the-art classical algorithms \cite{ronnow_defining_2014,mandra_pitfalls_2017,mandra_deceptive_2018,kowalsky_3-regular_2022,andrist_hardness_2023}.

The focus on finding the exact optimum may also be overly restrictive. Quantum annealers are heuristic solvers and may rarely return the optimal solution for large and difficult instances. For practical transport applications, the time required to obtain a solution within an acceptable optimality gap may therefore be more meaningful than the time required to recover the exact optimum. Munoz-Bauza and Lidar~\cite{munoz-bauza_scaling_2025} recently considered the time required to obtain a solution within $\epsilon$\% of optimality by redefining the success probability $p$ in equation~\eqref{TTS}. For a class of spin-glass problems, they found empirical evidence that quantum annealing had superior finite-size scaling to a state-of-the-art classical solver for optimality gaps above approximately $1$\%. However, these results concern problem instances with little direct relation to transport engineering and do not guarantee an asymptotic scaling advantage as the problem size approaches infinity.

The current absence of quantum advantage should not dissuade transport researchers from exploring annealing applications, but it does warrant caution. Future hardware will undoubtedly improve in scale, connectivity, and accuracy, and even a polynomial improvement in runtime could provide substantial practical benefits. However, the ultimate performance of annealers will depend strongly on the structure of the problem, the density of its QUBO formulation, the overhead introduced by minor embedding, and the strength of competing classical algorithms. Claims of advantage must therefore compare against state-of-the-art classical methods and account for the complete workflow, including embedding, parameter selection, repeated sampling, and classical pre- and post-processing.

\subsection{Critical summary}

Quantum hardware has now entered a commercial stage with multiple providers offering on-demand cloud access to small-scale gate-based processors and quantum annealers. Transport researchers can undertake proof-of-concept studies on these platforms to explore how different algorithms behave on real devices. A critical summary of current hardware and algorithmic performance metrics is presented in Table~\ref{tab:summary}.

On current gate-based hardware, ``late-stage'' quantum algorithms with provable asymptotic advantage over classical methods remain restricted to trivial instance sizes. In contrast, VQAs such as the QAOA can be implemented in small optimisation problems and in some cases achieve optimal or near-optimal solutions. However, there is presently no compelling evidence that these approaches systematically outperform state-of-the-art classical algorithms, and any indications of advantage are confined to carefully chosen or highly structured instances. In comparison, quantum annealers offer much larger raw qubit capacities and can encode small-to-moderate problem instances (especially when used in hybrid workflows), but their long-term prospects for delivering a robust quantum advantage are still theoretically and empirically uncertain. Nevertheless, hardware capabilities have improved rapidly, and expectations for the next 5--10 years are high. Continued progress in coherence times, connectivity, and especially error-correction techniques is likely to expand the scale and realism of transport problems that can be meaningfully tackled on quantum devices. The following section provides guidance for transport researchers looking to exploit these developing quantum capabilities and critically assess claims of quantum advantage.

\begin{table}[ht]
\centering
\caption{Summary of major computing architectures and their current applications and performance.}
\label{tab:summary}
\begin{tabular}{|c|c|c|c|}
\hline
 
  & \multicolumn{2}{c|}{\textbf{Gate-based quantum computer}} 
  & \textbf{Quantum annealer} \\
\hline
\begin{tabular}{c}Typical device size\\(2025)\end{tabular}
  & \multicolumn{2}{c|}{\begin{tabular}{c} 50--1000 physical qubits \\
  Quantum volume $\leq 25$ \end{tabular}}
  & 4,400+ qubits \\
\hline
Algorithm type 
  & Quantum 
  & Hybrid VQAs 
  & Quantum annealing \\
\hline
Application scope 
  & General 
  & Mainly optimisation 
  & Optimisation \\
\hline
\begin{tabular}{c}Expected\\performance\\enhancement\end{tabular}
  & Polynomial to exponential 
  & Likely polynomial 
  & Possibly polynomial \\
\hline
Onset of advantage 
  & \begin{tabular}{c}Long-term\\(10 years +)\end{tabular}
  & \begin{tabular}{c}Intermediate\\(5--10 years?)\end{tabular}
  & \begin{tabular}{c}Intermediate\\(5--10 years?)\end{tabular} \\
\hline
\end{tabular}
\end{table}

\section{Quantum solutions pipeline}
\label{sec:pipeline}

\begin{figure}[!htbp]
    \centering
\includegraphics[width=0.95\linewidth]
    {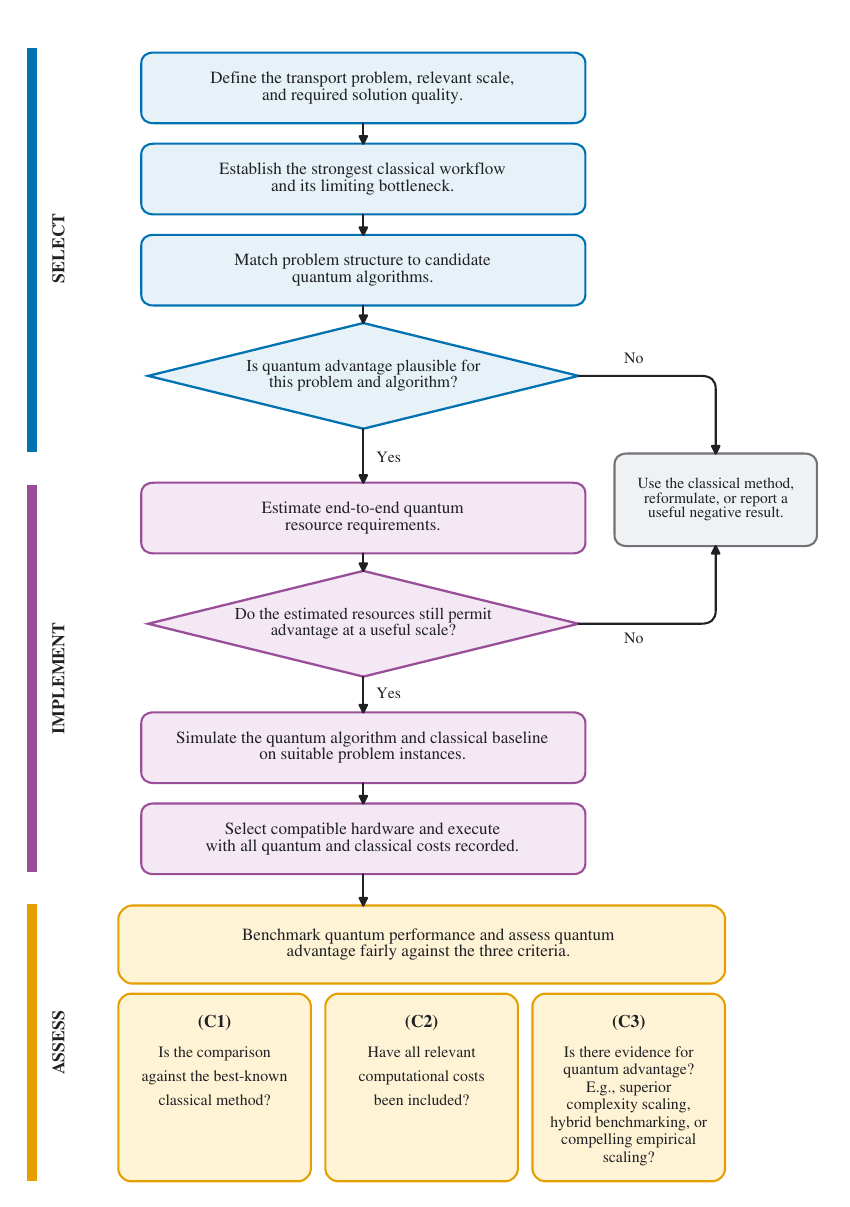}
    \caption{Proposed pipeline for developing and assessing quantum solutions for transport problems. Stage 1 selects a suitable problem and algorithm by identifying the classical computational bottleneck and assessing whether quantum advantage is plausible. Stage 2 estimates the required resources before simulation and hardware execution. Stage 3 benchmarks the complete quantum workflow fairly against the three criteria discussed in Section~\ref{sec:assessing_advantage}. If advantage is no longer plausible at either decision point, the problem formulation or algorithm should be reconsidered.}
    \label{fig:pipeline}
\end{figure}

The functional purpose of a quantum computer is to obtain computational advantage over the best-known classical algorithms for a given problem. As demonstrated above, achieving this objective is difficult due to the small set of known quantum algorithms and the limitations of existing and near-term hardware. These realities constrain the set of transport-related problems which are compatible with quantum acceleration. Indeed, many transport problems may be simply unsuitable for quantum computing and it is therefore important to demarcate use cases where quantum technology can potentially provide benefit. To be clear, no end-to-end practical quantum advantage has yet been demonstrated for a transport application, and transport researchers should not expect existing devices to improve a problem of practical interest. This does not mean that transport applications should be ignored until fault-tolerant hardware exists. The immediate task is to identify problems which may eventually benefit, improve the quantum methods for this problem, and estimate the scale and hardware requirements at which an advantage could become possible. Negative results are also useful where they narrow this problem scope. This section therefore considers a three-stage pipeline: choosing a suitable problem and an appropriate algorithm, implementation through quantum simulation or quantum computation, and then assessing performance fairly. The complete pipeline is presented graphically in Figure~\ref{fig:pipeline}.

\subsection{Choosing a suitable problem and algorithm}
\label{sec:choosing_problem}

There are two general approaches to choosing a suitable problem. The first is to take a problem with a well-motivated quantum formulation, such as Max-Cut using QAOA, and identify where this problem structure occurs within a transport-relevant context. The second approach is to begin with the transport problem and assess whether quantum methods are suitable to overcome the relevant computational bottlenecks. For stage 1 of the pipeline we focus on the second approach, as it is more relevant to the transport researcher and their knowledge domain.

\paragraph*{Establish problem, best classical baseline, and bottleneck} 

The first step is to define the computational task for the transport problem. This includes the practically relevant problem size, the required solution quality and time frame, and whether feasibility or optimality must be certified. Based on these requirements, the transport researcher should establish the state-of-the-art method for solving the computational problem classically. This could include heuristic methods like simulated annealing or tabu search, mixed integer programming (MIP), constraint programming, dynamic programming, etc. The relevant baseline is whichever complete workflow provides the strongest performance for the required transport problem. If classical methods are sufficient to solve the problem to the required accuracy and time frame, then quantum methods are not necessary. 

Otherwise, identify the problem's computational bottleneck: the accuracy, time, memory, or energy constraint which makes the classical solution unsatisfactory. The computational bottleneck will generally arise from the interaction between the problem structure, instance scale, and classical solution method. For example, simulated annealing may struggle when the combinatorial search space grows rapidly, feasible solutions occupy only a small fraction of this space, or the objective landscape has no exploitable structure. Alternatively, a MIP formulation may produce an extremely large branch-and-bound tree due to weak relaxations, symmetry, or constraints which couple decisions across the entire problem. Dynamic programming may become impractical because the number of states (and hence the memory required to store them) grows exponentially. The purpose of this step is to identify precisely why the strongest classical method is inadequate.

\paragraph*{Identify candidate quantum algorithms}

The next step is to identify quantum algorithms which could conceivably address the computational bottleneck. Algorithms should be chosen according to the source of classical difficulty, rather than merely according to whether the problem can be encoded for them. The aim is to target the bottleneck while retaining as much useful problem structure as possible. Table~\ref{tab:bottleneck_algorithm} summarises several broad correspondences between classical bottlenecks and candidate quantum methods. For example, QAOA or quantum annealing may be worth investigating when the bottleneck involves searching through a binary optimisation problem with manageable constraints. Grover-style search or quantum branch-and-bound may instead be relevant where a classical solver has exploited the available problem structure but is still left with an enormous and weakly guided search. Quantum linear algebra or Fourier transform-based methods may be relevant where solving linear systems or performing spectral calculations constitutes a substantial part of the computational workload. Transport researchers can consult the Quantum Algorithm Zoo for further inspiration\cite{jordan_quantum_2025}. This is a broad catalogue of established quantum algorithms and their theoretical speedups.

\begin{table*}[t]
\centering
\caption{Potential quantum methods associated with different classical computational bottlenecks.}
\label{tab:bottleneck_algorithm}
\small
\begingroup
\renewcommand{\arraystretch}{1.25}
\setlength{\tabcolsep}{6pt}
\begin{tabularx}{\linewidth}{
    @{}
    >{\raggedright\arraybackslash}p{0.25\linewidth}
    >{\raggedright\arraybackslash}p{0.25\linewidth}
    Y
    @{}
}
\toprule
\textbf{Classical bottleneck} &
\textbf{Possible quantum direction} &
\textbf{Initial plausibility question} \\
\midrule

Difficult binary quadratic optimisation &
QAOA or quantum annealing &
Is the QUBO representation natural and compact, and can feasibility be imposed without excessive auxiliary variables or penalties? \\

Large residual search or search tree &
Grover-style search or quantum branch-and-bound &
Is classical runtime dominated by the number of candidate solutions or search-tree nodes which must be examined, and would reducing this number materially reduce the total computation? \\ \\

Large, structured linear systems &
HHL or another quantum linear-systems algorithm &
Can the matrix be represented efficiently, is the system sufficiently well-conditioned, and is a compact property of the solution acceptable rather than the complete classical solution vector? \\

Phase or periodicity estimation &
Quantum phase estimation or another QFT-based algorithm &
Can the required result be encoded and recovered as a compact phase or periodicity estimate within a broader quantum algorithm, rather than by using the QFT as a direct replacement for a classical Fourier transform? \\

Exponential dynamic-programming state space &
Quantum dynamic programming or quantum search within the recurrence &
Can the recursive structure be retained, and does the proposed speedup survive the cost of constructing, storing and accessing the required subproblem information? \\

\bottomrule
\end{tabularx}
\endgroup
\end{table*}

\paragraph*{Assess whether quantum advantage is plausible}

The final step in Stage 1 is to assess whether quantum advantage is sufficiently plausible to justify implementation. A hard proof or detailed simulation is not essential at this point; it is sufficient to establish a credible mechanism through which quantum leverage could arise. This is generally easier for quantum algorithms with analytically characterised complexity scaling than for heuristic approaches such as QAOA or quantum annealing. We therefore separate the two cases and provide guidance through reference to real examples.

First consider quantum algorithms with a clear complexity scaling such as those presented in Table~\ref{tab:algorithms}. A crude initial indicator of advantage can be obtained by substituting the quantum algorithm or subroutine into the relevant classical workflow and comparing the resulting complexity against that of the strongest classical method for the same task. Importantly, this comparison must be made for the complete computational problem rather than for the accelerated operation in isolation. Amdahl's law provides a simple illustration. If a fraction $f$ of the classical runtime is accelerated by a factor $S$, the maximum end-to-end speedup is
\begin{equation}
    S_{\mathrm{total}}
    =
    \frac{1}{(1-f)+f/S}.
    \label{eq:amdahl}
\end{equation}
Even if the quantum subroutine were made arbitrarily fast, the total speedup would remain bounded by $1/(1-f)$. Hence, a substantial quantum improvement to one operation may provide little practical value if that operation represents only a small fraction of the complete workflow.

Considerable nuance is also required when interpreting the reported quantum complexity. The theoretical improvement may depend on sparsity, conditioning, precision, oracle access, QRAM, or other restrictive assumptions. There may also be costs which are implicit or absent in the stated complexity, such as state preparation, error correction, measurement and classical post-processing. Moreover, worst-case complexity can also be a poor indicator of performance on real instances. For example, Grover-style quantum search can reduce the number of oracle calls from $O(N)$ to $O(\sqrt{N})$. This may be practically meaningful where the required evaluations are reduced from approximately $10^{12}$ to $10^6$, but may remain useless for a reduction from $10^{100}$ to $10^{50}$. Ultimately, whether the predicted scaling manifests for a relevant problem distribution must eventually be assessed through simulation, analytical resource estimation, or execution on quantum hardware. This empirical assessment is considered in Stage 3 where we discuss ``hybrid benchmarking''.

The following two examples demonstrate good practice in applying this decision process. Firstly, Nannicini~\cite{nannicini} developed quantum subroutines for the simplex method for solving linear programs. This replaces the principal linear-algebra and pivot-selection operations in each simplex iteration with a comparable quantum algorithm. A complexity analysis was then performed, which demonstrated potential quantum advantage for certain instance classes. However, numerical simulations of the quantum-enhanced simplex method were then performed by Ammann~\textit{et al.}~\cite{ammann_realistic_2023} through hybrid benchmarking. These ultimately showed that the quantum runtime improvement did not manifest on practically relevant instances.

In contrast, Wilkening~\textit{et al.}\cite{wilkening_quantum_2025} developed a quantum search algorithm for the 0--1 knapsack problem termed Quantum Tree Generation (QTG). A simple analysis predicts that QTG may provide a polynomial improvement in search complexity. Indeed, hybrid benchmarking showed that QTG could obtain a speedup over the specialised COMBO solver for some instances above 600 variables. The method would also provide a substantial memory benefit. Hence, for both linear programming and the 0-1 knapsack problem, a quantum implementation was worth pursuing based on complexity arguments. However, in the case of the quantum-enhanced simplex algorithm, the favourable scaling did not guarantee quantum advantage.

Another consideration is whether the reported complexity survives the memory model assumed by the quantum algorithm. For example, quantum frameworks have been developed for both exponential-time~\cite{ambainis_quantum_2019} and polynomial-time dynamic programming~\cite{caroppo_quantum_2025}. Caroppo~\textit{et al.}~\cite{caroppo_quantum_2025}, for example, obtain an improved complexity for a quantum version of the Bellman--Ford algorithm, which is routinely used for shortest-path calculations in transport modelling. These frameworks assume that previously computed entries in a dynamic-programming table can be addressed coherently in superposition using QRAM. Consequently, their improved time complexity depends on the ability to construct, store and access this table without introducing an equivalent computational cost elsewhere. While the eventual development of scalable QRAM is not a physical impossibility\cite{jaques_qram_2025}, it should still factor into the plausibility analysis.

It is more difficult to establish whether heuristic quantum methods such as QAOA or quantum annealing could be beneficial. Unlike the algorithms considered above, these do not guarantee a scaling advantage over the strongest classical method. Their performance is instead contingent on the problem structure, instance distribution, formulation, parameter selection, and hardware. The No Free Lunch theorem provides useful context. Averaged across all possible objective functions, improved performance by one optimisation algorithm on one problem class is offset by poorer performance elsewhere\cite{wolpert_no_1997}. Hence, quantum advantage may vary on an instance-by-instance basis and this is challenging to quantify without benchmarking.

However, this ambiguity does not mean that QAOA or annealing is always worth implementing. A problem's compatibility with quantum approaches does not imply plausible quantum advantage. Almost any binary optimisation problem can be converted to QUBO form, and a small instance of almost any QUBO can be executed on a simulator or quantum processor. Neither observation indicates that the resulting method is useful. For QAOA and quantum annealing, QUBO is most attractive where the objective is already binary quadratic (such as Max-Cut and Ising models) or where most constraints describe pairwise conflicts which can be encoded directly in quadratic form. In contrast, routing, scheduling, packing, and network-flow problems often possess flow-conservation, capacity, logical, or recursive structure which are better exploited by specialised classical algorithms. While these problem types are not categorically unsuitable for QUBO methods, they are cumbersome to reproduce using a quadratic penalty model. This leads to a useful preliminary test: do competitive classical solvers for this problem adopt a QUBO formulation?  If not, quantum heuristic methods are likely not competitive either.

The travelling salesperson problem (TSP) provides a clear negative example. While the problem can be recast in QUBO form, a recent review by Smith-Miles~\textit{et al.}\cite{smith-miles_travelling_2025} argues convincingly that this formulation is weak compared against state-of-the-art classical solvers. They contend that quantum methods do not rescue the inefficiency inherent to the QUBO formulation in this context, but otherwise remain open to the possibility for quantum algorithms to accelerate established classical solvers. We argue that the same concern is likely to apply to many other routing contexts, which often possess their own dedicated heuristic formulations which dominate QUBO. In summary, QUBO formulations are not categorically unsuitable for transport problems; however, an otherwise uncompetitive formulation does not become useful just because it is solved using a quantum computer. This point is particularly relevant to the literature reviewed below, in which routing problems dominate amongst quantum-transport studies.

\subsection{Implementation}

The second stage is to implement the quantum algorithm for the transport problem. This proceeds through three interrelated steps: estimating the required computational resources, evaluating the algorithm through analytical methods or simulation, and selecting compatible hardware for execution.

\paragraph*{Resource estimation}

Resource estimation involves determining the approximate number of qubits, gates, and repetitions required to execute the algorithm. The goal of this step is to ascertain which simulation and hardware platform is best suited for the algorithm and whether or not quantum advantage remains plausible. For gate-based algorithms, the main quantities are the number of logical qubits, circuit depth, one- and two-qubit gate counts, and the number of circuit repetitions or shots. For algorithms intended for future fault-tolerant computers, the number and depth of non-Clifford gates should ideally be reported. For near-term implementation, the circuit should instead be compiled to the native gate set and connectivity of a candidate device (discussed below). This compilation may introduce additional gates and substantially increase circuit depth. Note that additional qubits may be required for data encoding, ancillary calculations, constraint evaluation, and uncomputation. 

For QAOA, the required resources are determined by the number of binary variables, the circuit depth $p$, the density of the QUBO matrix, and the number of shots and classical optimisation iterations. A dense QUBO matrix may require considerable routing overhead on hardware with limited qubit connectivity. Increasing $p$ also repeats the cost and mixer layers, potentially improving the solution whilst exposing the state to further noise. Resource estimates should therefore consider the complete variational procedure rather than the cost of a single circuit execution. Quantum annealing requires a different set of estimates. The QUBO problem must first be minor-embedded onto the hardware connectivity graph, and a single logical variable may consequently require a chain of several physical qubits. Relevant quantities therefore include the physical qubit count after embedding, chain lengths, coefficient ranges, chain strength, anneal duration, and number of reads.

Explicit direction for constructing a QUBO formulation and estimating resource costs is provided in Section~\ref{ssec:routing} of the literature review. We demonstrate how to minimise resources and maximise performance through a survey of routing problems. Once this process is complete it is essential to review whether quantum advantage remains plausible. For example, the resource estimate may indicate that the qubit or gate count scales too poorly with problem size for any advantage to remain practically meaningful. For example, a dense QUBO with $n$ variables already contains $O(n^2)$ interactions, so an $O(n^2)$ gate count per QAOA layer may be reasonable. In contrast, a gate count which scales with a much higher polynomial power (or even exponentially) is harder to justify.

\paragraph*{Simulation}

Resource estimates establish whether an implementation is conceivable, but do not show how the algorithm will perform. Before using quantum hardware, the researcher should evaluate the method using analytical calculations, classical simulation, or hybrid benchmarking. These approaches provide different forms of evidence and are useful at different problem scales.

Analytical evaluation is extremely powerful but often challenging to derive. It can provide the success probability, approximation quality, or computational scaling for problem sizes far beyond current hardware, although it is generally limited to particular instance distributions or circuit structures. For example, Boulebnane and Montanaro~\cite{boulebnane_solving_2024} derived analytical bounds for the performance of QAOA on random $k$-SAT instances. For random 8-SAT, QAOA at approximately 14 layers outperformed the scaling of the strongest classical solver tested. Deeper circuits were predicted to obtain more favourable scaling. However, analytical evaluations for QUBO problems are relatively rare and most research relies on small-scale circuit simulations. 

Quantum simulation involves emulating the quantum computer's operation on a classical device. One can simulate the effect of gate operations on a small quantum state and then model the outcome distribution following repeated quantum measurements. This can include contributions from error and quantum decoherence which would occur on real hardware. For gate-based quantum computing there are now many established simulation platforms, including IBM's Qiskit, Google's Cirq, and qsim. Ultimately, state-vector simulations are memory limited and can only handle a few dozen qubits. An $n$-qubit state vector contains $2^n$ complex amplitudes and requires approximately $16 \times 2^n$ bytes when double-precision complex numbers are used. A 30-qubit state therefore requires approximately 16~GB of memory, whilst a 40-qubit state requires approximately 16~TB. Tensor-network and stabiliser methods can simulate larger systems in some circumstances, but performance is circuit-dependent and does not overcome the general exponential scaling. Simulation should initially establish that the encoding and circuit produce the intended result under ideal conditions. Noise can then be introduced if desired. For optimisation problems, useful outputs include the probability of observing the optimal solution, the approximation ratio, fraction of feasible samples, and the distribution of sampled energies.

Hybrid benchmarking provides a powerful intermediate approach between analytical evaluation and simulation. It is particularly well-suited to quantum algorithms which must be evaluated at problem sizes beyond current simulation or hardware limits. Note that this method is not restricted to hybrid quantum--classical algorithms. Instead, the term ``hybrid'' here refers to the combination of classical execution with analytical or simulated estimation of quantum costs. This is achieved through performing the classical workflow and then recording quantities relevant to the target quantum algorithm, such as candidate evaluations or quantum subroutine calls. The recorded operations are then replaced by estimates of the corresponding quantum gate, query, or cycle counts. Cade~\textit{et al.}~\cite{cade_quantifying_2023} pioneered this approach when they considered potential quantum speedups to classical heuristics for the MAX-$k$-SAT problem. They classically simulated hill-climbing heuristics and then replaced their neighbourhood searches with rigorous query estimates for Grover search and quantum maximum finding. Despite having a favourable asymptotic scaling complexity, the inclusion of quantum search did not translate into significant overall speedups for the large instances they studied. Hence, hybrid benchmarking can expose practical limitations which are not apparent from analytical expressions alone.

\paragraph{Hardware execution}

If desired, the simulated quantum algorithms can then be executed on physical hardware. The current access model for quantum hardware is similar to high-performance classical computing. Users submit queued jobs to cloud servers and interact with cloud-hosted quantum devices through established SDKs and compilers. For example, vendor-native stacks include IBM's Qiskit on the IBM Quantum Platform, Rigetti's pyQuil on Quantum Cloud Services, and D-Wave's Ocean tools via the Leap portal. Alternatively, there also exist hyperscaler platforms which aggregate multiple hardware types such as Amazon Braket (IonQ, Rigetti, QuEra, etc.) with a Python SDK and direct OpenQASM 3 submission and Microsoft's Azure Quantum (IonQ, Quantinuum, Atom Computing, etc.) through their Q\# programming language. Cost models also mirror conventional cloud computing, with options for on-demand pay-per-use, subscriptions, and free/community tiers. Billing varies between platforms and may be metered for execution time or per shot.

The choice of hardware depends on user aims, problem scope, and budget. Gate-based computers provide the greatest algorithmic flexibility and can implement both purely quantum and hybrid algorithms. However, current qubit counts, connectivity and noise generally restrict transport-related QUBO demonstrations to small proof-of-principle instances. Quantum annealers can represent substantially larger QUBO formulations. D-Wave's current Advantage2 processor, for example, contains more than 4,400 physical qubits, although the number of logical variables that can be embedded depends strongly on the density and structure of the QUBO. Hybrid D-Wave solvers can accept considerably larger formulations by decomposing the problem across classical and quantum resources. This provides access to more realistic problem scales but practitioners should be forthright about the solver's classical components.

\subsection{Assessing quantum advantage fairly}
\label{sec:assessing_advantage}

As with all computational techniques, the performance of quantum algorithms must be benchmarked fairly. We therefore provide three criteria to guide assessment of quantum advantage: (C1) is the comparison made against the best-known classical method for the same task; (C2) have all relevant computational costs been included; and (C3) is there credible evidence of quantum advantage?

\paragraph*{(C1) Is the comparison against the best-known classical method?}

A quantum speedup is only meaningful relative to a competitive classical baseline. If a quantum algorithm obtains a speedup over an uncompetitive classical method, this is not a compelling comparison. It may establish a theoretical speedup over that particular algorithm, but does not show that the quantum method provides the best way to solve the underlying problem. The appropriate baseline depends on the computational task. An exact quantum algorithm should be compared with the strongest exact classical method, including the time required to certify optimality. A heuristic quantum algorithm should be compared at the same solution quality, feasibility rate, and probability of success.

\paragraph*{(C2) Have all relevant computational costs been included?}

It is vital that any comparison uses equivalent start and end points. Quantum execution time is only one component of the complete workflow. Depending on the algorithm, the total time may include classical preprocessing, data encoding, state preparation, oracle construction, circuit compilation (or QUBO embedding for annealers), quantum execution, repeated measurements, further classical optimisation, error mitigation, and post-processing. One possible decomposition is
\begin{equation}\label{time_costs}
    T_{\mathrm{total}} =
    T_{\mathrm{pre}}+
    T_{\mathrm{encode}}+
    T_{\mathrm{compile}}+
    T_{\mathrm{quantum}}+
    T_{\mathrm{classical}}+
    T_{\mathrm{post}}.
\end{equation}
Each term should include every repetition required to reach the specified solution quality and probability of success. For QAOA, for example, $T_{\mathrm{quantum}}$ includes the execution of all shots across every circuit evaluated during optimisation, whilst $T_{\mathrm{classical}}$ includes the procedure for angle optimisation (e.g., gradient descent). For quantum annealing, the number of reads, programming time, anneal duration, and readout time should be included alongside the cost of finding an embedding. Some costs can be amortised when appropriate. For example, this may apply when re-using a compiled circuit, embedding, or dataset across many problem instances.

\paragraph*{(C3) Is there evidence for quantum advantage?}

Finally, results should be analysed for credible evidence of quantum advantage. This evidence may vary in strength and form depending on the algorithm, problem context, and instance suite. While this review focuses primarily on improvements to runtime, advantage may also manifest through memory or energy. Regarding runtime, compelling evidence may include a combination of superior complexity scaling and numerical simulations, as in the study by Boulebnane and Montanaro~\cite{boulebnane_solving_2024} on boolean satisfiability. To further validate their analytically derived asymptotic scaling law, the authors perform quantum simulations across 10,000 instances. Wilkening~\textit{et al.}~\cite{wilkening_quantum_2025} provide similarly strong results in their hybrid benchmarking of QTG for the 0-1 knapsack problem. The agreement between the finite-size simulations and analytical prediction provides considerably stronger evidence than extrapolation from either result alone. Plausible evidence could include analytic complexity scaling, such as for exponential-time dynamic programming studies by Ambainis~\textit{et al.}\cite{ambainis_quantum_2019} or the TSP on bounded graphs by Moylett\cite{moylett_quantum_2017}. Here the distinction depends on the caveats of data encoding, readout, and practical run times mentioned above.

Finally, potential (but contested) advantage may include functional fitting to empirical results, but only in the case of specific problems following careful analysis. An example is Shaydulin~\textit{et al.}~\cite{shaydulin_evidence_2024} in their study of QAOA for the problem of Low Autocorrelation Binary Sequences (LABS). This is a challenging combinatorial optimisation for which exact solutions are known to $N=66$ and classical heuristics decay above approximately $N=200$. The authors perform simulated experiments up to $N=40$, with empirical fitting showing inferior complexity scaling to the best known classical method of memetic tabu search. However, if QAOA is combined with a quantum minimum finding algorithm (a variant of Grover-type search), the authors argue for a superior run-time advantage based on analytical grounds. Hence, the scaling analysis demonstrates potential for advantage only under speculative conditions. These include that the exponential scaling law holds above $N=40$ and that there is a practical cross-over between quantum and classical run times once the costs of the quantum minimum finding are accounted for.

Many results in the quantum transport literature perform a similar comparison between classical and quantum runtimes. Sometimes a scaling analysis is performed through functional fitting to empirical results. The quantum method is usually a VQA or quantum annealing and is compared to classical simulated annealing, tabu search, or MIP. These results must be interpreted carefully before a credible claim of quantum advantage can be made.  Firstly, NP-hardness alone does not imply that every instance is difficult. If an apparently NP-hard problem exhibits constant or linear runtime over the tested range, this may indicate that the instances are too small, that they do not possess the structure associated with difficult cases, or that a fixed computational overhead dominates the runtime. Benchmarks should ideally include practically relevant instances, recognised or provably hard instance suites, and variation in structural features such as density, constraint tightness, and network topology. 

Secondly, performance on a small number of toy instances is useful for validating an implementation but is rarely sufficient to estimate performance at practical scales. Many (but not all) transport-related optimisation problems that can be represented using fewer than 40 binary variables -- and hence approximately 40 qubits -- can be solved trivially by specialised classical methods. Consequently, the range which can be simulated quantum mechanically may be too small to reveal the scaling of either the quantum or classical algorithm. A convincing benchmark must eventually test instances that bite, and it may therefore be inappropriate to extrapolate solve times based on small instances alone.

\begin{figure}[t]
    \centering
    \includegraphics[width=0.5\linewidth]
    {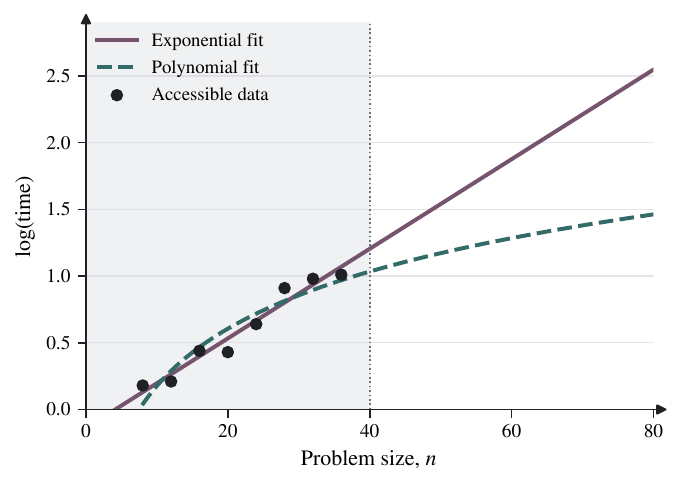}
    \caption{A common pitfall associated with extrapolating from small problem instances. Within the observed range ($n\leq40$, shaded grey), the synthetic data can be fitted reasonably by both an exponential model and a polynomial model. The models diverge substantially outside this range, and hence favourable scaling over small instances does not provide compelling evidence of asymptotic performance. The data are illustrative and do not represent measurements from a particular algorithm or device.
    }
    \label{fig:small_scale_ambiguity}
\end{figure}

Thirdly, any functional form used for extrapolation should be motivated either theoretically or empirically. Alternative forms should be considered when several models describe the observed range equally well. Figure~\ref{fig:small_scale_ambiguity} demonstrates this pitfall when fitting to a small number of data points. Here both polynomial and exponential scaling laws agree nearly perfectly for the instance sizes accessible to quantum simulators or hardware, and hence extrapolation beyond this point would be inappropriate without further information. Once again, Boulebnane and Montanaro~\cite{boulebnane_solving_2024} provide a useful example of good practice in their study of QAOA for random $k$-SAT where simulation and analytical results corroborate.

It is unlikely that current quantum hardware can demonstrate an experimental proof of advantage. However, a projected advantage is an extremely valuable result when supported by compelling scaling evidence or hybrid benchmarking performance. However, we acknowledge that cleanly demonstrating this is challenging, especially for heuristic solvers using QAOA or quantum annealing. Hence, the classification above is not meant to negate the value of proof-of-principle or negative results. The focus is instead on avoiding strong claims of advantage based on preliminary evidence. To this end, we now undertake a rigorous review of the transport literature through the lens of the quantum solutions pipeline. For each study in our corpus, we assess evidence of quantum advantage against the three criteria.

\section{Literature Review: Quantum Computing for Transport Engineering}
\label{sec:transporTpplications}

\subsection{Search Strategy and Study Selection}
\label{ssec:search_strategy}

This review follows the Preferred Reporting Items for Systematic Reviews and Meta-Analyses (PRISMA) 2020 guidelines~\cite{page_prisma_2020}. A systematic search was conducted in the Scopus and IEEE Xplore databases on 22 July 2026. It was not pre-registered and a single reviewer screened records. The search query targeted the intersection of quantum computing and transport research by combining quantum-computing terms with transport-specific terms. The complete query is reproduced in the ``Database search query'' section of the Supplementary Material. In addition to the database search, studies were identified through Google Scholar and forward/backward citation searching during a scoping phase following the systematic search.

Studies were included if they: (i)~applied quantum computing---gate-based, annealing, or hybrid quantum--classical---to a transport engineering problem; (ii)~were published in English between 2017 and 2025; and (iii)~appeared in a peer-reviewed venue (journal article, conference paper, or book chapter). The date of 2017 was chosen to coincide with the advent of cloud-based access to quantum hardware. Four exceptions to (iii) were made for industry-based case studies and preprint technical papers by quantum and transport-related companies. These are included to reflect emerging commercial interest in quantum technologies, but do not change the conclusions drawn from the review. Moreover, we, the authors, declare no affiliations or conflicts of interest. Studies were excluded if they: (a)~used only so-called ``quantum-inspired'' classical algorithms without quantum hardware or simulation; (b)~were review papers, study proposals, errata, or retracted publications; or (c)~were not sufficiently related to the transportation discipline (e.g., addressed warehouse management, battery charging, or inventory without a direct transport component).

\begin{landscape}

\begin{figure}
    \centering
    \includegraphics[width=0.84\linewidth]{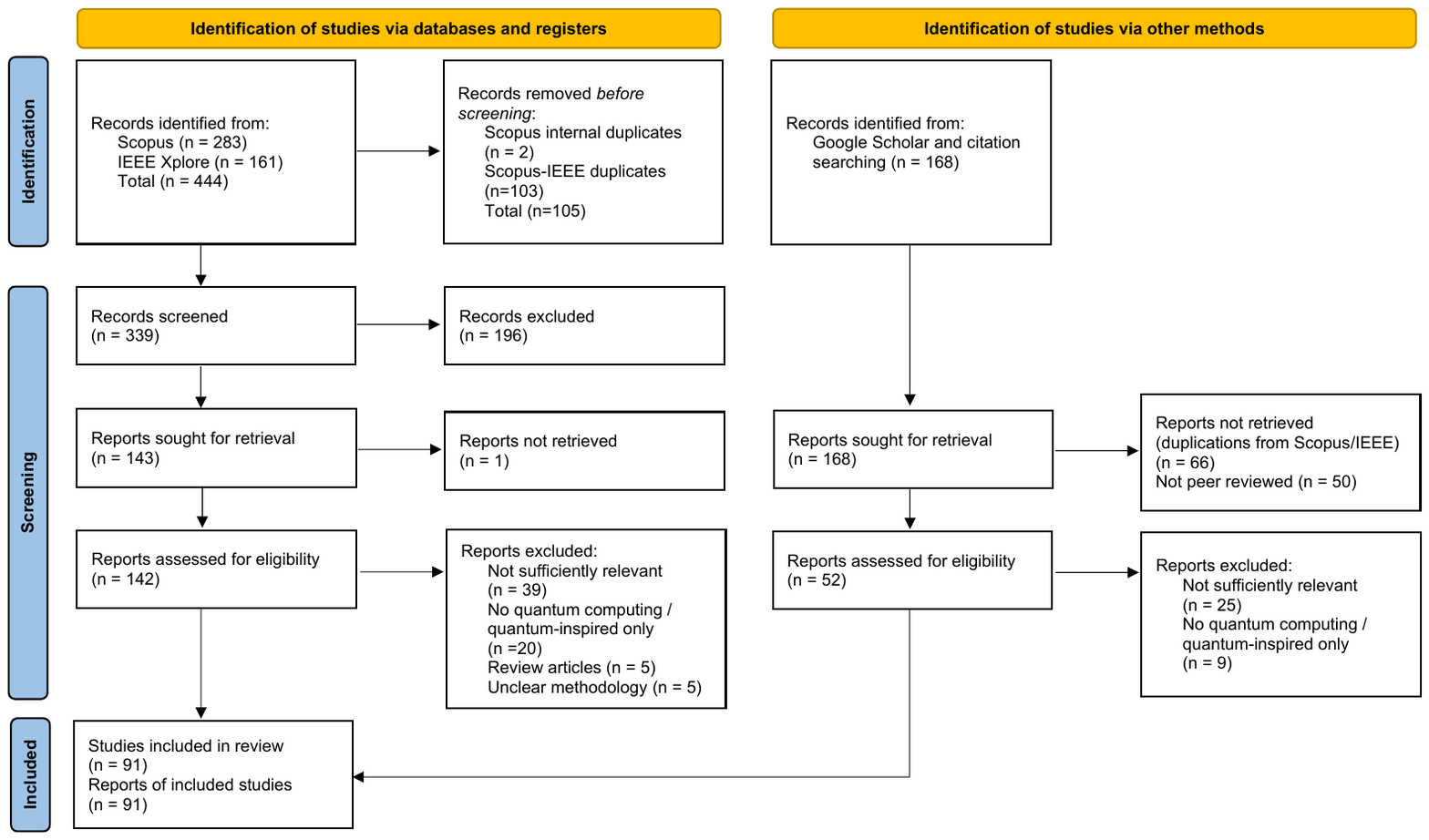}
    \caption{PRISMA 2020 flow diagram for the systematic review on quantum computing for transport-related studies\cite{page_prisma_2020}.}
    \label{fig:prisma}
\end{figure}
\end{landscape}

The Scopus and IEEE Xplore searches returned 444~records. After removing 105~duplicates, 339~records underwent title and abstract screening, of which 196~were excluded. Of the 143~reports sought for retrieval, one could not be retrieved, leaving 142~reports for full-text eligibility assessment. Of these, 73~were included in the review and 69~were excluded. Full-text exclusion reasons were: not sufficiently relevant to the review scope (39), no quantum computing or quantum-inspired only (20), reviews or perspectives (5), and unclear methodology (5). An additional 168~records were identified through Google Scholar and forward/backward citation searching. After removing 66~records that duplicated Scopus or IEEE Xplore results and 50~records that were not peer reviewed, 52~reports were assessed for eligibility. Of these, 18~were included and 34~were excluded: 25~were not sufficiently relevant and nine involved no quantum computing or were quantum-inspired only. In total, 91~studies were included in the review. The complete study selection process is summarised in the PRISMA flow diagram (Figure~\ref{fig:prisma}).

\clearpage

Several limitations of the review methodology should be acknowledged. First, the systematic database search was limited to Scopus and IEEE Xplore, and was supplemented by Google Scholar and citation searching. While Scopus and IEEE Xplore provide broad coverage of peer-reviewed engineering literature, some relevant studies indexed only in Web of Science or other databases may have been missed. Second, no formal protocol was pre-registered prior to conducting the review; however, the complete search query, eligibility criteria, and screening decisions are reported to ensure transparency and reproducibility. Finally, the exclusion category ``not sufficiently relevant to review scope'' involves a degree of subjective judgement regarding a study's contribution to the review narrative. Having established the eligible corpus, we next describe how each included study was evaluated and how its evidence for quantum advantage was assessed.

\subsection{Corpus evaluation method}

Each study in the corpus was assessed through two frameworks. Firstly, a metadata analysis identified the following criteria: year and publication type, key transport focus and sub-focus, problem formulation (e.g., QUBO, neural network), quantum method (e.g., annealing, QAOA), evaluation and platform (e.g., D-Wave, classical simulation), data setting (i.e., the provenance of the instances solved), largest demonstrated instance, and quantum scale (e.g., number of qubits or logical variables). This information is compiled in Table~S2A in the Supplementary Material and discussed below in subsection~\ref{ssec:overview}.  Secondly, a quantum advantage assessment was based on the three criteria identified in subsection~\ref{sec:assessing_advantage}. For each criterion, the studies are evaluated as either \textit{N -- Does not meet}, \textit{P -- Partially meets}, or \textit{M -- Meets}. We maintain strongly that these assessments are not indicative of a study's quality, but reflect our subjective interpretation of their reported results against the criteria. Furthermore, in performing these rankings, we are not suggesting that any authors have themselves made claims of quantum advantage. Our broad reasons for each ranking are as follows, with further details presented in Table~S2B of the Supplementary Material.

For C1, studies evaluated as N either do not report a comparison to classical methods or compare to a classical method which is underspecified, trivial, or highly non-competitive  (e.g., brute-force enumeration). Studies evaluated as P indicate a relevant, non-trivial classical comparator was executed, providing an informative reference point. However, it was not established as competitive with leading problem-specific methods (e.g., simulated annealing or standard MIP implementation). Studies evaluated as M both report a classical method and justify it as state-of-the-art. For C2, an assessment of N indicates that no timings are reported or the timings have been significantly under-reported. An assessment of P indicates that some aspects of the timing are reported, but other contributions may not be reported. For example, annealing studies may report the total anneal time, but not the embedding time or classical processing time. An assessment of M indicates that full end-to-end times are reported and ideally broken down into relevant components. For theoretical studies which develop complexity scaling arguments, we instead assess C2 based on the inclusion of all relevant algorithmic costs rather than a material run-time.

For C3, an assessment of N indicates that no plausible or compelling evidence of advantage has been reported. For example, this rating applies to proof-of-concept studies demonstrated on a single or small instance suite, or to studies with no (or insufficient) timings reported. It is also applied to studies where empirical scaling evidence is documented, but there is no common baseline for the reported timings. For example, the quantum method is run to an approximate solution quality whereas the classical method is run to optimality. We reiterate that the rating N does not mean quantum advantage is impossible, but that it has not been sufficiently demonstrated within the study. An assessment of P indicates a potential quantum advantage. These studies may report an empirical scaling law which is superior to a strong classical competitor (such as in the LABS study by Shaydulin~\cite{shaydulin_evidence_2024}), but the extrapolation at scale is contentious. A ranking of M indicates plausible or compelling evidence for advantage. The former includes analytical studies with superior complexity scaling, but no crossover point of quantum advantage has yet been demonstrated. The latter includes a compelling argument indicating the scale at which quantum advantage occurs.

\subsection{Overview}
\label{ssec:overview}

\begin{figure}
    \centering
    \includegraphics[width=1\linewidth]{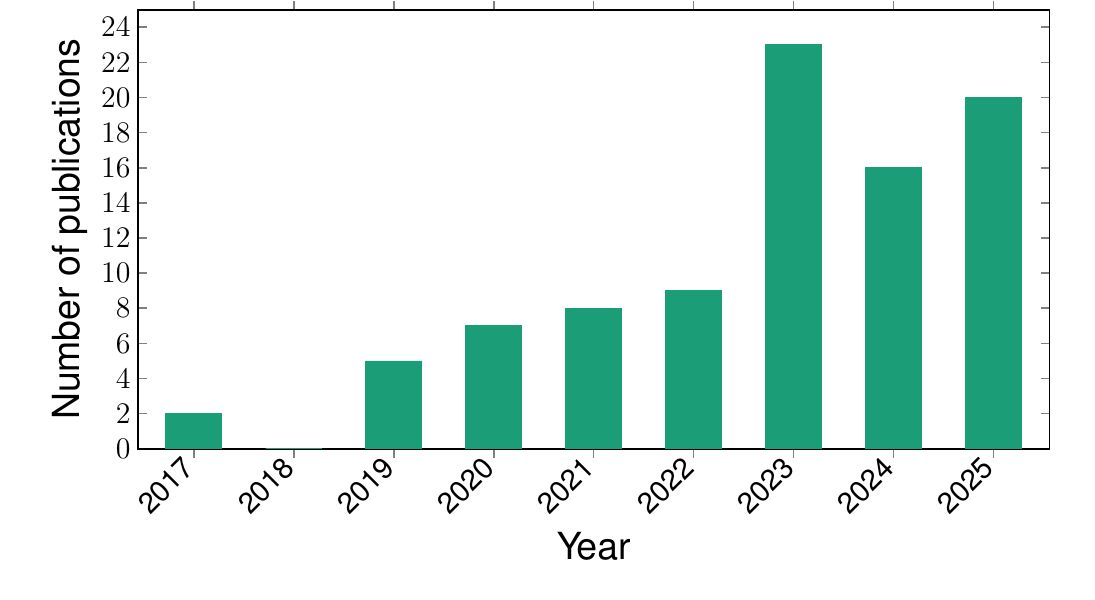}
    \caption{Number of transport-related quantum publications per year as included in this review.}
    \label{fig:bar_chart}
\end{figure}

Analysis of the 91-study corpus reveals seven transport focus domains solved using a diverse set of quantum algorithms. These are documented in Table~\ref{tab:apps}, which catalogues each study's focus as routing, scheduling, transport network design, traffic signalling, computer vision, traffic forecasting, and drive-cycle analysis. Routing is the most popular transport focus, accounting for approximately 2/3 of all studies. Most studies (69\%) use a QUBO formulation to model the transport problem. An emerging set of studies is focused on QML, integrating quantum layers in computer vision studies or traffic forecasting. Figure~\ref{fig:bar_chart} displays the year of publication for studies within the corpus through a histogram. Broadly, interest in quantum computing amongst transport researchers has grown healthily each year. Although we do not include works beyond 2025 in this review, the diversity and extent of published works and preprints in 2026 suggest that this growth is continuing in both the range of transport problems considered and the variety of quantum approaches being explored\cite{arxiv1,arxiv2,arxiv3,arxiv4,arxiv5,arxiv6,arxiv7,arxiv8,arxiv9}.

Table~\ref{tab:corpus_architectures} catalogues the quantum hardware and algorithmic methods used within the corpus. These reveal a near-even split between annealing and gate-based approaches (47 vs 46). Practically all annealing studies used D-Wave. Here direct quantum annealing was used to solve problems up to an average of 132 logical variables. For problems of practical size and those with complex constraints, practitioners have used D-Wave hybrid solvers to handle up to approximately 40,000 logical variables. In contrast, gate-based studies largely relied on classical simulation using IBM Qiskit, primarily to simulate VQAs such as QAOA. They were limited to proof-of-demonstration studies on toy problems up to 15 qubits. A small number of studies (8) further validate their gate-based simulations using quantum hardware, primarily IBM devices. A minority of studies are theoretical only (6) and focus on complexity analysis of quantum algorithms.

The following subsections provide an in-depth review of the literature in each transport focus category. We then distil our analysis by evaluating key cross-cutting insights and limitations across the entire literature corpus.

\begin{table*}[tbp]
\centering
\caption{Summary of quantum algorithms and their applications within the transport literature. \textsuperscript{$\dagger$}  This column also includes closely-related variants of QAOA such as those using constraint-preserving mixers. \textsuperscript{$\ddagger$} Also includes adiabatic quantum computing.}
\label{tab:apps}
\resizebox{\textwidth}{!}{%
\begin{tabular}{lcccccccc}
\toprule
 & 
 & \multicolumn{3}{c}{Hybrid quantum} 
 & \multicolumn{1}{c}{Quantum annealing\textsuperscript{$\ddagger$}} 
 & \multicolumn{3}{c}{Quantum algorithms} \\
\cmidrule(lr){3-5}\cmidrule(lr){6-6}\cmidrule(lr){7-9}
Transport focus
  & Number of studies
  & VQE 
  & QAOA\textsuperscript{$\dagger$} 
  & QML
  &  
  & search 
  & QFT 
  & QBN \\
\midrule
Routing
  & 60
  & \cite{Fagiolo_2025,harwood_formulating_2021,leonidas_qubit_2024,li_quantum_2024,mohanty_analysis_2023,palackal_quantum-assisted_2023,qian_comparative_2023} 
  & \cite{azad_solving_2023,bennett_quantum_2021,bentley_quantum_2022,finzgar_quark_2022,fitzek_applying_2024,harwood_formulating_2021,Hussain_2025,li_quantum_2024,Maciejunes_2025,palackal_quantum-assisted_2023,qian_comparative_2023,ruan_quantum_2020,spyridis_variational_2023,xie_feasibility-preserved_2024,Zawalska_2023}   
  & \cite{Li_2024,mohanty_solving_2024,Moosavi_2025,XU_2025}
  & \cite{arino_sales_adiabatic_2023,borowski_new_2020,niu_applying_2025,Ciacco_2025,Codognet_2024,dixit_quantum_2024,Dong_2020,feld_hybrid_2019,finzgar_quark_2022,haba_routing_2025,He_2024,Holliday_2024,irie_quantum_2019,jain_solving_2021,Kieu_2019,le_quantum_2023,Li_2023,Na_e_cz_Charkiewicz_2025,neukart_traffic_2017,osaba_solving_2024,papalitsas_qubo_2019,salehi_unconstrained_2022,Silva_2021,sinno_analyzing_2025,Tambunan_2023,Vargas_2024,villar-rodriguez_analyzing_2022,weinberg_supply_2023,yarkoni_quantum_2020,yarkoni_solving_2021}  
  & \cite{Bai_2025,De_Andoin_2023,Hussain_2025,liu_solving_2025,makhanov_quantum_2024,Markevich_2019,moylett_quantum_2017,Sato_2023,Sharma_2025,Tszyunsi_2023}  
  & --  
  & --  \\
Scheduling
  & 14
  & \cite{mohammadbagherpoor_exploring_2021,q-ctrl_accelerating_2025}  
  & \cite{dalyac_qualifying_2021,grange_design_2024,kea_leveraging_2023,svensson_hybrid_2023,vikstal_applying_2020} 
  & --
  & \cite{bickert_optimising_2023,domino_quadratic_2022,domino_quantum_2023,koniorczyk_solving_2025,stollenwerk_quantum_2020,xu_high-speed_2023}
  & \cite{grange_design_2024}  
  & --  
  & --  \\
Transport network design
  & 6
  & \cite{Rao_2022} 
  & -- 
  & --
  & \cite{Chandra_2022,dixit_quantum_2023,gabbassov_transit_2022,Nourbakhsh_2024,okada_joint_2026,Rao_2022} 
  & --  
  & --  
  & --  \\
Traffic signalling
  & 5
  & --  
  & -- 
  & --
  & \cite{hussain_optimal_2020,inoue_traffic_2021,inoue_traffic_2024,marchesin_improving_2023,shikanai_quadratic_2025}  
  & --  
  & --  
  & --  \\
  Computer vision
  & 4
  & --
  & --
  & \cite{innan_qnn-vrcs_2025,Kang_2025,kuros_traffic_2023,meghanath_qdcnn_2025}
  & --
  & --
  & --
  & -- \\
  Traffic forecasting
  & 2
  & -- 
  & --
  & \cite{schetakis_quantum_2025}
  & --
  & --
  & --
  &  \cite{harikrishnakumar_forecasting_2023} \\
Drive-cycle analysis
  & 1
  & -- 
  & -- 
  & --
  & -- 
  & --  
  & \cite{dixit_quantum_2022}  
  & --  \\
\bottomrule
\end{tabular}}
\end{table*}

\begin{table}[!htbp]
\centering
\caption{Quantum-computing methods, demonstrated scale, and major platforms in the corpus. Numbers in brackets are study counts for the indicated category. Some studies use multiple quantum methods and simulation/hardware platforms. Not all studies report qubit or logical variable scale.}\label{tab:corpus_architectures}
\small
\setlength{\tabcolsep}{4pt}
\setlength{\extrarowheight}{1pt}
\renewcommand{\arraystretch}{1.10}
\begin{tabularx}{\textwidth}{@{}
    >{\raggedright\arraybackslash}p{0.16\textwidth}
    >{\raggedright\arraybackslash}p{0.22\textwidth}
    >{\centering\arraybackslash}p{0.14\textwidth}
    >{\centering\arraybackslash}p{0.14\textwidth}
    >{\raggedright\arraybackslash}X
@{}}
\toprule
\multirow[c]{2}{*}{\textbf{Architecture}}
  & \multirow[c]{2}{*}{\textbf{Method}}
  & \multicolumn{2}{c}{\textbf{Maximum scale}}
  & \multirow[c]{2}{*}{\textbf{Major platforms}} \\
\cmidrule(lr){3-4}
  & & \textbf{Mean} & \textbf{Median} & \\
\midrule
\multirow[t]{2}{*}{\textbf{Annealing (47)}}
  & Quantum annealing (31)
  & 132 logical variables & 83 logical variables
  & D-Wave (28); MATLAB/PyTorch (1); Python simulation (1) \\
\cmidrule(lr){2-5}
  & Hybrid annealing (21)
  & 42,023 logical variables & 2,659 logical variables
  & D-Wave managed hybrid solver (19); D-Wave with qbsolv (3) \\
\midrule
\multirow[t]{3}{*}{\textbf{Gate-based (46)}}
  & Classical simulation (41)
  & 16 qubits & 15 qubits
  & IBM Qiskit (24); PennyLane (7); MATLAB (2); others (13) \\
\cmidrule(lr){2-5}
  & Quantum hardware (8)
  & 23 qubits & 11 qubits
  & IBM (8); IonQ (1); Rigetti (1) \\
\cmidrule(lr){2-5}
  & Theoretical (6)
  & -- & --
  & -- \\
\bottomrule
\end{tabularx}
\par\vspace{0.25em}
\begin{minipage}{\textwidth}
\footnotesize\raggedright
\end{minipage}
\end{table}

\begin{table}[htbp]
\centering
\caption{Distribution of evidence-assessment ratings in the corpus by primary transport focus. Each cell reports N/P/M counts, followed by row percentages in parentheses. N = no; P = partly; M = meets. C1 assesses comparison against the best relevant classical method, C2 assesses complete computational costs, and C3 assesses evidence for quantum advantage, including defensible scaling evidence.}\label{tab:corpus_evidence}
\small
\setlength{\tabcolsep}{3.5pt}
\setlength{\cmidrulewidth}{0.20pt}
\renewcommand{\arraystretch}{1.12}
\begin{tabularx}{\textwidth}{@{}
    >{\raggedright\arraybackslash}X
    r
    >{\centering\arraybackslash}X
    >{\centering\arraybackslash}X
    >{\centering\arraybackslash}X
@{}}
\toprule
& & \multicolumn{3}{c}{Rating counts and row percentages, N/P/M} \\
\cmidrule(lr){3-5}
Primary focus
    & $n$
    & C1: classical comparator
    & C2: computational costs
    & C3: quantum advantage \\
\midrule
Whole corpus
    & 91
    & \makecell[c]{55 / 32 / 4\\(60.4 / 35.2 / 4.4\%)}
    & \makecell[c]{57 / 30 / 4\\(62.6 / 33.0 / 4.4\%)}
    & \makecell[c]{89 / 0 / 2\\(97.8 / 0.0 / 2.2\%)} \\
\cmidrule(l{4pt}r{4pt}){1-5}

Routing
    & 60
    & \makecell[c]{44 / 14 / 2\\(73.3 / 23.3 / 3.3\%)}
    & \makecell[c]{43 / 15 / 2\\(71.7 / 25.0 / 3.3\%)}
    & \makecell[c]{59 / 0 / 1\\(98.3 / 0.0 / 1.7\%)} \\
\cmidrule(l{4pt}r{4pt}){1-5}

Scheduling
    & 13
    & \makecell[c]{7 / 5 / 1\\(53.8 / 38.5 / 7.7\%)}
    & \makecell[c]{5 / 6 / 2\\(38.5 / 46.2 / 15.4\%)}
    & \makecell[c]{12 / 0 / 1\\(92.3 / 0.0 / 7.7\%)} \\
\cmidrule(l{4pt}r{4pt}){1-5}

Transport network design
    & 6
    & \makecell[c]{3 / 3 / 0\\(50.0 / 50.0 / 0.0\%)}
    & \makecell[c]{3 / 3 / 0\\(50.0 / 50.0 / 0.0\%)}
    & \makecell[c]{6 / 0 / 0\\(100.0 / 0.0 / 0.0\%)} \\
\cmidrule(l{4pt}r{4pt}){1-5}

Traffic signalling
    & 5
    & \makecell[c]{0 / 5 / 0\\(0.0 / 100.0 / 0.0\%)}
    & \makecell[c]{1 / 4 / 0\\(20.0 / 80.0 / 0.0\%)}
    & \makecell[c]{5 / 0 / 0\\(100.0 / 0.0 / 0.0\%)} \\
\cmidrule(l{4pt}r{4pt}){1-5}

Computer vision
    & 4
    & \makecell[c]{0 / 4 / 0\\(0.0 / 100.0 / 0.0\%)}
    & \makecell[c]{3 / 1 / 0\\(75.0 / 25.0 / 0.0\%)}
    & \makecell[c]{4 / 0 / 0\\(100.0 / 0.0 / 0.0\%)} \\
\cmidrule(l{4pt}r{4pt}){1-5}

Traffic forecasting
    & 2
    & \makecell[c]{1 / 1 / 0\\(50.0 / 50.0 / 0.0\%)}
    & \makecell[c]{1 / 1 / 0\\(50.0 / 50.0 / 0.0\%)}
    & \makecell[c]{2 / 0 / 0\\(100.0 / 0.0 / 0.0\%)} \\
\cmidrule(l{4pt}r{4pt}){1-5}

Drive cycle analysis
    & 1
    & \makecell[c]{0 / 0 / 1\\(0.0 / 0.0 / 100.0\%)}
    & \makecell[c]{1 / 0 / 0\\(100.0 / 0.0 / 0.0\%)}
    & \makecell[c]{1 / 0 / 0\\(100.0 / 0.0 / 0.0\%)} \\
\bottomrule
\end{tabularx}
\par\vspace{0.35em}
\begin{minipage}{\textwidth}
\footnotesize\raggedright
\end{minipage}
\end{table}

\subsection{Routing}
\label{ssec:routing}

\paragraph{Overview}

The dominant application of quantum computing in the transport literature is for routing. This sustained interest can largely be attributed to three reasons. Firstly, routing has a practical impact on minimising costs and travel times in a variety of transport contexts, including freight shipping, air travel, and emergency services. Secondly, the QUBO form is versatile enough to handle many different problem formulations. Thirdly, most routing problems are NP-hard and exact solutions may be computationally intractable for large urban networks\cite{pessoa_generic_2020}. While heuristic algorithms can often obtain efficient solutions to similarly sized problems within 1--2\% of the optimum route\cite{arnold_efficiently_2019}, even small enhancements to this accuracy would provide drastic reductions in cost, time, and environmental impact at scale. Hence, routing problems may benefit tremendously from quantum acceleration given future advances in hardware.

Transport engineers have explored routing through a diverse set of use cases, algorithms, and hardware, with frequent collaboration between quantum computing experts and industry partners. Most literature addresses variants of the TSP and the more general vehicle routing problem (VRP). In brief, given a list of customers at different geographical locations and a set of delivery vehicles, assign routes for each delivery vehicle that minimise some overall transportation cost (e.g., distance, money, emissions). The basic constraints are that all vehicles begin and end their route at a depot, each customer is visited only once, and there exist no subtours. Variants of the problem include constraints on vehicle capacity, timing of deliveries, and depot number, all of which may be expressed in QUBO form through appropriate cost and penalty terms. Practically every major variant of the VRP and TSP has been explored in the literature using different problem formulations and quantum solution methods. A list of the major variants with corresponding studies is presented in Table~\ref{tab:routing_variants}, whereas other work has considered stochastic time-dependent shortest path routing\cite{dixit_quantum_2024}, the multicommodity network flow problem\cite{niu_applying_2025}, and routing of aeroplanes\cite{makhanov_quantum_2024} and aerial drones\cite{haba_routing_2025}. Clearly, the space has been thoroughly explored.

The principal research gap is now finding evidence that the resulting quantum methods are competitive at meaningful scales. In this regard, the assessment of the three evidence criteria reveals sobering results. As demonstrated in Table~\ref{tab:corpus_evidence}, 73\% of all routing studies are rated N for C1, indicating that they largely failed to compare to a classical method. Similarly, 72\% of studies did not report any timing components whatsoever. Overall, only 2 studies met C1 and C2. This indicates a significant limitation in reporting amongst routing studies. Only 1 study was rated as providing plausible evidence of quantum advantage, a theoretical algorithm for the TSP by Moylett~\textit{et al}\cite{moylett_quantum_2017}. The following section investigates the evidence assessment for routing in detail. This reveals the prevalence of QUBO formulations, and we therefore continue by analysing different approaches to constructing these formulations and assessing their effectiveness.

\begin{table*}[t]
\centering
\caption{Major variants of the TSP and VRP considered in the literature.}
\label{tab:routing_variants}
\begin{tabularx}{\textwidth}{lXX}
\toprule
\textbf{Variant} & \textbf{TSP} & \textbf{VRP} \\
\midrule
Standard
& \cite{jain_solving_2021,villar-rodriguez_analyzing_2022,finzgar_quark_2022,ruan_quantum_2020,qian_comparative_2023,Dong_2020,Silva_2021,Li_2023,He_2024,Codognet_2024}
& \cite{azad_solving_2023,mohanty_analysis_2023,borowski_new_2020,mohanty_solving_2024,arino_sales_adiabatic_2023,Na_e_cz_Charkiewicz_2025,Maciejunes_2025} \\

Capacitated
& ---
& \cite{bentley_quantum_2022,palackal_quantum-assisted_2023,bennett_quantum_2021,borowski_new_2020,sinno_analyzing_2025,xie_feasibility-preserved_2024,Vargas_2024,Holliday_2024} \\

Time windows
& \cite{salehi_unconstrained_2022,papalitsas_qubo_2019}
& \cite{irie_quantum_2019,harwood_formulating_2021,weinberg_supply_2023,leonidas_qubit_2024} \\

Heterogeneous fleet
& ---
& \cite{fitzek_applying_2024,osaba_solving_2024} \\

Selective routing
& \cite{le_quantum_2023}
& --- \\

Weighted segments
& ---
& \cite{Tambunan_2023} \\

Collision avoidance
& ---
& \cite{li_quantum_2024} \\
\bottomrule
\end{tabularx}
\end{table*}

\paragraph{Evidence criteria assessment}

Authors have adopted three broad quantum approaches to solving routing problems. (1) Direct conversion of a MIP formulation into a monolithic QUBO with constraints implemented via penalty terms (26 studies, e.g., \cite{azad_solving_2023,salehi_unconstrained_2022}); (2) structure-aware quantum optimisation that avoids (or reduces) monolithic penalty encodings by preserving feasibility, compressing the representation, retaining explicit constraints, or decomposing the problem into quantum and classical components (20 studies, e.g. \cite{bennett_quantum_2021,bentley_quantum_2022}); (3) gate-based quantum search algorithms that use techniques such as Grover search, minimum finding, or quantum backtracking (9 studies, e.g., \cite{moylett_quantum_2017}). Studies with approach (1) and (2) solve their optimisation problem using VQAs or annealing. When comparisons are made to classical methods, it is typically to Gurobi or CPLEX solvers of a corresponding arc-based MIP or simulated annealing/tabu search applied to the QUBO formulation. Studies with approach (3) are typically theoretical and develop a time-complexity scaling for a gate-based algorithm. These approach types are noted for each study in Table S2A.

Consistent across study approaches (1) and (2) is the absence of comparisons with the strongest available classical routing methods. Across 45 relevant studies, 33 were rated as N for C1, 11 as P, and only 1 as M. Those rated as P typically use a direct arc-based MIP solved with Gurobi or CPLEX. This is a sensible exact baseline. These solvers improve the submitted model through presolve, primal heuristics and cutting planes, but remain general purpose methods applied to the chosen formulation. However, leading exact approaches go further by reformulating the problem over complete routes, generating useful columns dynamically through pricing, and incorporating routing-specific cuts and branching rules within branch-price-and-cut\cite{poggi_new_2014,pecin_improved_2017,pecin_enhancements_2017,pessoa_generic_2020}. A direct MIP comparison can therefore establish correctness and baseline performance, but does not represent the practical frontier of exact classical optimisation. The same distinction applies to heuristic comparisons. Simulated annealing or tabu search on the QUBO can show whether a quantum method improves upon classical solution of the chosen encoding, but not whether it is competitive with the strongest methods for the underlying routing problem. Leading classical heuristics may instead use frameworks such as hybrid genetic search and problem-specific adaptive large neighbourhood search, whose representations and search operators are designed around the routing constraints\cite{vidal_diversity_2013,vidal_hgs_2022,ropke_alns_2006,azi_alns_2014}.

Consequently, most of the studies that adopted approaches (1) and (2) did not meet C1. This omission does not invalidate the results obtained on the generally small instances considered in these routing studies, as a direct MIP implementation or annealing should recover and prove the same optimum as a tailored exact solver. However, it does limit conclusions about classical competitiveness and scaling. If the direct arc-based MIP begins to stall as instance size increases, this may reflect the formulation and the plug-and-play solution approach rather than the practical limit of classical routing optimisation.

Although several type (1) and (2) studies report promising solver-level speed or scaling results, none yet provides evidence sufficient to establish practical end-to-end quantum advantage. Of the 46 studies in these categories, only 11 report solver runtime at multiple different problem scales. An informative example is Dixit~\textit{et al.}~\cite{dixit_quantum_2024}, who adopt a type (2) approach to the problem of stochastic time-dependent shortest path routing. They compare the runtime of CPLEX with the D-Wave hybrid solver and report that the former has exponential time scaling with problem size whereas the latter has linear time scaling. Moreover, the reported hybrid-solver runtimes are smaller than the CPLEX runtimes for many of the larger instances trialled. Across 18 instances containing up to 147,600 variables, the authors report a maximum 17-fold solver-time speedup relative to CPLEX. These reported speed and scaling results are important. However, the evidence is not yet sufficient to establish practical end-to-end quantum advantage. Namely, the two methods are not consistently compared at the same solution quality. The hybrid solver obtains results which exceed the CPLEX objective by 0.47--4.61\%. Additionally, the embedding time is not disclosed and the CPLEX solver attempts to prove optimality whereas the hybrid solver does not. The reported linear trend should also be interpreted as a solver-budget result. This is because the hybrid sampling time was set to be linear in the scale variables, rather than measured as the time required to reach a fixed solution-quality target. The remaining studies which report runtimes are limited by similar challenges, either comparing solutions between solvers to different standards, omitting critical embedding costs, or making comparisons to classical methods which are not established as state-of-the-art.

Niu~\textit{et al.}~\cite{niu_applying_2025} provide a further informative comparison for a multicommodity network-flow model. Here they evaluate D-Wave's hybrid solver against single-threaded CPLEX on instances containing up to 192,000 variables. For the larger instances, they report hybrid-solver runtimes 4.5--13 times shorter and, within the selected time limits, often lower-cost solutions. They also find CPLEX superior for smaller instances and note that it tends to provide better solutions when allowed more time. These are important solver-level results at comparatively large problem scales. They do not yet establish end-to-end quantum advantage because the methods are not compared at a common solution-quality or stopping target, the cost of the embedding and pre-processing is not included, CPLEX is not shown to represent the strongest specialised multicommodity-flow workflow, and the quantum contribution of the proprietary hybrid service is not isolated.

Despite these limitations, type (2) approaches appears better positioned than type (1) to produce compelling evidence of quantum advantage. The central problem with type (1) is that enforcing constraints in the QUBO objective becomes a penalty calibration problem. The penalties must be large enough to exclude infeasible routes, but if they are too large the constraint terms dominate the route-cost objective. The difference can become hard to resolve and finding the smallest penalty that guarantees an exact QUBO reformulation is itself NP-hard\cite{alessandroni_alleviating_2025}. This practical difficulty was identified by Smith-Miles~\textit{et al.}~\cite{smith-miles_travelling_2025} in their review for the TSP. The problem is likely to become more acute for richer VRPs, where capacity, scheduling, assignment, and other interacting constraints introduce additional penalties to calibrate.  This becomes burdensome in the type (1) approach, whereas strong classical methods often retain and exploit this structure explicitly to their advantage. While type (2) approaches do not guarantee quantum advantage, they can alleviate this weakness by preserving feasibility, retaining explicit constraints, or decomposing the problem rather than forcing its full structure into a monolithic penalty model.

Type (3) studies pursue a different form of evidence. They derive asymptotic speed-ups for gate-based algorithms rather than benchmarking near-term hardware. The strongest example is Moylett~\textit{et al.}~\cite{moylett_quantum_2017}, who apply quantum backtracking to established exact algorithms for the weighted TSP on bounded-degree graphs. The study was rated M for all three criteria. It compares against the best-known exact algorithms reported for the same graph classes (C1). It includes the principal algorithmic costs of exploring the backtracking tree, evaluating the reduction and branching routines, performing binary search over route cost, controlling the failure probability, and reducing higher-degree instances (C2). Finally, it proves improved scaling for an arbitrary number of graph vertices \(n\), from \(O^*(1.232^n)\) to \(O^*(1.110^n\operatorname{polylog}L)\) for degree-3 graphs and from \(O^*(1.692^n)\) to \(O^*(1.301^n\operatorname{polylog}L)\) for degree-4 graphs, where \(L\) is the maximum edge cost (C3). Despite this success, the result remains theoretical and somewhat restrictive. Namely, it assumes bounded vertex degree and independent branching decisions. The construction also fails to provide a speed-up above degree 6 and assumes efficient and coherent access to the graph and edge weights. From a practical perspective, one must also account for data encoding, reversible implementation of the required subroutines, readout, circuit resources, and fault tolerance. These costs do not alter the exponential scaling, but materially affect runtimes and therefore the finite-size crossover at which quantum advantage occurs.

\paragraph{Design and implementation of quantum routing formulations}

The prevalence of QUBO-based approaches in the routing literature warrants closer examination of how these formulations are constructed, implemented, and optimised. Their performance varies considerably with the underlying mathematical formulation, treatment of constraints, qubit encoding, quantum algorithm and hardware used. The effects of these choices are illustrated below through a series of case studies.

\subparagraph{Problem formulation}

First, we compare two different problem approaches to the CVRP solved with quantum annealers. Table~\ref{tab:routing} summarises the method and results of Feld~\textit{et al.}\cite{feld_hybrid_2019} and Sinno~\textit{et al.}\cite{sinno_analyzing_2025} for similarly sized problem instances (i.e., number of customers and vehicles). Feld~\textit{et al.} use a two-phase heuristic to decompose the CVRP into two sub-problems: a clustering phase (which reduces to the knapsack problem) and a routing phase (the TSP). The clustering phase is solved classically to assign a subset of customers to each vehicle. The route for each vehicle within this subset is then solved using a quantum annealer. This approximation drastically reduces the problem size and number of constraints in the QUBO formulation. In contrast, Sinno~\textit{et al.} adopt a monolithic approach which treats the CVRP in its entirety. Moreover, they also have access to D-Wave's constrained quadratic model which uses classical heuristics to guide feasible solution regions to the quantum annealer.

The results of Feld~\textit{et al.} were obtained on an earlier generation of D-Wave hardware (2000Q) than those of Sinno~\textit{et al.} (Advantage). Moreover, Feld~\textit{et al.} performed only 10 annealing runs per problem instance compared to Sinno~\textit{et al.}'s 100 annealing runs. One may therefore expect Sinno~\textit{et al.} to obtain significantly better results for similarly sized problems. However, this is not the case. As Table~\ref{tab:routing} demonstrates, Feld~\textit{et al.} obtain solutions which deviate by 4--10 times less from the best-known result than Sinno~\textit{et al.} Even though the two-phase heuristic is an approximation of the CVRP, it requires drastically fewer qubits and coupling terms to implement. Hence, the problem becomes significantly more tractable for current hardware and better quality solutions are obtained. Decomposing challenging problems into separate (but still NP-hard) sub-problems can therefore be a viable strategy for improving performance on near-term hardware.

Even without decomposition, the problem formulation can have a dramatic impact on the quantum resource requirements and therefore solution quality. This has been clearly demonstrated by Harwood~\textit{et al.}~\cite{harwood_formulating_2021} in their investigation of the VRP with time-windows for the maritime inventory routing problem. This problem extends the VRP by requiring deliveries be made to each customer within specific time windows. For maritime shipping, the travel times are typically much longer than the time windows and so relatively long time horizons are required to maximise benefits from optimisation. Harwood~\textit{et al.} explored this problem through multiple QUBO formulations which encode different decision variables. It is informative to understand how to construct these formulations and assess their overall resource costs.

For example, three of the formulations were termed route-based, arc-based, and sequence based. The route-based formulation uses a classical pre-processing step to determine all valid routes between a set of customers (i.e., routes that satisfy the timing constraints). The decision variables then encode which vehicle takes which route, and the annealer minimises the combined route costs accordingly. The arc-based formulation considers the travel cost between individual pairs of customers (termed arcs). The time horizon is discretised, and the decision variables encode whether a vehicle uses a given arc between two different time steps. In the sequence formulation, vehicles visit different customers in a sequence. The decision variables encode whether a given vehicle travels to a given customer at a given point in the sequence. While the problem formulations differ, in principle all should be capable of producing the same optimal solution.

\begin{figure}
    \centering
    \begin{subfigure}[b]{0.47\textwidth}
        \includegraphics[width=\textwidth]{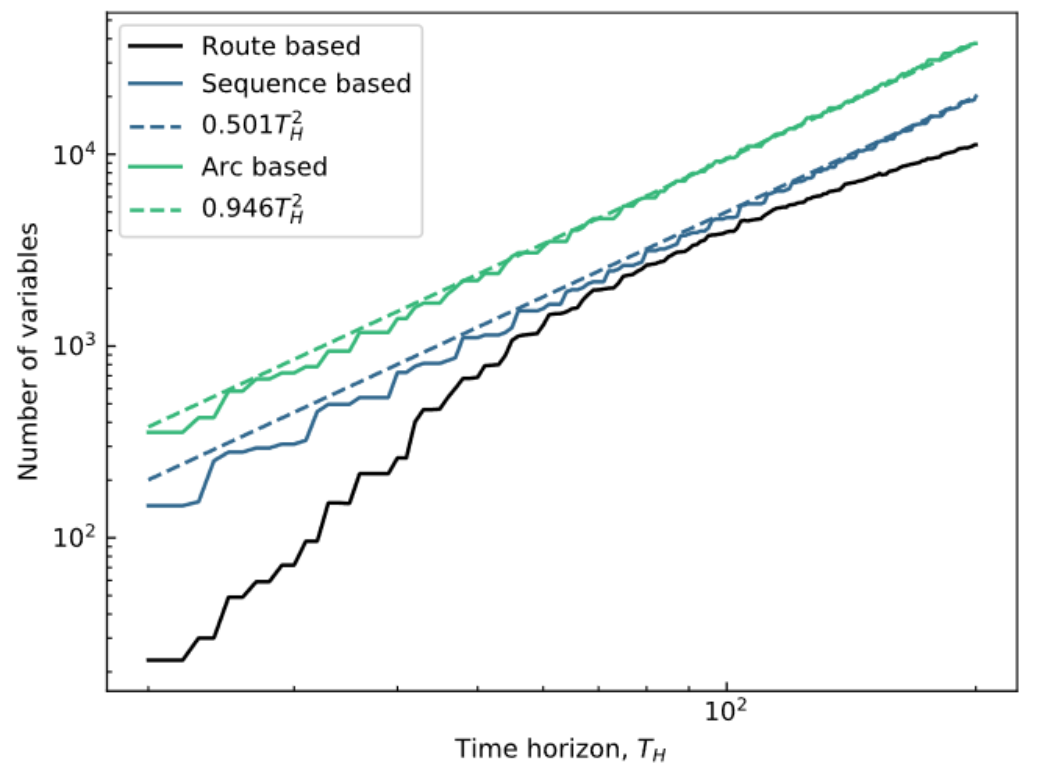}
        \caption{}
        \label{fig:harwood_a}
    \end{subfigure}
    ~ 
    \begin{subfigure}[b]{0.47\textwidth}
        \includegraphics[width=\textwidth]{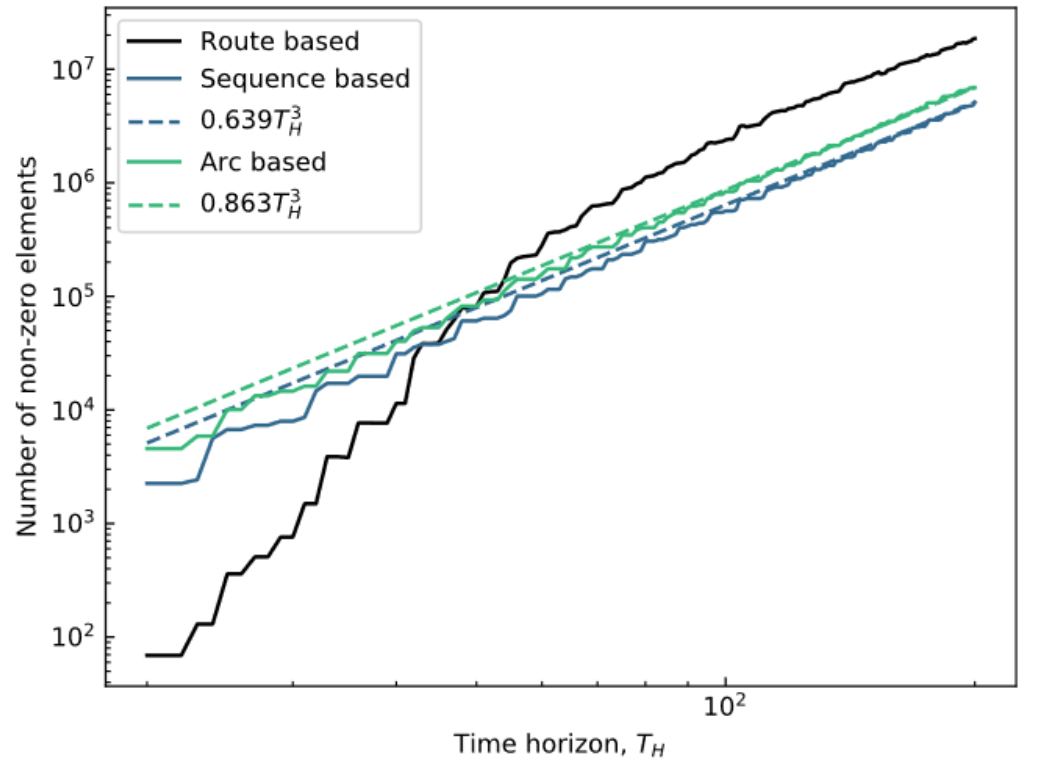}
        \caption{}
        \label{fig:harwood_b}
    \end{subfigure}
    
\caption{Resource requirements for the VRP with time windows as a function of the time horizon (i.e., maximum time length considered in the simulations). Harwood~\textit{et al.} consider different QUBO formulations termed route based, sequence based, and arc based. These require different numbers of (a) decision variables and (b) non-zero elements in the QUBO matrix. For VQAs, this directly corresponds to the number of qubits and control gates, whereas on a quantum annealer it corresponds to the number of qubits and chain length. The scaling of resource requirements with problem size is non-trivial and the QUBO formulation must be tailored to the specific quantum hardware. Reproduced from Harwood~\textit{et~al.}~\cite{harwood_formulating_2021}
under the terms of the Creative Commons Attribution 4.0 International License (CC BY 4.0).}\label{fig:harwood_resource_requirements}
\end{figure}

Harwood~\textit{et al.} then consider scaling of the resource requirements for these three formulations with increasing time horizon. Recall from Section~\ref{par:HA} that these resources are the number of logical qubits needed to encode the decision variables and the number of non-zero elements in the QUBO matrix. These resource requirements are depicted in Figure~\ref{fig:harwood_resource_requirements} for a realistic problem instance. The route-based formulation requires the fewest decision variables and therefore qubits; however, this compactness comes at the cost of a relatively expensive classical pre-processing of all valid routes. It is therefore more suitable for near-term hardware with limited qubit capacity than the arc and sequence formulations. The route-based formulation also has the fewest coupling terms for small time horizons, but becomes more expensive than the other formulations for longer time horizons. This corresponds to larger chain lengths for quantum annealers and increased gate depths for VQAs. In summary, the problem formulation has a non-trivial impact on the quantum resource requirements which varies with problem scale. For each problem, a trade-off must be made between the number of decision variables and non-zero matrix elements, and the formulation must be tailored to specific quantum hardware.

\begin{table*}[tbp]
\centering
\caption{Selected CVRP and TSP results reported using gate-based hardware or hybrid annealing services.}
\label{tab:routing}
\begin{tabular}{l *{9}{c}}
\toprule
 & \multicolumn{3}{c}{Feld~\textit{et al.} (2019)\cite{feld_hybrid_2019}}
 & \multicolumn{3}{c}{Sinno~\textit{et al.} (2023)\cite{sinno_analyzing_2025}}
 & \multicolumn{3}{c}{Palackal~\textit{et al.} (2023)\cite{palackal_quantum-assisted_2023}} \\
\midrule
Problem
  & \multicolumn{3}{c}{CVRP}
  & \multicolumn{3}{c}{CVRP}
  & \multicolumn{3}{c}{TSP} \\
Approach
  & \multicolumn{3}{c}{Two-phase heuristic}
  & \multicolumn{3}{c}{Monolithic}
  & \multicolumn{3}{c}{Two-phase heuristic} \\
Encoding
  & \multicolumn{3}{c}{Node-based one-hot}
  & \multicolumn{3}{c}{Edge-based one-hot}
  & \multicolumn{3}{c}{Node-based one-hot} \\
Device
  & \multicolumn{3}{c}{D-Wave 2000Q}
  & \multicolumn{3}{c}{D-Wave Advantage}
  & \multicolumn{3}{c}{IBM 27-qubit} \\
Algorithm
  & \multicolumn{3}{c}{Hybrid annealing}
  & \multicolumn{3}{c}{Hybrid annealing}
  & \multicolumn{3}{c}{VQE + NFT optimizer} \\
\addlinespace
\midrule
Customers
  & 33 & 51 & 76
  & 32 & 60 & 80
  & 4  & \ 5  & 6    \\
Vehicles
  & 4  & 5  & 7
  & 5  & 9  & 10
  & -  & \ -  & -    \\
\makecell[tl]{Solution quality\\(\% deviation from\\best-known result)}
  & 2.0  & 6.9   & 9.7
  & 24.0 & 36.1  & 43.0
  & 12.0   & \ 7.0     & 27.0  \\
\bottomrule
\end{tabular}
\end{table*}

\subparagraph{Qubit encoding}

Even for similar problem formulations, the qubit encoding can make a significant impact on resource requirements. Consider as an example the node-based QUBO formulation for the TSP. Two possible ways to represent the decision variables are the one-hot and binary formulations. For the standard node-based one-hot formulation\cite{lucas_ising_2014},
\begin{equation}\label{TSP-OH-DV}
x_{i,p} = \begin{cases}
1, \ \text{ if city $i$ is visited $p^\text{th}$}, \\
0, \ \text{ otherwise.}
\end{cases}
\end{equation}
If the distance between city $i$ and $j$ is $d_{ij}$, then the cost function is given by
\begin{equation}\label{TSP-OH-cost}
C(x) = \sum_{p=1}^N\sum_{i,j=1}^N d_{ij}x_{i,p}x_{j,p+1},
\end{equation}
for a tour with $N$ cities. The constraints are that each city is visited once in the tour, and that only one city is visited at a time. If the decision variables~\eqref{TSP-OH-DV} are expressed as an $N\times N$ matrix, these constraints are equivalent to each row and column having only one non-zero entry -- hence the naming convention `one hot'. Note that there are $N^2$ decision variables and therefore $N^2$ qubits are required to encode a problem of this size. Moreover, we see that the cost function includes $N^3$ different quadratic terms, corresponding to $\mathcal{O}\left ( N^3 \right )$ two-gate operations in order to implement QAOA.

Now consider a different qubit encoding by expressing the index of each city visited in the tour through binary\cite{glos_space-efficient_2022,bentley_quantum_2022}. Define $\ell = \lceil \log_2 (N) \rceil$. If city $i\in\{1,\ldots,N\}$ is visited at the $p^\text{th}$ position in the tour, we encode the zero-based city label $i-1$ through the binary string
\[
\mathbf{x}^{(p)} = [ x^{(p)}_0, x^{(p)}_1, \ldots, x^{(p)}_{\ell-1} ],
\]
such that
\[
i-1=\sum_{k=0}^{\ell-1}2^k x^{(p)}_k.
\]
For example, for a total of $16=2^4$ cities, if the sixth city is visited at the third position in the tour, then $i-1=5$ and
\[
\mathbf{x}^{(3)} = [1, 0, 1, 0].
\]
All decision variables may then be compactly expressed in a single vector of size $N\ell=N\lceil\log_2(N)\rceil$ as
\begin{equation}
\mathbf{x} = [ \mathbf{x}^{(1)} | \mathbf{x}^{(2)} | \ldots | \mathbf{x}^{(N)}].
\end{equation}
The price of this compact encoding is a more complicated cost function. In the same spirit as~\eqref{TSP-OH-cost}, we write
\begin{equation}\label{TSP-B-cost}
C(x) = \sum_{p=1}^N\sum_{i,j=1}^N d_{ij}\delta_{i,p}\delta_{j,p+1},
\end{equation}
where $\delta$ is defined as
$$\delta_{i,p} = \begin{cases}
1, \ \text{ if city $i$ is visited in position $p$}, \\
0, \ \text{ otherwise.}
\end{cases}.$$
One simple way to realise this using the binary encoding is by checking each bit pair-wise. Let $b_{i,k}\in\{0,1\}$ denote the $k^\text{th}$ bit, with $k=0$ the least significant bit, in the binary expansion of $i-1$, so that
\[
i-1=\sum_{k=0}^{\ell-1}2^k b_{i,k}.
\]
Then
\[
\delta_{i,p} = \prod_{k=0}^{\ell-1} \left [ 1- \left (x^{(p)}_k-b_{i,k} \right )^2\right ].
\]

While the cost function~\eqref{TSP-B-cost} is quadratic in $\delta$, it is of order $2\ell$ in the basis of decision variables.\footnote{The order is not $4\ell$ because for binary variables we have $\left ( x^{(p)}_k \right )^2=x^{(p)}_k$.} This is a higher-order binary optimisation (HOBO) formulation, and therefore execution on a quantum annealer will require `quadratisation' of the higher-order terms through the costly introduction of ancillary qubits. Alternatively, the higher-order terms in~\eqref{TSP-B-cost} can be performed on a gate-based computer through a series of CNOT gates. Using sophisticated control schemes, this can require on the order of $O(N^3 \log N)$ two-qubit gates\cite{glos_space-efficient_2022}.

In summary, one can represent the same problem formulation using different encoding schemes. A binary encoding reduces the size of the qubit register compared to a one-hot encoding at the cost of more gate operations. This may be preferable for near-term quantum computers with limited qubit registers. However, other encodings are also possible, such as the domain wall encoding\cite{chancellor_domain_2019}, and these have different advantages and trade-offs depending on the specific problem formulation and available hardware.

\subparagraph{Choice of algorithm}

\begin{figure}
    \centering
    \includegraphics[width=1.0\linewidth]{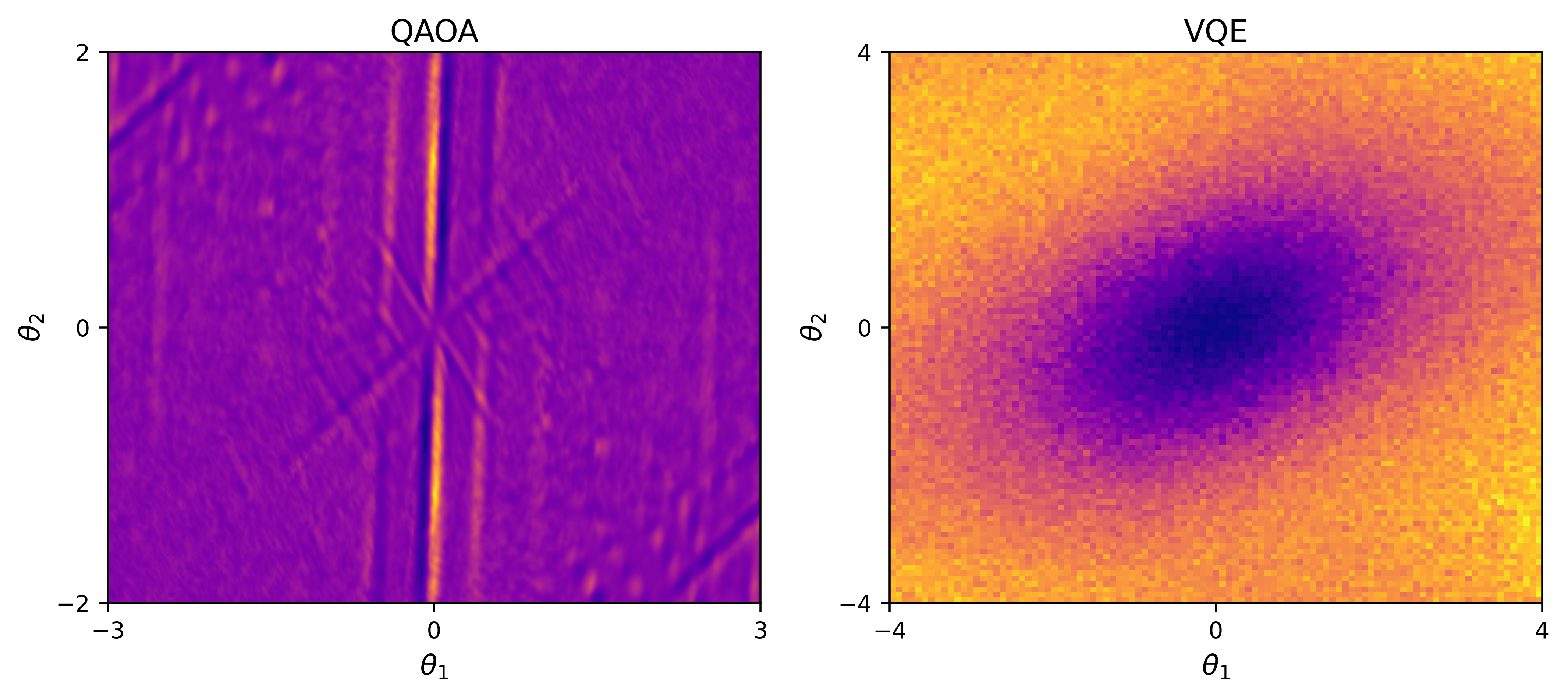}
    \caption{Density plot of the loss landscapes for QAOA (left) and hardware-efficient VQE ansatz (right) applied to the TSP. Plots have been reproduced following the methods outlined in Palackal~\textit{et al.}~\cite{palackal_quantum-assisted_2023} for a 4-node instance. This includes $p=5$ layers for QAOA and a VQE ansatz given by a single layer of single-qubit X and Z rotations followed by nearest-neighbour controlled-X entangling rotations. Results have been obtained through classical simulations by slicing a two-dimensional plane randomly oriented in the parameter space of 10 (QAOA) and 27 (VQE) dimensions. The QAOA landscape is more structured, with multiple local minima and valleys with high curvature. These regions are surrounded by a barren plateau with very small curvature. The VQE landscape has some radial symmetry with features that resemble noise. Palackal~\textit{et al.}~\cite{palackal_quantum-assisted_2023} minimise their loss landscapes using a variety of classical optimisers and conclude that gradient-free optimisers, particularly Powell and NFT, perform best.}
    \label{fig:palackal_LL}
\end{figure}

Finally, for gate-based devices the hybrid algorithm influences the solution quality and therefore guides the choice of classical optimisation technique. This is because the solution landscape produced by QAOA and different VQE ansatzes can differ considerably. Palackal~\textit{et al.}~\cite{palackal_quantum-assisted_2023} have demonstrated this explicitly for the TSP through density plots of the loss landscape. We have replicated their methodology and produced comparable results as depicted in Figure~\ref{fig:palackal_LL}. These reveal that the QAOA landscape has significantly more structure than their chosen hardware-efficient VQE ansatz (consisting of a single layer of single-qubit X and Z rotations
followed by nearest-neighbour controlled-X entangling rotations). The QAOA landscape is anisotropic with valley-like regions of sharp curvature and multiple local minima. Outside of these valleys there is effectively no curvature and the landscape can be characterised as a barren plateau. This makes optimisation challenging because many measurement shots are required to estimate the gradient of the cost function. In contrast, the VQE landscape has some radial symmetry, low curvature, and fluctuations that resemble noise. Note that these landscapes are specific to the QAOA circuit depth, the mixing Hamiltonian, and the specific VQE ansatz used.

The landscape is critical to consider when tailoring the specifics of each VQA. For the QAOA landscape, it may be advantageous to use classical heuristics to obtain a so-called ``warm start'' for the hybrid algorithm. For example, through an initial coarse sampling of the landscape or approximate solution to the TSP it is possible to guide initial parameters for QAOA away from barren plateaus. Conversely, the fluctuations in the VQE landscape may require more measurement shots than usual to provide a good representation of the gradient. Regardless, for both VQE and QAOA multiple authors have highlighted the importance of the classical optimisation algorithm in the parameter update step. They note that gradient-free optimisers (such as SPSA and COBYLA)\cite{azad_solving_2023,mohanty_analysis_2023,palackal_quantum-assisted_2023} provide better convergence to optimal solutions. This enhancement is often attributed to better handling of rugged landscapes as seen in the QAOA landscape of Figure~\ref{fig:palackal_LL}.

\paragraph{Performance}

The current performance of quantum algorithms for standard implementations of the TSP and CVRP is underwhelming. Table~\ref{tab:routing} presents several representative performance metrics for quantum annealers and gate-based computers. For example, Palackal~\textit{et al.}~\cite{palackal_quantum-assisted_2023} perform VQE on a 27-qubit IBM quantum computer to solve the TSP for six customers and obtain a route which is 27\% longer than the optimal solution. Two recent reviews on the TSP by Osaba~\textit{et al.}\cite{osaba_systematic_2022} and Smith-Miles~\textit{et al.}\cite{smith-miles_travelling_2025} demonstrate that across the board, quantum computers fail to find optimal or near-optimal solutions to relatively simple TSP instances with small numbers of cities ($\lesssim10$). These poor results can partly be attributed to hardware limitations, with small qubit volumes limiting the size of problem instances and errors/decoherence limiting accuracy.

While current performance may seem discouraging, it is important to clarify that published results for routing problems do not necessarily represent the state-of-the-art for quantum computing. The translation of quantum computing from academia to industry has led to the development of proprietary algorithms and services which greatly improve performance. For example, Q-CTRL's Fire Opal software enhances circuit fidelity on NISQ hardware through optimised gate decompositions and automated error-mitigation techniques. In combination with a proprietary hard-mixer to encode constraints, they claim a 22-fold reduction in relative error and enhanced circuit performance compared to conventional QAOA implementations of the CVRP\cite{bentley_quantum_2022}. Q-CTRL has now applied their optimised methodology in multiple practical and industry settings, including scheduling for part of London's train network\cite{q-ctrl_accelerating_2025}. They claim optimal scheduling of 26 trains over 18 minutes of real scheduling data.

While it is difficult to assess specifics of Q-CTRL's methodology and solution quality, it is undeniable that these hardware-optimised implementations significantly improve performance. This can be inferred through drastic increases in solution quality reported for other binary optimisation problems\cite{sachdeva_quantum_2024,mcgeoch_comment_2024}. While performance metrics will undoubtedly improve with advances to both theory and hardware, this does not necessarily guarantee utility of quantum computers for routing problems. To reiterate the argument in the review by Smith-Miles~\textit{et al.}~\cite{smith-miles_travelling_2025}, state-of-the-art classical heuristics for the TSP are extremely efficient and QUBO formulations -- executed with or without quantum algorithms -- are unlikely to be competitive.

\paragraph{Towards practical implementation}

The challenge now facing transport engineers is to identify routing problems which are both of practical interest and quantum compatible. Significant work has explored optimal QUBO formulations, algorithms, and hardware for the relatively simple routing problems discussed above. However, industry needs are significantly more complicated than the standard TSP or CVRP with time windows. Practical problems are complex. They may involve varying transport modes embedded within larger supply chains, worker schedules, and dynamic customer demands amongst a myriad of other requirements. All these factors require additional qubits or gate depths which are crippling for current and near-term hardware. Even for standard industry problems, such as the dynamic and time-scheduled\cite{irie_quantum_2019} multi-depot CVRP, the number of constraint terms can quickly become unwieldy and thereby degrade solution quality.

Weinberg~\textit{et al.}~\cite{weinberg_supply_2023} provide an excellent case study on adapting quantum algorithms to the complexity of real-life routing problems. The authors develop a HOBO formulation to describe a realistic supply chain for the Aisin Corporation, a Japanese automotive manufacturer. Limitations in existing quantum hardware make a monolithic treatment of all nodes and vehicles infeasible, and hence the authors adapt an iterative approach where each vehicle route is considered separately. The optimal route is obtained for a given vehicle, and then the HOBO formulation is updated to reflect the routing requirements for the remaining vehicles (i.e., the terms in the cost function are updated). The route for the subsequent vehicle is then optimised. Consequently, the problem becomes tractable while still maintaining limited interactions between vehicle routes. Weinberg~\textit{et al.} optimise the supply chain for 74 vehicles in this manner using the D-Wave hybrid solver and obtain 99.39\% demand satisfaction. While the authors do not compare their results to state-of-the-art classical solvers, their work presents a viable means for adapting near-term hardware to the complex requirements of industry.

Finally, quantum technologies have also been used for minimising traffic congestion through vehicle re-routing. The original formulation developed by Neukart~\textit{et al.}~\cite{neukart_traffic_2017} considered 418 taxis travelling between the city centre and Beijing airport. The original routes of these taxis were obtained through GPS data and the goal was to identify alternative routes for each taxi which led to a decrease in overall network congestion. This was achieved through a QUBO cost function which calculates the squared value of all vehicles on a given road link, then sums this value over the entire network. The optimal solution therefore reduces overall congestion by spreading vehicle routes over the entire network. Yarkoni~\textit{et al.}~\cite{yarkoni_quantum_2020} then adopted this formalism for real-time routing. Assisted by classical data collection techniques and an app service, buses at the 2019 Web Summit conference in Lisbon, Portugal were re-routed live to minimise congestion. These works were performed using the D-Wave hybrid service.

\subsection{Scheduling}

\paragraph{Overview}

Scheduling and timetabling problems have received significant attention from academia and industry looking to leverage quantum technology. From the 13 scheduling studies identified in the literature review, existing domains of interest are trains/rolling stock (7), aircraft (4), and electric vehicle charging (2). Broadly speaking, these works aim to allocate and sequence vehicles or services to timing, capacity, maintenance, and safety constraints. They consider a variety of operational contexts from real-time or disruption response scheduling, short-horizon planning, and strategic long-term planning. All but one of the works adopt a QUBO formulation. Similar to routing problems, authors have identified that practical problems require inclusion of onerous constraint terms which inhibit performance. They have adapted to this limitation through classical pre- and post-processing stages which maximise computing resources. We first review the quantum advantage evidence criteria before discussing results for train and aircraft scheduling in more detail.

\paragraph{Evidence criteria assessment}

The scheduling studies cover a variety of focuses and scenarios, requiring more individualised assessments of the evidence criteria. We first consider the QUBO studies. For C1, 7 studies were rated as N and 5 as P. The three studies by Domino~\textit{et al.}~\cite{domino_quadratic_2022}, Domino~\textit{et al.}~\cite{domino_quantum_2023}, and Koniorczyk~\textit{et al.}~\cite{koniorczyk_solving_2025} address variants of the railway dispatching problem and were assigned a rating of P. This broadly consists of adjusting routes and schedules following train delays, and is a complex problem partly due to the competition for physical resources (e.g., blocks, junctions), density of traffic, and requirement for fast solutions in real-time. All three studies formulate or solve an equivalent MIP model classically. While these are relevant baselines for the small instances solved, they are not established as the state-of-the-art.

Two studies by Vikstål~\textit{et al.}~\cite{vikstal_applying_2020} and Svensson~\textit{et al.}~\cite{svensson_hybrid_2023} consider airline scheduling (the tail-assignment problem) through branch-and-price. This is broadly considered the state-of-the-art classical method, and involves pricing to generate feasible aircraft routes while a restricted master problem (set-partitioning) selects a combination of routes that covers the flights. Nonetheless, these studies are assessed as N for C1 as QAOA is applied to the set-partitioning subproblem alone, and does not consider dedicated classical solvers solely within this context.

For C2, 5 studies are rated as N, 6 as P, and 1 as M. Those rated as P chiefly suffered from similar limitations to the routing studies. Namely, aspects of the quantum run-times were reported, but failed to account for embedding or critical pre- and post-processing stages. This is a significant omission, as many of the scheduling studies discussed below rely on classical processing to reduce the complexity of QUBO formulations for suitability with NISQ-era hardware. The exception is Koniorczyk~\textit{et al.}, who present a rigorous run-time scaling with problem size against their CPLEX solves. Ultimately, this demonstrated that hybrid annealing was not yet competitive with classical approaches. Consequently, none of the QUBO-based scheduling studies demonstrate a compelling claim to quantum advantage. 

Finally, Grange~\cite{grange_design_2024} performs a theoretical study which considers train scheduling within the broader computational problem of finding exact job sequences over constrained resources. These results are primarily derived for abstract single-machine and flow-shop scheduling problems rather than a specific railway formulation. The approach classically computes partial solutions using dynamic programming and applies quantum minimum finding to search over balanced combinations of these solutions. It meets C1 by comparing its running-time bounds directly with cited exact classical algorithms for the same scheduling problems and solution requirements. It meets C2 because the asymptotic analysis includes all relevant algorithmic costs, with the explicit caveat that advantage requires QRAM access. Finally, it meets C3 by analytically reducing an \(O^*(2^n)\) classical dynamic-programming scaling to \(O^*(1.728^n)\) for several single-machine problems (where $n$ is the number of jobs), with related improvements for three-machine flow-shop scheduling. While these results show that quantum advantage is plausible, Grange does not identify a crossover point at which the improved asymptotic scaling would provide a practical benefit for railway scheduling.

\paragraph{Rail scheduling}

Rail scheduling has been considered in two different settings: pre-operational planning and real-time dispatching after disruption. For the former, compatibility with quantum methods is generally obtained by reducing the original problem to a selection among classically prepared options. For example, Xu~\textit{et al.}~\cite{xu_high-speed_2023} first formulate a detailed space--time model incorporating train movements, stops, station capacity, headways and maintenance. The corresponding QUBO representation is highly complex with many constraint terms, and the authors report that the solutions were almost always infeasible. Hence, they reframe the problem by considering the largest mutually compatible subset of historical trajectories whose times, stops, and paths are known to be feasible. The resulting QUBO is significantly simpler and attains higher quality results. Bickert~\textit{et al.}~\cite{bickert_optimising_2023} consider the allocation of fixed passenger trips to rolling stock while satisfying maintenance requirements. Maintenance is enforced exactly in their classical constraint-programming model but only approximately through penalties in the QUBO. This requires trips that violate maintenance limits to be removed during post-processing. On the larger instances, the D-Wave hybrid solver consequently retains fewer trips and requires substantially longer than their classically implemented constraint-programming solver. Hence, these two examples demonstrate how quantum compatibility can be achieved on NISQ-era devices through classical pre- or post-processing.

Domino~\textit{et al.}~\cite{domino_quadratic_2022,domino_quantum_2023} and Koniorczyk~\textit{et al.}~\cite{koniorczyk_solving_2025} directly address train dispatching. Specifically, they address how to re-time and reorder train trips following a disturbance so that infrastructure conflicts are avoided and secondary delay is minimised. Domino~\textit{et al.} demonstrate the cost of retaining operational detail. Their multi-track formulation is natively a HOBO problem, which must be converted to a QUBO through the introduction of many auxiliary variables. However, they show that a simpler single-track formulation can remain quadratic by assigning stations unlimited capacity and rejecting capacity-violating schedules afterwards. Across both studies, direct annealing produces feasible schedules only for the smallest cases and deteriorates rapidly with problem size. Koniorczyk~\textit{et al.} develop a substantially richer fixed-route rescheduling model which is solved through hybrid annealing. They compare its performance with a MIP formulation using CPLEX and find that hybrid annealing is generally slower. Moreover, for the cases in which hybrid annealing is faster, substantial losses in objective quality are incurred.

\paragraph{Aircraft scheduling}

Aircraft studies have addressed three distinct scheduling problems. A substantial investigation has been performed by Stollenwerk~\textit{et al.}~\cite{stollenwerk_quantum_2020} who developed a QUBO formulation for de-conflicting aircraft trajectories. Using 984 real flights over the North Atlantic Ocean, they consider the optimal way to delay aircraft takeoffs while avoiding any spatial conflicts. For compatibility with NISQ-era hardware, potential conflicts and independent groups of interacting flights are identified classically before each resulting problem is mapped to QUBO. This is a useful decomposition of a realistic dataset, but route construction and conflict detection remain outside the quantum computation. Indeed, only departure delays are optimised, and the largest reported cases demonstrate successful embedding on quantum annealers rather than compelling quantum advantage.

Mohammadbagherpoor~\textit{et al.}~\cite{mohammadbagherpoor_exploring_2021} explore the optimisation of airline gate scheduling through a collaboration with Delta Air Lines. This work considers the ideal assignment of planes to gates by minimising the distance travelled by passengers between connecting flights. They first produce a one-hot encoding which requires $N\times K$ decision variables, for $N$ total flights and $K$ gates. However, this was reduced to $N\times \lceil \log_2(K) \rceil $ variables through reformulation into a graph colouring problem. The authors perform VQE simulations for a 27-qubit IBM quantum computer on problem instances up to five planes and eight gates. They successfully demonstrate how the compact encoding reduces the qubit requirements, circuit depth, and run times.

Vikst{\aa}l~\textit{et al.}~\cite{vikstal_applying_2020} and Svensson~\textit{et al.}~\cite{svensson_hybrid_2023} develop a compelling hybrid approach for the tail assignment problem which exploits the individual strengths of classical and quantum solvers. In branch-and-price, the pricing problem iteratively generates aircraft routes that satisfy connection, maintenance, and other operational requirements, while the restricted master problem selects a combination of these routes that covers every flight. This provides a natural hybrid boundary to divide the classical and quantum labour. The complicated scheduling constraints remain within classical pricing, leaving a comparatively simple exact-cover or costed set-partitioning problem for QAOA. Svensson~\textit{et al.} develop this hybrid architecture and propose QAOA as a primal heuristic for the restricted master problem. Specifically, the quantum solver aims to accelerate pruning of the search tree by supplying improved integer bounds on target nodes. The ultimate speedup provided by this approach is likely to be modest. Svensson~\textit{et al.} only expect a constant-factor speedup in the overall branch-and-price process and state that a polynomial speedup is uncertain.

\subsection{Traffic signalling}

\paragraph{Overview}

Five different works have considered network-level control of traffic signals using quantum and hybrid annealing\cite{hussain_optimal_2020,inoue_traffic_2021,marchesin_improving_2023,inoue_traffic_2024,shikanai_quadratic_2025}. These studies encode a signal mode at each intersection using binary variables and repeatedly re-optimise those modes within an evolving traffic model. A QUBO formulation is used to coordinate modes between adjacent intersections. Individual terms in the Hamiltonian aim to reward localised metrics such as discharging queues, minimising downstream congestion, or the promotion of successive green signals. Performance is then evaluated through network-level metrics such as throughput or total emissions. Results on small-scale networks (both realistic and synthetic) indicate that quantum signal control can produce network-level operations which are comparable to other simple control methods. However, it remains questionable whether QUBO formulations can attain meaningful network-level outcomes at scale and across diverse network topologies. It is also unclear whether quantum hardware improves solution quality or speed rather than simply reproducing a controller that classical methods can solve.
\paragraph{Evidence criteria assessment}

For C1, all five traffic signalling studies were assessed as P. These works share a common problem approach wherein local interactions are encoded into a QUBO matrix. Finding the optimum network configuration is therefore a combinatorial optimisation problem. In this sense, all five works approach this optimisation using strong classical heuristic methods such as tabu search or simulated annealing. However, from the overall perspective of signal control, it is necessary to compare the QUBO approach to state-of-the-art methodologies. This deficiency is discussed in the review below.

For C2, four studies are assigned P as they report some useful timing information, but timing boundaries, classical hardware and tuning, overheads, stopping conditions, or matched solution-quality targets remain incomplete. Inoue~\textit{et al.}~\cite{inoue_traffic_2021} report no elapsed time and are rated as N. None of the studies demonstrated a compelling case for quantum advantage and hence were rated as N for C3. Overall, the evidence therefore demonstrates the feasibility of QUBO-based controllers, and in some cases, their improvement over simple local coordination. However, they do not demonstrate that quantum annealing improves the complete closed-loop control workflow.

\paragraph{Implementation and analysis of quantum traffic signal control}

Traffic signal control is an innovative and promising application for quantum annealing. The QUBO matrices are based on local interactions between adjacent intersections and are therefore relatively sparse. Consequently, they embed efficiently onto annealing network topologies and are therefore good candidates for quantum advantage. The challenge ultimately lies in extracting utility. Defining a cost function which optimises a value of practical interest (such as vehicle wait time or throughput) has proven non-trivial. Specifically, it is difficult to formulate these vehicle-centred cost functions directly given that the decision variables are signal modes. Instead, the values of practical interest must be captured indirectly through the clever design of cost functions involving local interactions between signal modes of neighbouring intersections.

Researchers have designed these local interactions through two different approaches. In the first, a linear term favours signal modes that maximise vehicle throughput while a quadratic term incentivises ``green waves'' between neighbouring intersections (i.e., increases the probability of uninterrupted progression through consecutive green lights)\cite{hussain_optimal_2020,marchesin_improving_2023,shikanai_quadratic_2025}. The second approach aims to minimise flow bias at each intersection by equalising queue lengths across competing signal modes and further penalises frequent switching of modes\cite{inoue_traffic_2021,inoue_traffic_2024}. Both approaches enforce the obvious constraint that only a single mode may be active at any given intersection. 

The effectiveness of the traffic signal optimisation is then evaluated through time-dependent simulations. Vehicles are loaded into either simplified or real-world networks using macroscopic flow models\cite{hussain_optimal_2020,inoue_traffic_2021,marchesin_improving_2023} or detailed microscopic simulations\cite{inoue_traffic_2024,shikanai_quadratic_2025}. The density of vehicles at each intersection defines parameters in the Ising model, which is subsequently solved using annealing or classical algorithms to determine the optimal signal pattern for the entire network. While it is also possible to use gate-based VQAs, this has not yet been performed. Vehicles advance through the network according to the traffic flow model over a chosen time-step. The updated traffic state defines new parameters in the QUBO model and the signal pattern is optimised accordingly as the simulation repeats. Performance of the optimised signal control is then assessed using metrics such as average vehicle speed, total wait times, and CO${}_2$ emissions.

\begin{figure}
    \centering
    \includegraphics[width=0.75\linewidth]{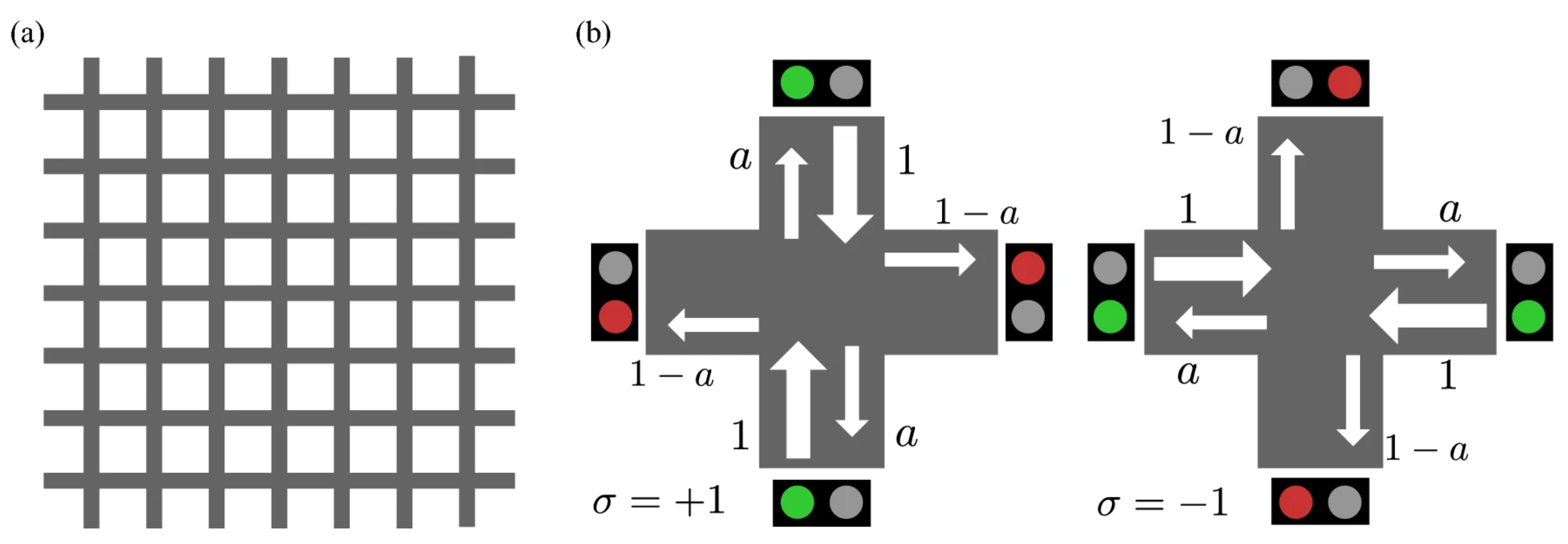}
    \caption{Basic structure of the Ising model developed by Inoue~\textit{et al.} The decision variables $\sigma$ encode different signal modes for intersections arranged in a simple grid. These signal modes describe traffic flow in the north-south or east-west direction. Vehicles can travel straight through the intersection with probability $a$ or turn left with probability $1-a$. Reproduced from Inoue~\textit{et~al.}~\cite{inoue_traffic_2021} under the terms of the Creative Commons Attribution 4.0 International License (CC BY 4.0).}
    \label{fig:inoue_fig_1}
\end{figure}

Inoue~\textit{et al.}~\cite{inoue_traffic_2021} developed one of the first Ising models for signal optimisation and provide a good representation for the workflow of the other studies. As presented in Figure~\ref{fig:inoue_fig_1}, they consider a $50\times50$ grid of intersections with two possible signal modes describing north-south and east-west flow. In their simple model, cars may travel straight through the intersection with probability $a$ and turn left with probability $1-a$, though this assumption can be generalised in a straightforward manner. The decision variables describe the two modes at each intersection and the cost function seeks to minimise local flow bias while penalising frequent mode switching. This formulation is highly compatible with quantum computers because each intersection interacts only with its nearest neighbours. Hence, the quadratic cost matrix becomes sparser as the network size increases.

The authors use the D-Wave 2000Q to solve for the optimal configuration of traffic modes at each time step. This annealer has only 2,048 qubits and therefore it is necessary to decompose the 2,500 intersections in Inoue~\textit{et al.}'s $50\times50$ grid into smaller sub-networks. This procedure could be performed through D-Wave's hybrid solver, but the authors opt for a direct approach using the classical METIS algorithm. This algorithm clusters adjacent intersections based on their mutual interaction strengths. Intersections within each sub-network interact strongly whereas interactions between different sub-networks are weaker. This approximation reduces the larger QUBO problem into many sub-problems which the quantum annealer can then solve individually. For example, Inoue~\textit{et al.}~\cite{inoue_traffic_2021} divide their $50\times50$ grid into 42 sub-networks.

\begin{figure}
    \centering
    \includegraphics[width=0.9\linewidth]{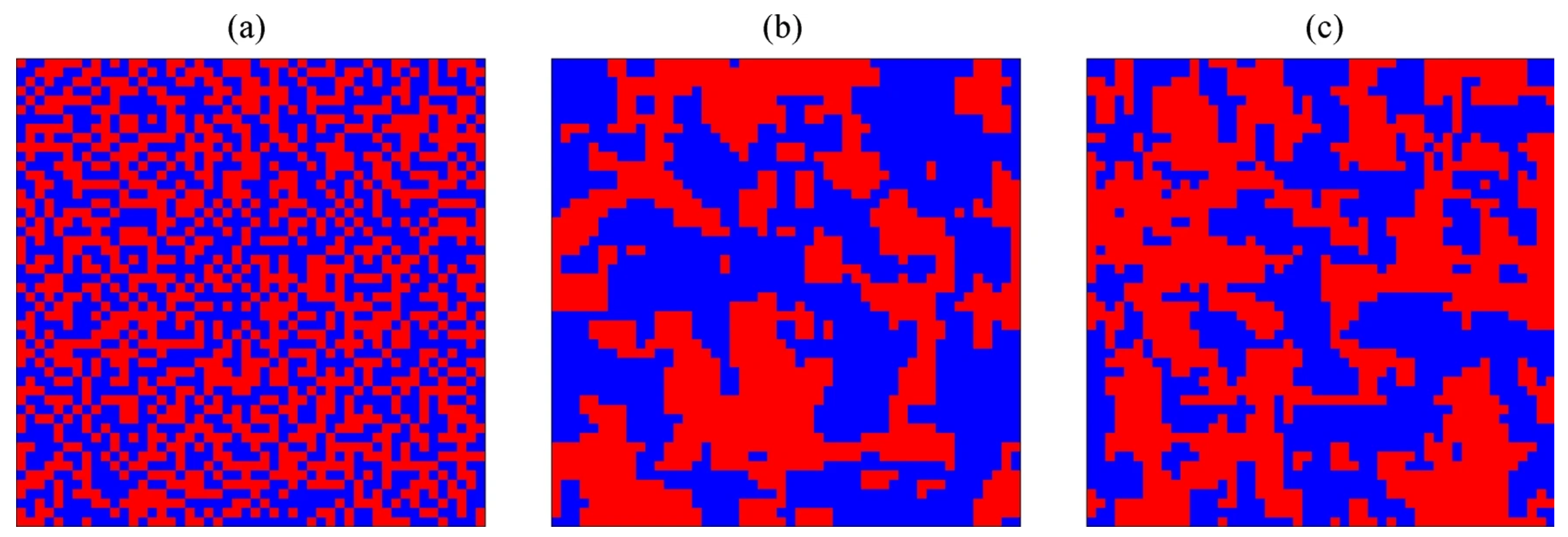}
    \caption{Graphical representation of the optimal signal modes as identified by Inoue~\textit{et al.} The different colours represent the signal state at each intersection, with blue indicating the east--west mode and red the north-south mode. Their results compare basic local control in (a) and the QUBO model solved via simulated and quantum annealing in (b) and (c), respectively. Global control using the Ising model can reduce traffic congestion through long-range synchronisation of signal modes. Reproduced from Inoue~\textit{et~al.}~\cite{inoue_traffic_2021} under the terms of the Creative Commons Attribution 4.0 International License (CC BY 4.0).}
    \label{fig:inoue_fig_2}
\end{figure}

Figure~\ref{fig:inoue_fig_2} presents the optimised signal models as determined by Inoue~\textit{et al.} The different colours represent the signal state at each intersection, with blue indicating the east--west mode and red the north-south mode. Their results compare basic local control in (a) and the QUBO model solved via simulated and quantum annealing in (b) and (c), respectively. Unsurprisingly, the Ising model produces synchronised signal modes with long-range correlations whereas the local control is effectively random. Other studies using different Ising models likewise show that this signal synchronisation enhances vehicle throughput and outperforms simple control methods such as fixed cycles\cite{hussain_optimal_2020,marchesin_improving_2023,inoue_traffic_2024,shikanai_quadratic_2025}. However, the extent of this enhancement varies depending on network load and traffic model parameters. Moreover, these studies consistently find quantum annealers perform on par with classical methods such as simulated annealing and Tabu search in identifying optimal signal patterns.

The overall benefit of QUBO models relative to sophisticated methods of signal control has yet to be demonstrated. Their performance has not been compared to intelligent approaches in urban networks such as model predictive control\cite{hao_mpc_signal_2018}, max-pressure algorithms\cite{levin_max_pressure_2020,liu_nmp_2024,GHOSH2026105804}, or machine learning\cite{bie_marl_signal_2024,yazdani_ivpl_2023,li_cooperative_perception_2024}. This is a critical shortcoming  given that such approaches have been extensively investigated and, in some cases, deployed or evaluated in operational traffic-control environment. Nor has a comparison been made between different formulations of the Ising model itself. Global control via an Ising model may indeed offer advantages for certain network configurations, but the challenge lies in formulating an optimal QUBO cost function. The impact of local interactions between intersections on the overall network behaviour is difficult to predict. Hence, while the cost functions defined in the above studies are well founded, there is no reason to presume they are ideal.

This calls for a systematic approach to identify the optimal cost function for a given network topology and load. Machine learning techniques are well-poised to address this challenge. Alternatively, it is not necessary to rely on a QUBO formulation to leverage quantum advantage. The cost function may be encoded into Heisenberg-type or other fermionic Hamiltonians\cite{powell_introduction_2010} and optimised using hybrid algorithms. This may even be advantageous, because alternative Hamiltonians may possess more degrees of freedom and therefore encode richer interactions within the network. To summarise, future work should explore a wider set of cost functions and seek to demonstrate performance enhancement over established signal control methods. 

\subsection{Transport Network design}

\paragraph{Overview}

Transport network design involves optimising the locations, capacity, or service offerings of infrastructure subject to budgetary costs or efficiency metrics. These optimisation problems are considered suitable for quantum computing as they often involve a combinatorial search over binary decision variables. We have identified six network design problems which adopt a QUBO formulation and are solved primarily through quantum and hybrid annealing. These include general frameworks for optimising the design of road networks\cite{dixit_quantum_2023}, bus stops\cite{gabbassov_transit_2022}, bike share stations\cite{Nourbakhsh_2024}, and electric vehicle charging stations\cite{okada_joint_2026,Chandra_2022,Rao_2022}. In each study, the  decision variables encode whether to develop, modify, or remove network infrastructure at a given geographic location. Researchers have adopted bi-level frameworks\cite{dixit_quantum_2023}, approximation techniques\cite{gabbassov_transit_2022}, and methods of Bayesian inference\cite{okada_joint_2026} to reduce model complexity for compatibility with existing and near-term hardware. We assess the three evidence criteria and then consider the individual studies in further detail.

\paragraph{Evidence criteria assessment}

Three studies were assigned N for both C1 and C2, while the remaining three were assigned P. Several report encouraging solver-level performance, but none provides evidence sufficient under all three criteria to establish practical end-to-end quantum advantage. The N ratings reflected the absence of a sufficiently competitive classical comparison and usable computational timings. The P studies reported relevant classical comparisons and some timing information, but did not establish that their baselines were state-of-the-art or evaluate the methods at matched solution quality and computational budgets. In particular, the reported timings often omitted important workflow costs or did not measure the time required to reach a common solution-quality target. These limitations are especially informative for interpreting the results of Dixit and Niu~\cite{dixit_quantum_2023} and Rao and Sodhi~\cite{Rao_2022}, discussed below.

\paragraph{Network infrastructure}

Dixit and Niu~\cite{dixit_quantum_2023} used quantum and hybrid annealing within a bi-level framework for transport network design. Their algorithm determines which links should be expanded for greater capacity given a fixed budget. Firstly, the lower-level problem is solved using classical means to determine link flows for a given origin-destination demand assuming user equilibrium. Using these link flows, the upper-level problem optimises the link upgrades that minimise total system travel time subject to budget constraints. The upper-level problem is expressed through a QUBO formalism where the decision variables encode which links to upgrade. However, their QUBO formulation is highly general and could be adapted for free-flow times or even the creation/removal of links. Across benchmark networks containing up to 6,674 links, the authors report that D-Wave's hybrid solver was up to 11 times faster than tabu search; on the largest instance, the hybrid solution was also almost 2\% better.

These are important reported solver-level speed and scaling results. However, they do not yet establish practical end-to-end quantum advantage because critical workflow costs and a common solution-quality target are absent. The study was assigned P for C1 because tabu search is a relevant comparator, but is not established as the strongest method for the complete bilevel problem. For C2, the study was assigned P because the authors compare classical and hybrid annealing runtimes across ten network sizes derived from public benchmarks. They report linear runtime scaling with instance size for hybrid annealing and exponential scaling for tabu search. However, the reported comparison excludes the critical costs of QUBO construction and embedding as well as the end-to-end runtime of the classical lower-level traffic assignment. In addition, the solver budgets were selected using D-Wave Ocean's QPU-access-time estimator, without reporting the time required to reach a common convergence or solution-quality target.

Gabbassov~\cite{gabbassov_transit_2022} formulates bus-stop retention as selecting the $p$ facilities that maximise population accessibility while limiting competition between nearby stops. Applied to Vancouver's B20 route, the method suggests that approximately 40\% of stops could be removed while maintaining geographic coverage. Although classical heuristics are used within the hybrid annealing solution process, no competitive classical-only benchmark or usable runtime comparison is reported. The study was therefore assigned N for both C1 and C2.

Nourbakhsh~\textit{et al.}~\cite{Nourbakhsh_2024} use quantum annealing to assign Toronto bike-share stations to three clusters based on location, capacity, available bikes, and surrounding population. The study was assigned P for C1 because it compares runtime with K-means, an established clustering method. C2 was also rated P because several timing components are reported, but embedding is excluded from the favourable comparison. The authors acknowledge that including this cost makes the classical method faster. The result is therefore best viewed as a real-data clustering demonstration rather than evidence of computational advantage.

\paragraph{Electric vehicle charging station placement}

Three works employ quantum methods to optimise the placement of electric charging stations. Chandra~\textit{et al.}~\cite{Chandra_2022} used quantum annealing to minimise the distance of chargers from pre-defined points of interest while spreading their locations sufficiently far apart. To do so, they develop a QUBO formulation with 3 objectives (minimise distance to points of interest, maximise distance from existing chargers, maximise spread of new chargers) plus a cardinality constraint. They introduce an entropy-based modification to the distance minimisation (softmax over distances) to prevent clustering near a few points of interest. The key novelty is a hybrid annealing and genetic algorithm pipeline. The D-Wave quantum annealing provides an initial discrete solution, which then seeds a continuous genetic algorithm for local refinement. This hybrid pipeline improves performance relative to annealing alone and a randomly initialised genetic algorithm. However, the comparison primarily evaluates the benefit of quantum seeding rather than performance against a competitive charger-placement method, and no wall-clock timings are reported. The study was therefore assigned N for C1 and C2.

Rao and Sodhi~\cite{Rao_2022} consider placement of charging stations in a multi-stakeholder formulation. A two-stage hybrid approach is used to optimise station placement while considering both the transport network (demand capacity) and power grid (voltage deviation, power loss). In the first stage, a classical particle swarm is used to optimise the continuous power grid variables. Namely, this stage determines how to distribute charging station load across buses to minimise voltage deviation, power loss, and demand-load mismatch. This yields a constraint on the number of stations required per bus. In the second stage, they construct a QUBO whose solution decides which specific locations to build. The objective function minimises total cost of ownership and demand-capacity difference subject to the per-bus allocation constraint. They implement this on both a quantum annealer (D-Wave Advantage) and a variational quantum circuit (using the Cirq simulator).

The study was assigned P for C1 because simulated annealing solves the same QUBO and a binary lightning-search algorithm provides an additional heuristic comparison. These are useful baselines, but neither is established as state-of-the-art or evaluated under clearly matched stopping conditions. Rao and Sodhi also provide the most informative timing breakdown of the six studies and demonstrate concretely the classical cost associated with QUBO construction. For a 2,000-variable instance, QUBO construction required approximately 5,000 seconds compared with 5 seconds for the hybrid annealing solver and 30 seconds for simulated annealing. The workflow is therefore dominated by classical QUBO construction, and the total hybrid and simulated annealing times are nearly identical once this common cost is included. However, the study did not report the time required for embedding, and hence was assigned P for C2.

Finally, Okada~\textit{et al.}~\cite{okada_joint_2026} developed a QUBO-based approach to simultaneously route electric vehicles and optimise the location of charging stations. This problem is particularly challenging as the optimal route is strongly influenced by charger location. Ideally, a QUBO formulation would include multiple decision variables to describe vehicle route and charger positions, but such a model is highly complex and likely intractable on existing quantum hardware. Consequently, Okada~\textit{et al.} develop an innovative bi-level framework which iteratively solves for both the ideal route and charger placement. A classical Bayesian procedure learns the ideal charger-site locations through quantum annealing. The much larger routing QUBO is then solved by a GPU annealer while battery feasibility is checked classically. Only one synthetic 20-location instance was considered and the quantum and simulated annealing methods produce similar results. No runtime was reported and hence the study was assigned N for both C1 and C2. Whereas the preceding domains formulate transport decisions primarily as combinatorial optimisation problems, the next group uses QML for predictive tasks in computer vision and traffic forecasting.

\subsection{Computer vision and traffic forecasting}

\paragraph{Overview}

Quantum machine learning has been explored within the transport literature in the contexts of computer vision and traffic forecasting. Quantum-enhanced computer vision has focused on classification of traffic lights\cite{innan_qnn-vrcs_2025}, road signs\cite{kuros_traffic_2023}, number plates\cite{Kang_2025}, as well as the analysis of shadow, lane, and steering direction\cite{meghanath_qdcnn_2025}. Traffic forecasting has considered short-term vehicle flow at loop detectors\cite{schetakis_quantum_2025} and bike-share demand\cite{harikrishnakumar_forecasting_2023}. Across these applications, the quantum component is generally a small layer or subroutine embedded within a substantially larger classical workflow for detection, compression, discretisation, optimisation, or post-processing. All six studies execute their quantum circuits through classical simulation. Overall, the works in this section demonstrate that quantum methods can successfully be incorporated into broader classical pipelines. However, it is still debatable whether the quantum component improves predictive accuracy, training/inference cost, or scaling relative to a strong task-matched classical method.

We note that the studies identified in this review represent a relatively narrow subset of the potential applications of QML within intelligent transport systems. Indeed, much of the wider QML literature may be relevant to transport perception and decision support. This includes methods for object detection, semantic segmentation, anomaly detection, LiDAR analysis, and multimodal fusion of camera, radar, and LiDAR data~\cite{rattan_quantum_2025,pandharipande_sensing_2023}. Moreover, quantum sensing technologies could eventually augment these pipelines at the data-acquisition stage~\cite{gallego_torrome_advances_2024}. The small corpus identified here therefore reflects the early stage of transport-specific QML research rather than the overall scope of intelligent transport systems.

\paragraph{Evidence criteria assessment}

All four computer-vision studies were assigned P for C1. Each performs a relevant classical comparison on at least some similar data. Conventional convolutional neural networks are tested for traffic light\cite{innan_qnn-vrcs_2025} and sign classification\cite{kuros_traffic_2023}, the number plate classifier is evaluated after removing its quantum layer\cite{Kang_2025}, and a classical deep neural network is tested for shadow/lane/steering analysis\cite{meghanath_qdcnn_2025}. However, these classical models are incompletely specified or matched in their tuning and computational budgets. None are established as the strongest available method for its task. Only the number plate study by Kang~\textit{et al.}~\cite{Kang_2025} partially meets C2 because it reports run time for the quantum layer, but all other costs are omitted. The other three studies are rated N because they do not report runtimes.

For traffic forecasting, the flow count study by Schetakis~\textit{et al.}~\cite{schetakis_quantum_2025} partially meets C1. They compared each quantum forecasting model with a corresponding classical neural network using the same traffic data and evaluation procedure. However, the authors note that stronger graph-based forecasting methods exist but are not tested. They partially meet C2 because they report epoch and batch counts but omit any end-to-end runtimes. The bike-share study by Harikrishnakumar~\textit{et al.}~\cite{harikrishnakumar_forecasting_2023} is rated N for both C1 and C2 because the classical methods serve only as components or checks within the forecasting pipeline. They do not consider an end-to-end classical alternative and no computational timings are reported.

\paragraph{Computer vision}

Recent work on QML for transport-related image classification has produced mixed results. Innan~\textit{et al.}~\cite{innan_qnn-vrcs_2025} designed a QNN to classify traffic light images into red, yellow, and green as part of a vehicle--road cooperation system. Their simulated model uses angle encoding to map downsampled image pixels to rotation angles on qubits. The data is then transformed through a PQC formed through a series of entangling layers. This QNN is evaluated alongside simpler quantum baselines based on $U U^\dagger$ classifiers as well as classical neural networks trained on two different traffic-light datasets. The authors report that the QNN achieves accuracies of 97.42\% and 84.08\% on the datasets, outperforming their classical counterpart at comparable input resolutions. Moreover, the QNN showed reasonable robustness to several noise channels, though within a relatively small-scale experimental regime.

Meghanath~\textit{et al.}~\cite{meghanath_qdcnn_2025} developed a QNN for detecting shadows in road images to inform safer navigation of autonomous vehicles. Their approach was to process patches of images using a classical pipeline and then use a $UU^\dagger$-style quantum circuit to compare pixel intensities against references. This quantum shadow-detection layer is embedded into a broader ``quantum deep convolutional neural network'' and complemented by a small PQC for steering-direction classification. The latter encodes lane geometry into a variational circuit trained via classical optimisation. Across both tasks, the authors report that the quantum-enhanced approach yields higher accuracy and markedly lower theoretical detection time than several classical thresholding and texture-based methods. However, the implementation remains at very small quantum scale and these gains assume idealised gate-operation and noise models.

By contrast, Kuros and Kryjak~\cite{kuros_traffic_2023} present a more conservative assessment in their study of traffic sign recognition. Here they develop a hybrid ``quanvolutional'' network and compare it directly to a purely classical deep convolutional neural network. Their architecture pre-processes traffic sign images and then applies quantum convolution layers built from small $2 \times 2$ patches encoded into qubits and transformed by shallow random circuits. The classical and quantum models are trained on the same data and evaluated across 43 different sign classes. While the hybrid QNN attains over 90\% accuracy, the classical neural network was clearly superior, achieving $99.6$\% accuracy with higher precision. In summary, these studies suggest that quantum-enhanced image classifiers can sometimes match or modestly outperform hand-picked classical baselines in constrained settings. This manifests only with careful engineering around simple encodings and small image sizes. However, they do not yet surpass strong deep learning models trained on realistic, large-scale benchmarks, and there is no evidence of broad advantage. 

\paragraph{Traffic forecasting}

Schetakis~\textit{et al.}~\cite{schetakis_quantum_2025} consider a different application of QML and focus on short-term traffic flow prediction. Their approach inserts small quantum layers into otherwise classical neural networks and considers whether this provides any advantage over standard architectures. They test two methods for predicting traffic counts from a roadside detector in Athens: (1) replacing a simple fully connected layer with a quantum layer, and (2) replacing a long short-term memory (LSTM)-style recurrent step with a quantum ``re-uploading'' layer that sees the input multiple times. The results are mixed. When swapped in for a basic dense layer, the quantum models generally perform worse than the classical counterpart. However, when used in place of the LSTM recurrence, the quantum version tends to train faster and achieve slightly better prediction accuracy on the traffic data. Consequently, quantum networks may offer modest gains in temporal forecasting tasks in scenarios where long-term structure matters.

Harikrishnakumar and Nannapaneni~\cite{harikrishnakumar_forecasting_2023} present a quantum Bayesian network to forecast bike-sharing demand. They first train two classical time-series models (a long short-term memory network and a Gaussian process regression) to predict hourly demand on New York's Citi Bike system. The continuous forecasts are discretised into four equal bins and structured into a simple three-node Bayesian network (two predictors feeding an aggregate demand node). This network is mapped to a quantum circuit by encoding the joint probability distribution into the amplitudes of a multi-qubit state. Specifically, each node's conditional probability table is realised through a series of controlled $R_{y}$ rotations on its qubits, with extra ancilla qubits used to decompose any multi-controlled rotations following the method in~\cite{borujeni_quantum_2021}. By measuring the qubit register in the computational basis and aggregating outcome frequencies, it is then possible to obtain the marginal distributions of the classical Bayesian network. The authors demonstrate this through Qiskit simulations of the  quantum circuit. They use 8,192 shots to obtain marginal probabilities that match the classical Bayesian network to within 2\%. While currently limited in scale, this work demonstrates how quantum Bayesian networks can be applied in the context of transport forecasting. The final application departs from both combinatorial optimisation and QML, using the QFT as a quantum signal-processing subroutine for drive-cycle analysis.

\subsection{Drive cycle analysis}

Dixit and Jian~\cite{dixit_quantum_2022} use the quantum Fourier transform (QFT) to analyse vehicle drive cycles. Specifically, they estimate the dominant frequency at which a vehicle alternates between acceleration and deceleration. This valuable information provides insight into fuel efficiency, emissions, and operational safety and has further applications in active control. They convert real acceleration data into a binary sequence and compare the resulting quantum frequency spectrum directly with the classical fast Fourier transform (FFT). An eight-qubit simulation of 256 time points closely reproduced the FFT spectrum. Next, a hardware experiment was performed on a 15-qubit IBM device using a four-qubit circuit to process 16 time points. The authors then develop a procedure to mitigate hardware noise through classical post-processing. By comparing classically simulated and experimental outputs, they calibrate corrections to the hardware results and discard frequency components below a noise threshold. The correction increased the reported correlation between the hardware and simulator probabilities from 0.59 to 0.86. On this basis, the authors argue that the QFT together with their stated post-processing retains \(O((\log N)^2)\) operation complexity and remains asymptotically faster than the \(O(N\log N)\) FFT even on noisy hardware.

The study meets C1 because the FFT is executed on the same signals and is the most suitable classical comparison. This is an important reported complexity advantage for a method demonstrated on actual quantum hardware. It should, however, be distinguished from an empirical end-to-end runtime advantage: the comparison is based on operation counts rather than measured wall-clock times and does not include state preparation, repeated measurement and readout, or the calibration cost of noise mitigation. Because these essential costs are neither included nor quantified, the study is rated N for C2. Under highly optimistic assumptions, the omitted input and output costs alone reduce any exponential improvement to at most a logarithmic advantage over the FFT. In the absence of explicit timing data over multiple problem scales, the study is therefore assigned N for C3. In summary, this work constitutes an important proof of concept for the application of quantum signal-processing methods to transportation data, but further analysis is required before advantage can be conclusively claimed.

\subsection{Review summary, cross-cutting insights, and limitations across applications}

Transport researchers have used quantum computers to explore a diverse set of use cases. They have done so by utilising the full gamut of available quantum algorithms, with a preference for solving QUBO formulations using VQAs or annealing. However, our systematic analysis of the corpus is sobering and indicates that application breadth has outpaced evidence. While several studies report encouraging solver-level speedups and scaling trends, these results do not yet establish end-to-end quantum advantage. Very few studies have demonstrated a plausible claim to quantum advantage. This shortcoming is critical because the central utility of quantum computers lies in their ability to deliver time, memory, or energy benefits over classical approaches. Without this advantage, the benefits of quantum computing for transport applications are limited and the long-term success of this growing subfield is doubtful.

The principal evidence deficit is methodological. As documented in Table~\ref{tab:corpus_evidence}, approximately 2/3 of studies in the corpus were rated as N for criteria C1 and C2. This indicates that the quantum transport literature is largely not making comparisons to classical methods and not adequately reporting run times. These studies are therefore incapable of establishing any compelling evidence for quantum advantage. Approximately 1/3 of studies were rated as P for criteria C1 and C2. For C1, the primary reason studies were allocated a rating of P was because they compared to a relevant baseline (such as simulated annealing or a generic MIP solver), but did not establish these methods as state-of-the-art. For C2, the primary reason for a P rating was inadequate reporting of end-to-end run times. For example, many annealing studies report the total annealing time, but fail to report the embedding or post-processing times. These costs must be accounted for as they can dominate the run-time. For example, the works by Rao and Sodhi~\cite{Rao_2022} and Vargas~\textit{et al.}~\cite{Vargas_2024} reveal the onerous costs for QUBO construction and embedding within their problem contexts. Only two studies were rated as M for C3, which were theoretical studies by Moylett~\textit{et al.}~\cite{moylett_quantum_2017} and Grange~\cite{grange_design_2024}. The quantum advantage in these cases is rated as plausible based on favourable scaling analysis, but not compelling as they rely on fast state encoding and do not estimate the problem scale at which quantum algorithms become superior. To summarise, most studies lack a competitive classical baseline, complete computational costs, or sufficiently broad scaling experiments.

A consistent challenge across the literature was encoding complex problem structure within quantum algorithms. Realistic transport systems are often rich with constraints, which may be due to regulations, safety requirements, capacity, timing, assignment, flow-conservation, or the inherent complexity of practical transport networks. These difficulties are most visible in routing and scheduling studies, where continuous variables are cumbersome to encode cleanly on quantum devices whereas strong classical methods retain and exploit problem structure explicitly. As argued in the routing section~\ref{ssec:routing}, simply encoding constraints as penalties within a monolithic QUBO can require onerous optimisation and is unlikely to be effective at scale. Consequently, authors have responded to this challenge through hybrid annealing, sophisticated constraint-handling schemes\cite{bennett_quantum_2021,bentley_quantum_2022}, or classical processing to feed only feasible solutions into the quantum algorithm\cite{xu_high-speed_2023,harwood_formulating_2021}. The risk with these approaches is that the quantum solution introduces further classical bottlenecks that degrade any potential quantum advantage, either through embedding costs, extensive gates, or combinatorial enumeration, respectively.

The most promising approaches therefore enhance classical frameworks through the natural introduction of quantum labour. Many classical methods already compartmentalise constraint handling through dedicated subproblems, and this provides the opportunity for quantum algorithms to target other computational bottlenecks. For example, Svensson~\textit{et al.}\cite{svensson_hybrid_2023} integrate QAOA within branch-and-price, leaving route generation and operational constraints within classical pricing. Additionally, Dixit and Niu~\cite{dixit_quantum_2023} separate classical traffic assignment from quantum-assisted network design in a bilevel framework. This provides a natural division of each computing paradigm's natural strengths, with classical methods handling the flow constraints and quantum methods tackling combinatorial search. While none of these approaches has yet demonstrated end-to-end practical advantage, they illustrate a more credible division of quantum and classical labour than a monolithic encoding. Hence, the key insight is to preserve the transport structure and assign a well-defined computational bottleneck to a quantum method. In the following section, we synthesise these insights into practical suggestions for transport practitioners and provide a perspective on the future of the field.

\section{Perspective}
\label{sec:future}

The preceding review reveals two distinct priorities for transport researchers. The first is a methodological change to reporting. Existing studies should be designed and their results disseminated in a manner such that potential quantum advantage can be assessed fairly. We therefore begin by distilling the cross-cutting limitations of the corpus into practical advice for selecting, implementing, and benchmarking quantum methods. Secondly, we look beyond the existing corpus to identify promising future directions. The current dearth of advantage should not dissuade the investigation of quantum enhancement in transport. We therefore provide a perspective on high-value research directions, including hybrid algorithms, long-term fault-tolerant applications, and QML.

\subsection{Practical advice for transport researchers}

The literature review revealed several recurring methodological limitations which prevent fair assessment of quantum advantage. These are incomplete reporting and benchmarking, insufficient attention to constraint-handling overheads, and limited evidence across meaningful problem scales. We therefore offer three practical recommendations that strengthen the application of the quantum-solutions pipeline developed in Section~\ref{sec:pipeline}. While these recommendations do not guarantee quantum advantage, they should provide a clearer basis for determining which approaches warrant further investigation.

\paragraph{1. Improve benchmarking and reporting}

Benchmarking and reporting limitations are a pervasive barrier to assessing quantum advantage. Approximately 2/3 of the corpus failed to report timings or a relevant classical comparator. It is plausible that several of these studies rated N for C1 and C2 could demonstrate eventual quantum advantage; however, this could not be fairly assessed -- there is simply not enough reported information to make the judgement. This is unfortunate, as the timings and classical benchmarks are relatively straightforward to assess and include within a manuscript. Our practical advice is therefore the following.

Researchers should identify and justify the strongest classical method used for comparison, including whether it is intended to provide an exact solution or a heuristic result. Classical and quantum methods should then be evaluated at a common solution quality, feasibility requirement, and probability of success. The end-to-end runtime should be separated into relevant components, including problem formulation, data encoding, embedding or compilation, quantum execution, repeated sampling, classical optimisation, error mitigation, and post-processing (e.g., see equation~\eqref{time_costs}). Hardware, software, parameter settings, stopping criteria, the number of instances and repetitions, and variability between runs should also be reported. If a cost is excluded because it can be amortised across repeated solves, this should be made explicit. Even if the results indicate that quantum advantage is not plausible, these reporting norms will drastically improve the quality and impact of research.

\paragraph{2. Treat constraint handling as part of the algorithm}

The density and complexity of constraints in transport problems are too important for constraint handling to be treated as a secondary implementation detail. We therefore recommend that emphasis be placed on maintaining feasibility as effectively as possible. Moreover, algorithmic approaches towards constraint handling should be evaluated as carefully as the optimisation routine itself. This is particularly important for QUBO formulations, as the only problems which have shown tentative evidence of quantum scaling are unconstrained\cite{boulebnane_solving_2024,shaydulin_evidence_2024}. On this basis, we recommend that researchers first seek formulations or subproblems which do not require penalty terms. In particular, researchers should avoid defaulting to a monolithic QUBO in which every operational constraint is absorbed into the objective function.

Fortunately, transport engineers have a rich set of alternatives to choose from. The CVRP provides a useful illustration of this diversity, including quantum walks over feasible routing solutions~\cite{bennett_quantum_2021}, constraint-preserving mixers~\cite{bentley_quantum_2022}, and decomposition into separate clustering and routing problems~\cite{feld_hybrid_2019}. More generally, Hadfield~\textit{et al.}~\cite{hadfield_quantum_2019} developed constraint-preserving mixers for several canonical optimisation problems, while constraints may also be retained within established hybrid frameworks such as branch-and-price~\cite{svensson_hybrid_2023}. Moreover, there is no shortage of new and emerging VQA variants for constrained optimisation which can be benchmarked\cite{herman_constrained_2023,abdul_rahman_feedback-based_2026}. The caveat is that all these approaches have associated costs. State preparation, indexing, mixer implementation, decomposition, classical optimisation, and feasibility checking can introduce substantial computational overheads. For example, Harwood~\textit{et al.}~\cite{harwood_formulating_2021} demonstrate how different formulations can produce markedly different Hamiltonian sizes and interaction structures for the VRP with time windows. These costs must therefore be measured and included within the end-to-end assessment.

Additionally, constraint enforcement does not necessarily have to be performed classically. Quantum circuits can instead be used to prepare a superposition over feasible solutions before the optimisation or search begins. This is the effective approach of Wilkening~\textit{et al.} for quantum tree generation applied to the 0-1 knapsack problem, where a quantum pre-processing step applies conditional \texttt{IF} logic to construct the feasible branches of the search space. Bai and Shang~\cite{Bai_2025} similarly prepare a uniform superposition over Hamiltonian cycles before using Grover-type minimum search for the TSP. This approach is also compatible with QAOA, and has been used to both prepare the initial feasible state and to construct a Grover-like mixer~\cite{christiansen_quantum_2025}. Importantly, quantum state preparation still has associated computational costs and is not necessarily superior to classical techniques.

\paragraph*{3. Move beyond proof-of-concept demonstrations}

The literature has now firmly established that transport problems can be formulated for, and executed on, quantum computers. Demonstrating this again provides progressively less insight. Future studies should instead be designed to determine whether a method has a plausible route to competitive performance. This requires a stated mechanism for advantage, comparison with the strongest relevant classical method, and experiments across multiple instances and problem scales. Recognised public benchmarks and realistic transport datasets should be used when possible. Synthetic instances remain useful for systematically varying instance features and identifying which characteristics influence quantum performance (e.g., network topology, constraint tightness, or QUBO density).

Current hardware will likely be too small to reveal meaningful scaling. Empirical results should therefore be augmented with classical simulation, analytical resource estimates, or hybrid benchmarking rather than extrapolated in isolation. Agreement between these different forms of evidence is substantially more persuasive than fitting a favourable scaling law to a few toy instances. Conversely, a negative result remains valuable where it shows that overheads eliminate a theoretical speedup or that a formulation is unlikely to become competitive.

\subsection{Intermediate-term hardware applications}

Irrespective of application, quantum computers have not yet delivered practical value for any problem relevant to transport research. The pressing question is therefore: when will quantum hardware become sufficiently mature to yield tangible benefits? Any answer is highly speculative and opinions differ greatly. In their case study of hybrid algorithms for quantum-enhanced rail scheduling, the company Q-CTRL estimates that practical quantum advantage will arrive as early as 2028\cite{q-ctrl_accelerating_2025}. There may be merit in this assessment, as empirical scaling of QAOA for Max-Cut hints that only hundreds of qubits will be required to achieve quantum advantage\cite{guerreschi_qaoa_2019}. In contrast, Harwood~\textit{et al.}~\cite{harwood_formulating_2021} estimate that tens of thousands of logical qubits will be required to solve real-world business problems associated with the maritime inventory routing problem. Consulting the roadmap predictions in Table~\ref{tab:hardware}, only IonQ predicts a device of this scale by 2030.

Regardless, VQAs and annealing are likely to remain the dominant quantum computing paradigm until fault tolerance is achieved. We therefore speculate that the strongest intermediate-term applications will achieve modest acceleration by leveraging quantum algorithms within broader classical frameworks. The foundations for this integration within the transport context have been established by Svensson~\textit{et al.}~\cite{svensson_hybrid_2023} with their QAOA-enhanced branch-and-price for the tail-assignment problem. Their method can be adapted to many mixed-integer and binary programs and scales naturally with hardware advances. Many applications which have already considered using QUBO approaches (routing, scheduling, network design, logistics) could conceivably be approached using Svensson~\textit{et al.}'s framework while including complex industry-relevant constraints.

Beyond branch-and-price, other authors have devised methodologies for hybrid integration of quantum algorithms into established optimisation frameworks. These include Benders\cite{Zhao2022HybridQuantumBenders,Paterakis2023HybridQuantumClassicalMulticutBenders,Leenders2024IntegratingQuantumAndClassicalComputing} and Lagrangian decomposition methods\cite{Yonaga2022QuantumOptimizationWithLagrangianDecomposition}, where the quantum solver is used for combinatorial optimisation while a classical solver handles additional constraints. Other formulations have taken the reverse approach to Svensson~\textit{et al.} and used quantum optimisation for the pricing subproblem\cite{coelho2023}. This is also a natural domain for quantum-based dynamic programming\cite{caroppo_quantum_2025,ambainis_quantum_2019}. Also unexplored in the transport context are quantum variants of genetic algorithms\cite{lahoz-beltra_quantum_2016}.

There is significant scope to integrate these hybrid pipelines into transport applications. For example, Cooper~\cite{cooper_exploring_2022} provides multiple suggestions for transport modelling in a short perspective paper, including network analysis, multi-objective routing, all-pairs shortest path computation, and model calibration. Moreover, optimisation of NP-hard combinatorial problems through mixed-integer linear programs forms the backbone of much contemporary transport research. Many of these explicitly acknowledge challenges associated with computational speed and memory which could be alleviated through quantum pipelines. Examples include variants of the VRP solved with branch-and-price\cite{sakarya_two_echelon_2025,bettinelli_bcp_2011,yang_bpc_2021}, electric bus charging and scheduling solved with branch-and-price\cite{zhou_electric_2024}, shortest path optimisation with dynamic programming\cite{shen_reliable_shortest_2020,chen_k_shortest_2020}, and rail/metro timetabling, scheduling, or rostering solved using adaptive large neighbourhood searches\cite{dong_alns_timetabling_2020}, Lagrangian relaxation\cite{tian_lagrangian_timetabling_2024}, ADMM\cite{yao_periodic_timetabling_2023}, and quadratic programming\cite{qi_joint_timetabling_2025}.

Ultimately, only modest enhancements to solution speed should be expected when using these hybrid approaches. Even when applied to the classical computational bottleneck, compartmentalisation of the quantum acceleration to a single aspect of the end-to-end solver pipeline will limit the overall speedup following Amdahl's law.  Nevertheless, exponential speedups are not required for practical value. Even polynomial improvements from quantum search or minimum-finding routines may be consequential at scale.

\subsection{Long-term hardware applications}

A more ambitious long-term direction is to move beyond compartmentalisation towards large coupled systems. The targets are computational bottlenecks arising from the interaction between multiple subsystems. This is highly relevant to connected transport networks and systems linking transport with energy, communications, or other infrastructure\cite{zhang_generalized_2011,lu_open_2019,wei_network_2018,zhou_analyzing_2021}.

Classical studies already demonstrate both the value and difficulty of preserving these interactions. For example, Lu~\textit{et al.}~\cite{lu_open_2019} studied a community model in which EV travel affected charging demand on the electricity network, and a communication system provided vehicles with routing information. Communication packet loss reduced the predicted morning-peak speed on some roads by up to 10.5\%, while jointly simulating the transport and communication systems changed the maximum predicted power drawn from the grid by 7\%. Hence, subsystems can display strong interdependencies which influence the performance of the whole network.

These strongly coupled systems can be particularly challenging to model. Some solvers tackle the problem recursively, with outputs from one system becoming inputs to another within a feedback loop until convergence occurs. For example, in coupled transport--power networks, route choices determine charging demand, while the resulting power flows and prices alter those choices. Traffic assignment and power flow must therefore be repeatedly reconciled to reach equilibrium\cite{wei_network_2018,zhou_analyzing_2021}. Similar dependencies and computational challenges arise across broader infrastructure systems. For example, Enayaty Ahangar~\textit{et al.}~\cite{enayaty_ahangar_modeling_2020} considered multi-layer networks in which a component can operate only if other networks can supply the services it requires. They found that deciding which components should operate and how those services should be delivered across the networks is strongly NP-hard.

Here the potential value of quantum computing would lie in treating the combined system as a whole rather than accelerating one isolated component of an otherwise classical workflow. However, merely encoding a complex system within a quantum device is not sufficient to provide advantage. It is still critical to consider whether preserving these detailed interactions provides practical value and, moreover, how to achieve this through efficient quantum encodings and oracles. The latter forms part of a broader quantum computing challenge, with recent work exploring QRAM for encoding classical data\cite{shen_bucket_2026} and oracle-sketching methods which avoid loading the complete dataset\cite{zhao_exponential_2026}.

A prime example of these challenges within the transport domain is bi-level traffic assignment and origin-destination demand estimation\cite{zhang_dta_calibration_2021,osorio_od_calibration_2019,ma_od_estimation_2018}. This is a notoriously difficult problem with significant computational bottlenecks which has so far been overlooked in the quantum literature. Standard deterministic user-equilibrium traffic assignment can be formulated as a convex optimisation problem under conventional assumptions and therefore is not, by itself, an obvious target for combinatorial quantum optimisation. The more compelling opportunity arises in bilevel formulations, where the assignment problem may need to be solved repeatedly within a difficult upper-level search over network design, demand or control decisions. Future quantum algorithms may be able to address this challenge if they can contend with the coupled nature of the problem; it is not sufficient to accelerate the upper or lower level alone. Applying the Stage 1 assessment from the quantum solutions pipeline in Figure 3, it is clear that a direct QUBO formulation is not a compelling approach at practical scale. Namely, continuous demands and link flows require discretisation while equilibrium and flow-conservation constraints introduce numerous auxiliary variables and penalty terms which are explicitly not quadratic. A more plausible direction is therefore to combine quantum search and quantum linear-algebra subroutines within a unified quantum framework. Doing so successfully would require a sophisticated oracle construction and continuous-variable representation. 

Finally, other speculative applications include the integration of quantum hardware into autonomous vehicles or intelligent transport systems. The current user model for quantum computing involves cloud-based access to dedicated mainframe devices. However, future innovations could enable access to relatively cheap, compact, and personal devices that operate in robust environments. For instance, diamond-based quantum computers can operate at room temperature and hold promise for rapid, offline signal processing\cite{oberg_bottom-up_2025}. Drawing from recent results using classical techniques, possible applications include trajectory planning of automated vehicles at intersections\cite{wang_cooperative_cav_2020,ma_trajectory_cav_2021}, in mixed-traffic environments\cite{shi_drl_cav_2021}, and during lane changes\cite{atagoziev_lane_change_2023}. We therefore encourage practitioners to consider on-demand quantum applications.

\subsection{Quantum machine learning}

There are significant opportunities for exploring QML within the broader transport context. QNNs could conceivably be applied to computer vision across the full spectrum of intelligent transport systems\cite{dilek_computer_2023}, origin-destination demand forecasting, anomaly detection, control policies for traffic signals and ramp meters, and queue-spillback prediction\cite{zantalis_review_2019}. Rather than operating as stand-alone models, QML components could be integrated into the broader perception-to-decision pipelines of intelligent transport systems. Reviews of quantum computing for autonomous vehicles and of machine learning for automotive perception describe pipelines in which data from cameras, radar, and LiDAR are processed for feature extraction, object detection and classification, sensor fusion, prediction, planning, and control\cite{rattan_quantum_2025,pandharipande_sensing_2023}. Within this architecture, quantum models could perform selected feature-extraction, similarity, classification, or decision-making tasks while the surrounding data management, simulation, and control processes remain classical. Quantum radar and quantum LiDAR offer a complementary longer-term direction at the sensing layer, potentially supplying new forms of environmental data for subsequent classical or QML analysis\cite{gallego_torrome_advances_2024}. Researchers can explore these applications using a variety of QML techniques, including QNNs, quantum kernel methods, quantum linear algebra methods, and quantum reinforcement learning\cite{peral-garcia_systematic_2024,chen_introduction_2024}. 

Despite this opportunity, researchers should still focus on identifying tangible benefits of QML with regard to speed or accuracy. It is important to acknowledge that noise, depth limits, and qubit counts on current hardware severely restrict the size and depth of QML models and that encoding large classical datasets into quantum states may erase any theoretical runtime advantage. Finally, systematic reviews note that robust, large-scale evidence of practical QML advantage on real classical datasets is still lacking, and that perceived advantages may vanish when compared to well-tuned classical baselines\cite{peral-garcia_systematic_2024}.

\section{Conclusion}
\label{sec:conclusion}

Quantum computing in transport has now largely completed its proof-of-demonstration phase. Across the 91 studies reviewed here, quantum methods have been applied to a diverse range of routing, scheduling, network-design, traffic-control, forecasting, and computer-vision problems. It is therefore clear that transport problems can be represented and executed using a broad range of quantum methods. What remains unresolved is whether doing so provides a practical benefit over the strongest available classical approach. A broad analysis of the literature is sobering and demonstrates scant evidence of quantum advantage. In many cases, the absence of a relevant classical comparison, complete computational costs, or evidence across meaningful problem scales prevents either a positive or negative assessment from being made. Consequently, the literature has established feasibility of implementation far more convincingly than competitiveness.

Hence, the immediate priority should be to move from demonstrations of quantum compatibility to tests of quantum benefit. The quantum solutions pipeline developed in this work provides a structured approach for doing so. Researchers should begin with the transport problem and its strongest classical solution workflow, identify the dominant computational bottleneck, and then determine whether a quantum algorithm provides a plausible mechanism for improving it. Particular attention must be paid to the auxiliary costs beyond just optimisation, including constraint handling, data encoding, parameter training, repeated sampling, and classical processing. Performance should then be assessed against a competitive classical method at a common solution quality and across instances capable of revealing meaningful scaling. In this emerging subfield, negative results can be highly valuable, especially where they show that an apparent speedup is eliminated by these wider costs.

The perspective developed in this work identifies two distinct horizons for subsequent research. In the near and intermediate term, the most credible opportunities arise from integrating quantum subroutines into established classical algorithms. Compartmentalisation will limit the total speedup according to Amdahl's law, but exponential speedups are not required for practical value and even polynomial improvements may be consequential at scale. Ambitious long-term opportunities include the treatment of large coupled systems and quantum machine learning, though their ultimate success may require broader advancements such as fast data encoding, efficient oracle construction, and fault-tolerant hardware.

In summary, while quantum computing can provide value to the transport domain, emphasis should be placed on formulating problems which can conceivably benefit from quantum advantage. We hope the suggestions made here help guide further advances in this burgeoning research field. For now, judicious optimism is warranted and the transport community should focus on developing well-posed quantum formulations, validating claims of advantage, and building the interdisciplinary expertise required to capitalise on quantum computing once it eventually matures.

\section*{Acknowledgements}

This project was funded by the Queensland Government through the Department of Environment, Tourism, Science and Innovation's (DETSI) Quantum 2032 Challenge Program. The program aims to accelerate the development of quantum-based innovations in sportstech and related fields, foster collaboration between Queensland's quantum research sector and industry, and showcase the state's quantum expertise on the global stage during the Brisbane 2032 Olympic and Paralympic Games, contributing to the lasting legacy of the Games. We thank Matthew McKague for helpful discussions regarding quantum computing.

\section*{Declaration of Competing Interests}

The authors declare no competing interests.

\begingroup
\raggedbottom     
\setlength{\emergencystretch}{2em}
\printbibliography
\endgroup

\clearpage
\newgeometry{margin=8mm}
\raggedbottom
\renewcommand{\theHtable}{supp.\arabic{table}}
\captionsetup{singlelinecheck=false}
\setcounter{page}{1}
\renewcommand{\thepage}{S\arabic{page}}
\setcounter{table}{0}
\setcounter{figure}{0}
\renewcommand{\familydefault}{\sfdefault}\normalfont
\setlength{\footskip}{15pt}
\hyphenation{Hari-krishna-kumar Moham-mad-bagher-poor}
\begin{refsection}
\input{supplementary_body.tex}
\end{refsection}
\end{document}

%% file: supplementary_body.tex
\begin{center}
{\LARGE\bfseries Supplementary material}\\[0.7em]
{\large Quantum computing for transport research: an introduction, systematic review, and perspective}\\[0.7em]
Lachlan Oberg, Paul Corry, Moji Ghadimi and Ashish Bhaskar
\end{center}

\noindent This supplementary material contains the quantum taxonomy used throughout the review (Table~S1), the study-level evidence tables for the final 91-study corpus (Tables~S2A and S2B), the complete database search query, and the completed PRISMA 2020 checklist for the submission manuscript.

\medskip
\begingroup
\small
\setlength{\tabcolsep}{6pt}
\renewcommand{\arraystretch}{1.15}
\renewcommand{\thetable}{S1}
\newcommand{\taxonomygroup}[2]{%
  \addlinespace[0.75em]
  \multicolumn{2}{@{}l@{}}{%
    \colorbox{black!8}{%
      \parbox{\dimexpr0.96\textwidth-2\fboxsep\relax}{%
        \strut\textbf{#1}\par
        {\footnotesize\itshape #2\strut}%
      }%
    }%
  }\\
  \addlinespace[0.35em]
}

\begin{longtable}{@{}>{\raggedright\arraybackslash}p{0.27\textwidth} >{\raggedright\arraybackslash}p{0.69\textwidth}@{}}
\caption{Quantum taxonomy of the principal concepts used in this review. Terms are grouped thematically.}
\label{tab:s1}\\
\toprule
\textbf{Term} & \textbf{Definition} \\
\midrule
\endfirsthead

\multicolumn{2}{@{}l}{\tablename~\thetable\ (continued)} \\
\toprule
\textbf{Term} & \textbf{Definition} \\
\midrule
\endhead

\midrule
\multicolumn{2}{r@{}}{Continued on next page} \\
\endfoot

\bottomrule
\endlastfoot

\taxonomygroup{1. Quantum states and measurement}{How quantum information is represented and observed.}
\textbf{Quantum state} & The complete mathematical description of a quantum system, written in ket notation as $|\psi\rangle$. Mathematically, a pure state is a normalised complex-valued vector. \\
\textbf{Qubit} & A two-level quantum system with state $|q\rangle=\alpha|0\rangle+\beta|1\rangle$, where $|\alpha|^2+|\beta|^2=1$. Unlike a classical bit, it may occupy any such linear combination of 0 and 1. \\
\textbf{Computational basis} & The standard basis labelled by classical bitstrings. For one qubit it is $\{|0\rangle,|1\rangle\}$; for $n$ qubits it contains $2^n$ states $|x\rangle$. \\
\textbf{Quantum amplitude} & A complex coefficient associated with a computational-basis state. Its squared modulus gives that state's measurement probability, while its phase affects quantum interference. \\
\textbf{Superposition} & A normalised linear combination of computational-basis states. An $n$-qubit state may have amplitudes over all $2^n$ possible bitstrings. \\
\textbf{Measurement (readout)} & The process that converts a quantum state into a classical result. By the Born rule, a basis state with amplitude $c_x$ is returned with probability $|c_x|^2$. Each execution produces one sample, or shot; repeated shots estimate probabilities and expectation values. \\

\taxonomygroup{2. Qubit hardware and reliability}{How qubits are physically realised and why present devices are error-prone.}
\textbf{Physical qubit} & An individual hardware realisation of a two-level quantum system. Physical-qubit count alone does not indicate how many reliable variables or logical qubits a device can support. \\
\textbf{Logical qubit} & An error-corrected unit of quantum information encoded across multiple physical qubits. Logical qubits are more reliable but impose substantial physical-resource and operational overheads. \\
\textbf{Decoherence} & The loss of useful quantum information through unwanted interactions between qubits and their environment. It limits how long a reliable computation can run. \\
\textbf{Noisy intermediate-scale quantum (NISQ)} & The present regime of processors with limited numbers of imperfect physical qubits and insufficient error correction to support large fault-tolerant computations. \\

\taxonomygroup{3. Gate-based computation and benchmarking}{How gate-based devices perform circuits and how their performance is assessed.}
\textbf{Unitary operation} & A reversible linear transformation $U$ satisfying $U^\dagger U=I$. It preserves the norm of the quantum state and hence total measurement probability. On a universal gate-based quantum computer, any unitary can be decomposed or approximated to arbitrary precision as a sequence of quantum gates. \\
\textbf{Quantum gate} & A reversible unitary operation acting on one qubit or a coupled pair. Two-qubit gates are usually noisier and often dominate circuit cost. Their physical implementation depends on the qubit architecture; examples include laser pulses, microwave pulses, and passive optics. \\
\textbf{Gate-based quantum computer} & A quantum computer that implements algorithms as sequences of one- and two-qubit logic gates, analogous to the circuit model of classical computation. \\
\textbf{Quantum circuit} & A representation of a quantum computation as ordered gates acting on qubit wires. The horizontal direction conventionally represents the passage of time. \\
\textbf{Circuit depth} & The number of sequential gate layers in a quantum circuit. Greater depth permits more complicated operations but exposes the state to more gate error and decoherence. \\
\textbf{Gate fidelity (infidelity)} & A measure of how closely an implemented gate matches its intended unitary operation. Infidelity is the corresponding error measure, commonly approximated as $1-\mathrm{fidelity}$. \\
\textbf{Quantum volume} & A hardware-agnostic benchmark combining gate fidelity, connectivity, and achievable circuit depth. It provides a rough measure of the overall performance of a gate-based quantum computer. \\

\taxonomygroup{4. Quantum algorithms and data handling}{How algorithms are structured and how data preparation and subroutine costs affect performance.}
\textbf{Quantum algorithm} & A prescribed sequence of quantum operations. It manipulates amplitudes of the quantum state so that a desired answer has a high probability of being observed upon measurement. \\
\textbf{State preparation} & The operations required to initialise qubits in the input state needed by an algorithm. Preparing arbitrary classical data can dominate the total computational cost. \\
\textbf{Asymptotic scaling} & How computational resources grow as the problem size approaches infinity. Favourable asymptotic scaling does not guarantee a practical advantage at finite problem sizes. \\
\textbf{Quantum oracle} & A coherent quantum circuit that evaluates a function as a reversible black-box operation. Oracle construction costs must be included when assessing an algorithmic speedup. \\
\textbf{Quantum random access memory (QRAM)} & A proposed memory architecture for querying classical data coherently in superposition. Large, fault-tolerant QRAM is not currently available, so input costs can erase apparent speedups. \\
\textbf{Hybrid algorithm} & A workflow that divides computation between gate-based quantum and classical processors, for example by embedding a quantum subroutine within a classical method or iteratively alternating between them, as in QAOA. \\
\textbf{Parameterised quantum circuit} & A quantum circuit containing gates controlled by continuous variables or angles. Its prescribed structure is termed the ansatz; a classical optimiser adjusts the parameters in variational algorithms such as VQE and QAOA. \\

\taxonomygroup{5. Quantum optimisation: encoding and annealing}{How optimisation objectives are encoded and solved through quantum annealing.}
\textbf{Hamiltonian} & An operator that represents the energy of a quantum system. In quantum optimisation it is constructed so that low-energy eigenstates correspond to low-cost solutions. A QUBO cost function is mapped to an Ising Hamiltonian whose lowest-energy eigenstate, termed the ground state, encodes a global minimum. \\
\textbf{Quantum annealing} & An optimisation procedure performed on a special-purpose analogue device called a quantum annealer. It evolves a system from an easily prepared Hamiltonian towards a QUBO or Ising problem Hamiltonian, then samples low-energy candidate solutions. \\
\textbf{Chain (quantum annealing)} & A connected group of physical qubits used to represent one logical variable during minor embedding. \\
\textbf{Hybrid annealing} & A workflow in which classical methods decompose and refine a large problem while a quantum annealer solves selected subproblems. D-Wave's hybrid solver is a proprietary example of such a workflow. \\

\end{longtable}
\endgroup

\clearpage
\begin{landscape}
\section*{Tables S2A and S2B: Study-level evidence tables}
\noindent\textbf{Assessment legend.} \colorbox{EvidenceM}{\strut\textbf{M}} = Meets the complete criterion\quad \colorbox{EvidenceP}{\strut\textbf{P}} = Partially meets; material elements remain missing or mismatched\quad \colorbox{EvidenceN}{\strut\textbf{N}} = Does not meet.

\medskip
\noindent\textbf{Routing approach note.} In Table S2A, 54 of the routing studies are labeled as (1), (2), or (3) depending on their approach as discussed in the main text. This numbering is included in the Formulation column. Six remaining studies adopted a QML, analogue, or other approach and are not labeled.

\renewcommand{\thetable}{S2A}
\setlength{\LTleft}{0pt}
\setlength{\LTright}{0pt}
\setlength{\tabcolsep}{2.4pt}
\renewcommand{\arraystretch}{1.15}
\footnotesize
\begin{longtable}{>{\RaggedRight\arraybackslash}p{2.55cm}>{\RaggedRight\arraybackslash}p{3.00cm}>{\RaggedRight\arraybackslash}p{3.65cm}>{\RaggedRight\arraybackslash}p{3.15cm}>{\RaggedRight\arraybackslash}p{3.75cm}>{\RaggedRight\arraybackslash}p{3.20cm}>{\RaggedRight\arraybackslash}p{3.35cm}>{\RaggedRight\arraybackslash}p{3.85cm}}
\caption{Study characteristics and implementation}\label{tab:s2a}\\
\toprule
\textbf{Study} & \textbf{Transport focus} & \textbf{Formulation} & \textbf{Quantum method} & \textbf{Evaluation and platform} & \textbf{Data setting} & \textbf{Largest demonstrated instance} & \textbf{Reported quantum scale} \\
\midrule
\endfirsthead
\multicolumn{8}{l}{\textit{Table S2A: Study characteristics and implementation (continued)}}\\
\toprule
\textbf{Study} & \textbf{Transport focus} & \textbf{Formulation} & \textbf{Quantum method} & \textbf{Evaluation and platform} & \textbf{Data setting} & \textbf{Largest demonstrated instance} & \textbf{Reported quantum scale} \\
\midrule
\endhead
\midrule
\multicolumn{8}{r}{\textit{Continued on next page}}\\
\endfoot
\bottomrule
\endlastfoot
\multicolumn{8}{c}{\textbf{Routing}} \\ \cmidrule(lr){1-8}
Ariño Sales and Palacios Araos (2023); journal article; peer reviewed~\cite{arino_sales_adiabatic_2023} & Routing — Last-mile CVRP. & Two-stage QUBO formulation. The first assigns customers to capacity-limited clusters, the second orders the visits within each cluster. Penalty terms act as constraints. (1) & Quantum annealing for both stages and hybrid annealing for clustering. Clustering problems too large for the annealer are first split into pieces by the classical qbsolv algorithm. & D-Wave Advantage quantum annealer, with classical qbsolv decomposition for clustering. & Public benchmarks. 14 standard vehicle-routing benchmark instances of varying size and customer distribution; exact customer counts were not reported in the paper. & Largest solved instance inferred as 199 customers (CMT05 or CMT10 instances), but not explicitly reported as such. & Embedding from problem variables into physical qubits is not reported. \\ \midrule
Azad et al. (2023); journal article; peer reviewed~\cite{azad_solving_2023} & Routing — VRP. & QUBO for vehicle routing with penalties enforcing visit and depot constraints. The model omits subtour elimination, so disconnected routes are possible. (1) & QAOA with a standard X mixer and SPSA, COBYLA, Powell, or BFGS parameter optimisation. Depth p ranges from 1 to 40 across experiments. & Classical simulation using IBM Qiskit. & Synthetic. 3 worked examples plus 150 random instances for each of 3 small problem classes. & 5 locations and up to 3 vehicles. & Up to 20 simulated qubits. \\ \midrule
Bai and Shang (2025); journal article; peer reviewed~\cite{Bai_2025} & Routing — TSP. & Feasible Hamiltonian cycles are generated in superposition using a successor-permutation encoding. QFT arithmetic computes route weights and a threshold oracle marks lower-cost tours. (3) & Gate-based quantum algorithm combining quantum dynamic programming for tour generation with Grover-style minimum search. & Analytical complexity analysis plus noiseless classical simulation using IBM Qiskit. & Synthetic. 7 small networks of 4-8 cities with hand-chosen integer distances. & 8 cities. & 13-30 simulated qubits across 4-8 cities; compiled circuit depth not reported. \\ \midrule
Bennett et al. (2021); journal article; peer reviewed~\cite{bennett_quantum_2021} & Routing — CVRP. & Every feasible route plan is indexed as one quantum state, and a classical algorithm computes each plan's cost including depot restocking. (2) & Quantum-walk optimisation. This is a VQA where a classical optimiser tunes the circuit parameters. & Classical simulation run on Pawsey supercomputing facilities. & Synthetic. 1 instance with 8 delivery locations and random costs and demand. & 8 delivery locations plus a depot, spanning 394,353 feasible route plans. & 19 simulated qubits to index the route plans. \\ \midrule
Bentley et al. (2022); preprint; not peer reviewed~\cite{bentley_quantum_2022} & Routing — CVRP. & Compact binary encoding of visit orders with feasibility enforced via constrained mixer. The full implemented model is not disclosed. (2) & QAOA using Q-CTRL's binary encoding and proprietary mixer. & Executed on 7-qubit IBM quantum processor, enhanced using Q-CTRL's Fire Opal compilation, control, and readout-correction stack. & Synthetic. Toy model, exact details not reported. & 2 passenger nodes plus a depot for 1 vehicle. & A 2-qubit circuit with 4 two-qubit gates. \\ \midrule
Borowski et al. (2020); conference paper; peer reviewed~\cite{borowski_new_2020} & Routing — VRP and CVRP. & Position-indexed QUBOs with visit-once penalties. This is integrated with 2 hybrid variants which cluster orders classically or split tours into capacity-feasible vehicle routes. (2) & Quantum annealing and hybrid annealing. & D-Wave and D-Wave hybrid solver in conjunction with qbsolv, classical clustering, and route partitioning. & Synthetic. 51 artificially generated orders on a Belgium/OpenStreetMap road network. Public benchmarks. 9 Christofides CVRP benchmarks. & Up to 200 orders on hybrid solver. Direct annealing solved up to 6 orders and 1 vehicle. & Embedding from problem variables into physical qubits is not reported. \\ \midrule
Chence et al. (2025); journal article; peer reviewed~\cite{niu_applying_2025} & Routing — Multicommodity network flow for parcel transportation. & Mixed-integer network-flow model with flow-conservation and vehicle-capacity constraints, converted to QUBO form. (2) & Hybrid annealing. & D-Wave hybrid solver. CPLEX ran single-threaded on an i7 laptop. & 10 parcel-logistics problems of increasing size. Provenance unclear. & Instance with 191,886 variables and 80,538 constraints. & Embedding from problem variables into physical qubits is not reported. \\ \midrule
Ciacco et al. (2025); conference paper; peer reviewed~\cite{Ciacco_2025} & Routing — Steiner TSP. & Time-indexed integer linear program with depot, visit, and flow-conservation constraints. Converted to QUBO. An arc-removal preprocessing step shrinks the model from a standard to reduced formulation. (2) & Quantum annealing and hybrid annealing, each applied to the standard and the reduced formulation. & D-Wave hybrid solver. Gurobi and classical simulated annealing ran on desktop. & Synthetic. 1 publicly archived sequence of random networks with 4-19 vertices, 70\% of them being required stops. & A 19-vertex instance required 91,965 variables in the standard formulation and 52,585 variables in the reduced formulation. & 208 logical variables for direct annealing. Embedding from problem variables into physical qubits is not reported. \\ \midrule
Codognet (2024); conference paper~\cite{Codognet_2024} & Routing — TSP. & 2 alternative QUBO encodings of the same tour problem. Standard one-hot and a newer domain-wall form, with visit-once penalty constraints. (1) & Hybrid annealing. & D-Wave hybrid solver. Classical GPU-based annealing as comparison. & Public benchmarks. 11 standard travelling-salesperson instances (TSPLIB) with 16-52 cities. & 52 cities (approximately 2,700 binary variables), solved only by the classical digital annealers; up to 13 cities with the hybrid solver. & Embedding from problem variables into physical qubits is not reported. \\ \midrule
Dixit et al. (2024); journal article; peer reviewed~\cite{dixit_quantum_2024} & Routing — Scenario-based stochastic time-dependent shortest path routing. & Binary time-expanded network-flow model with path-consistency and no-waiting constraints. MILP converted into QUBO for compatability with annealing. (2) & Quantum annealing and hybrid annealing & D-Wave for quantum and hybrid annealing. CPLEX solver used on same QUBO formulation run on an i7 desktop. & Synthetic. 1 hand-built 6-node example (solved with quantum annealing) and 18 randomly generated grid networks with 10-20 traffic scenarios each (solved with hybrid annealing). & A 100-node grid with 180 links, 20 scenarios, and 40 time intervals, formulated with approximately 148,000 binary variables. & 42 logical variables for the smaller quantum annealing run. Embedding from problem variables into physical qubits is not reported. \\ \midrule
Dong and Huang (2020); journal article; peer reviewed~\cite{Dong_2020} & Routing — TSP. & Ising Hamiltonian model with 1 variable per city-position pair. Valid tours are maintained by coordinated spin moves rather than penalty terms. (2) & Quantum Monte Carlo simulations of quantum annealing. & Python based classical simulations of quantum annealing. & 2 point sets; size and origin not reported. & Not clearly reported. & Qubit count not reported. \\ \midrule
Fagiolo and Vescera (2025); conference paper; peer reviewed~\cite{Fagiolo_2025} & Routing — TSP. & Permutation-encoded VQA. Continuous parameters optimised by Powell and circuit architecture optimised by simulated annealing. (2) & VQA whose circuit structure is redesigned by classical simulated annealing and whose parameters are tuned classically. & Classical simulation using IBM Qiskit. & Synthetic. 9 instances: 3 random symmetric distance matrices at each of 5, 6, and 7 cities. & 7 cities. & Simulated up to 13 qubits. \\ \midrule
Feld et al. (2019); journal article; peer reviewed~\cite{feld_hybrid_2019} & Routing — CVRP. & Decomposition with classical clustering to produce a single QUBO model per vehicle tour. Penalties assign each customer to exactly one position in the route. (1) & Hybrid annealing for the tour subproblems. The classical qbsolv routine splits each tour QUBO into 20-variable pieces for the annealer and merges the results. & D-Wave 2000Q accessed through the cloud. 20-microsecond anneal times. Runtime comparison was performed through a laptop. & Public benchmarks. 17 instances: 5 TSP instances of 14-38 cities and 12 standard vehicle-routing instances with 22-199 locations. & A 199-location vehicle-routing benchmark, though every piece sent to the annealer was capped at 20 variables by the classical decomposition. & 20 logical variables. Embedding from problem variables into physical qubits is not reported. \\ \midrule
Finžgar et al. (2022); conference paper; peer reviewed~\cite{finzgar_quark_2022} & Routing — Robot path planning, TSP variant. & QUBO models for the robot path and TSP constraints. (1) & Quantum annealing and QAOA. & D-Wave annealers and classical simulation of QAOA. & Synthetic. 1 simplified industrial robot-sealing problem and 1 reduced TSP benchmark. & Routing reached 14 TSP locations on hardware, but no valid tour was found above 10. & Routing used 16-196 logical annealing variables. Embedding from problem variables into physical qubits is not reported. Up to 25 simulated qubits for QAOA. \\ \midrule
Fitzek et al. (2024); journal article; peer reviewed~\cite{fitzek_applying_2024} & Routing — Heterogeneous VRP. & QUBO model in which binary variables encode each truck's visiting order and carried load, with penalty terms enforcing routing and capacity constraints. (1) & QAOA. & Classical simulations with 4 alternative classical routines tuning the circuit parameters. & Synthetic. 3 small constructed delivery problems with 3 or 4 cities and 1 or 2 trucks. & 3 cities served by 2 trucks. & 11-21 simulated qubits across the 3 cases. \\ \midrule
Garcia de Andoin et al. (2023); conference paper; peer reviewed~\cite{De_Andoin_2023} & Routing/Scheduling — Electric vehicle charging and routing problem. & Constrained multi-level quantum (qudit) model of per-vehicle movement, charging, and battery rules. The shared grid-power limit is handled by a separate classical stage. (3) & Grover amplitude amplification samples feasible partial trajectories for each vehicle. A classical greedy-tree procedure combines them into a global solution. & Analytical and numerical study. & Synthetic. Toy scenario on a 4-node network with specified prices and charge limits. & 4 vehicles, 4 time steps, and 4 locations. & 48 multi-valued logical variables encoded as qudits. \\ \midrule
Haba et al. (2025); journal article; peer reviewed~\cite{haba_routing_2025} & Routing/Scheduling — Urban air mobility fleet management. & QUBO for maximum weighted independent set problem. Select at most 1 candidate route per flight request while forbidding any pair of conflicting routes. (1) & Quantum annealing. & D-Wave quantum annealer inside a classical fleet-management simulation which classically repaired and refined solutions. & Synthetic. Flight requests on a simulated Singapore air-corridor network; no historical or live demand data. 629 total instances. & Simulations with up to 50 flight requests per 30-second step; the largest single quantum-solved problem had 50 candidate routes. & 5-50 logical variables per annealer problem. Embedding from problem variables into physical qubits is not reported. \\ \midrule
Harwood et al. (2021); journal article; peer reviewed~\cite{harwood_formulating_2021} & Routing — VRP with time windows. & Vehicle-routing models with visit-once and time-window constraints. This is expressed as several alternative QUBOs. A continuous-arrival-time variant is split into binary and continuous subproblems. (2) & VQE (depth-1 RY ansatz) and QAOA (p=1 or p=3), embedded within classical optimisation and sequential-convex ADMM. & Classical simulation using IBM Qiskit. & Public benchmarks. Modified maritime inventory-routing benchmark (MIRPLIB family) varied by time horizon. & 3 customers plus a depot was the largest problem actually solved by the quantum algorithms. Quantum resources were counted for larger instances. & 16 simulated qubits. \\ \midrule
He (2024); conference paper; peer reviewed~\cite{He_2024} & Routing — TSP. & QUBO for annealing plus a graph neural network trained on a QUBO-derived loss. (1) & Simulated quantum annealing. The companion neural-network pipeline and its local search are entirely classical. & Classical simulation using MATLAB and PyTorch. & Public benchmarks. Euclidean TSP benchmarks at 20, 50, and 100 cities. & 100 cities for the overall pipeline, but each quantum-simulated subproblem covered only 10 cities. & Up to 100 simulated qubits. \\ \midrule
Holliday et al. (2024); conference paper; peer reviewed~\cite{Holliday_2024} & Routing — CVRP. & A classical route-based tabu-search heuristic for the CVRP. A one-hot TSP QUBO is subsequently used to re-sequence individual fixed-customer routes to refine solutions. (2) & Quantum annealing. & D-Wave quantum annealer. & Public benchmarks. 13 instances: 7 standard vehicle-routing problems (50-199 customers) and 6 multi-depot mixed-fleet cases (10-200 locations). & 200 locations and 23 vehicles in the largest multi-depot case; only 1 route at a time was passed to the annealer. & Embedding from problem variables into physical qubits is not reported. \\ \midrule
Hussain et al. (2025); conference paper; peer reviewed~\cite{Hussain_2025} & Routing — TSP. & Cluster-first, route-second heuristic. Cluster cities, and then independently solve a permutation QUBO TSP within each cluster. The resulting subtours are then joined through another classical procedure. & QAOA and Grover search. & Classical simulation run on desktop. & Synthetic. 1 random 4-city example and 1 random 16-city map manually split into 4 clusters; no benchmark or real data. & A 16-city tour assembled classically from 4 separately solved 4-city clusters. & 24 simulated qubits for the 4-city Grover search and a 16-qubit budget for each QAOA cluster. \\ \midrule
Irie et al. (2019); conference paper; peer reviewed~\cite{irie_quantum_2019} & Routing — CVRP with time windows. & QUBO with arrival variables indexed by time, customer, and vehicle, and penalties for travel-time, single-visit, capacity, vehicle-state, and time-window rules. (1) & Quantum annealing. & D-Wave 2000Q, using 10,000 anneals at 20 microseconds each for a single demonstration run. & Synthetic. 1 hand-constructed 6-customer delivery example with illustrative time windows. & 6 customers and 3 vehicles over a 2-hour horizon. & Approximately 83 logical variables. Embedding from problem variables into physical qubits is not reported. \\ \midrule
Jain (2021); journal article; peer reviewed~\cite{jain_solving_2021} & Routing — TSP. & Edge-based QUBO. Subtour constraints not enforced. (1) & Quantum annealing. & D-Wave Advantage. Classical comparisons ran on a laptop. & Synthetic. 8 randomly generated cost matrices at 3 problem sizes of 8, 9, and 10 cities. & 10 cities were attempted but only 1 of 8 cases returned any solution. 8 cities was the largest size solved reliably. & 28-45 logical variables. Embedding from problem variables into physical qubits is not reported. \\ \midrule
Kieu (2019); journal article; peer reviewed~\cite{Kieu_2019} & Routing — TSP. & Adiabatic quantum computing Hamiltonian. The lowest-energy state is designed to encode the shortest valid tour. Penalty operators exclude invalid routes. & Analytical adiabatic quantum-computation algorithm. The system evolves towards a Hamiltonian encoding the shortest tour. & Theoretical. & None. & No instances solved. & N/A. \\ \midrule
Le et al. (2023); conference paper; peer reviewed~\cite{le_quantum_2023} & Routing — Selective TSP. & QUBO with route, budget, and subtour constraints encoded through penalty and slack terms. (1) & Quantum annealing. & D-Wave 2000Q. & Synthetic. 3 small instances with prizes and travel costs drawn randomly from small integers; no benchmark or real network data. & 7 cities were attempted without reaching the optimum; 6 cities was the largest instance reported solved to the claimed optimum. & 16-35 logical variables, embedded into 84-418 physical qubits. \\ \midrule
Leonidas et al. (2024); journal article; peer reviewed~\cite{leonidas_qubit_2024} & Routing — VRP with time windows. & QUBO selecting from classically pre-generated routes, with penalties enforcing that each destination is covered exactly once (set-partitioning). (2) & Hardware-efficient VQA. & Classical simulation for parameter training plus direct execution of the final circuits on IBM, IonQ, and Rigetti quantum processors. & Synthetic. 4 instances described as based on realistic shipping scenarios; underlying data were not reported. & 62 destinations with 3,964 candidate routes, run on hardware as a 13-qubit compressed circuit. & 5-13 qubits. \\ \midrule
Li and Mazumder (2023); conference paper;~\cite{Li_2023} & Routing — TSP. & Adaptive three-stage method: classical Johnson completion and vertex trimming, residual position-indexed TSP QUBO or Grover search, then classical route reconstruction. (2) & Quantum annealing. & D-Wave, compared against classical dynamic programming (Held-Karp algorithm). & Public benchmarks. 9 TSP and VRP instances, truncated to graphs of 4-20 locations. & 20 locations. The quantum stage received only a classically reduced graph of unreported size. & Logical variables or qubits not reported. \\ \midrule
Li and Ruan (2024); conference paper; peer reviewed~\cite{Li_2024} & Routing — TSP. & Hybrid transformer neural network with small quantum circuits inside its encoder, trained by reinforcement learning. A mask enforces 1 visit per city. & Quantum self-attention network (parameterised quantum circuit). & Classical simulation. & Synthetic. 5-city instances with uniformly random coordinates; approximately 51,000 training samples per cycle over 100 epochs. & 5 cities only, with a 4-qubit simulated quantum circuit. & 4 simulated qubits. \\ \midrule
Li et al. (2025); journal article; peer reviewed~\cite{li_quantum_2024} & Routing — VRP with collision avoidance. & QUBO with customer-visit, depot, subtour-elimination, delay-choice, and bridge-conflict constraints. Some constraints are added iteratively only after infeasible solutions appear. (2) & QAOA (p=1,2) and VQE, optimised with COBYLA. & Classical simulation using IBM Qiskit. & Synthetic instances: 1,000 random 4-node routing cases plus 2 small constructed bridge-collision examples on artificial maps. & 2 5-node, 2-vehicle routing problems solved 1 at a time. These were decomposed rather than solved as a single combined problem. & 17 simulated qubits for the joint model and 20 for each stepwise routing solve. \\ \midrule
Liu et al. (2025); journal article; peer reviewed~\cite{liu_solving_2025} & Routing — VRP. & Constrained binary model with depot, visit, and subtour-elimination conditions checked inside a quantum search. (3) & Grover adaptive search using a quantum dictionary, inverse QFT, feasibility/threshold oracle, amplitude amplification, and classical threshold updates. & Classical simulation using pyQPanda, run on Intel i9. & Synthetic. 4 randomly generated networks of 3 or 4 locations with 1 or 2 vehicles. & 4 locations and up to 2 vehicles, using 12 route decision variables. & Reported register sizes implies 22-29 simulated qubits. \\ \midrule
Maciejunes et al. (2025); conference paper; peer reviewed~\cite{Maciejunes_2025} & Routing — VRP. & 2-level decomposition of VRP into TSP. QUBO over directed route legs with visit and depot penalties. The published model omits subtour elimination, so disconnected routes can satisfy it. (1) & QAOA. & Classical simulation using IBM Qiskit. & Synthetic. 7-13 locations and 2-5 vehicles. & A 13-location, 5-vehicle problem was represented but never solved whole, instead decomposed into per-vehicle circuits of at most 12 simulated qubits. & 12 simulated qubits (6 after further cutting techniques). \\ \midrule
Makhanov et al. (2024); conference paper; peer reviewed~\cite{makhanov_quantum_2024} & Routing — 3D flight path optimisation. & Weighted directed-acyclic-graph shortest path. Classical Dijkstra replaces classical minimisation with Dürr–Høyer algorithm. (3) & Quantum-enhanced Dijkstra search. A Grover-style quantum minimum-finding step replaces the classical unstructured search. Graph construction and route bookkeeping remain classical. & Classical simulation, IBM quantum processor with Q-CTRL Fire Opal. & Public benchmarks. 9 flight-plan routes, synthetically perturbed, with a simplified Airbus A320 cruise-fuel model. & Route graphs of up to 674 nodes and approximately 4,000 edges were processed with the hybrid pipeline. & Search simulations used up to 8 simulated qubits and hardware with 4 physical qubits. \\ \midrule
Markevich and Trushechkin (2019); journal article; peer reviewed~\cite{Markevich_2019} & Routing — TSP. & Branch-and-bound search over city visiting orders, pruned with a spanning-tree lower bound. (3) & A quantum branch-and-bound built on nested Grover-style search. & Theoretical for quantum branch-and-bound. Numerical simulations for classical depth-first branch-and-bound. & Synthetic. 20 randomly placed cities, with 100 problems generated for each of 3 tour-length thresholds. & 20 cities. & Not reported and ancilla costs are unreported. \\ \midrule
Mohanty et al. (2023); journal article; peer reviewed~\cite{mohanty_analysis_2023} & Routing — VRP. & QUBO with penalties on visit, depot, and subtour constraints. (1) & VQE. & Classical simulation using IBM Qiskit, with 4 different classical optimisers and 5 synthetic noise models. & Synthetic. Toy models for 3 and 4 locations. & 4 locations, encoded in 12 simulated qubits. & 12 simulated qubits. \\ \midrule
Mohanty et al. (2024); journal article; peer reviewed~\cite{mohanty_solving_2024} & Routing — VRP. & QUBO over directed city-to-city links with visit and depot penalties. This is converted to an Ising model whose weights feed quantum feature-encoding circuits. & Quantum support vector machine trained using VQE. & Classical simulation using IBM Qiskit. & Synthetic. Random edge weights for a set of 3- and 4-city graphs. & 4 cities. & 6 and 12 simulated qubits with compiled circuit depths from approximately 48-800. \\ \midrule
Moosavi and Farooq (2025); conference paper; peer reviewed~\cite{Moosavi_2025} & Routing — Capacitated pickup and delivery problem with time windows for last-mile routing. & Binary quadratic routing model with visit, capacity, time-window, and precedence constraints, recast as sequential decisions. & Quantum reinforcement learning. A parameterised quantum circuit estimates action values inside a classical deep Q-learning loop. & Classical simulation using PennyLane. & Synthetic. Randomly generated locations, demands, and time windows in a 1-kilometre square. 10,000 training instances at a single problem size. & 1 problem size of 10 locations. & 11 simulated qubits. \\ \midrule
Moylett et al. (2017); journal article; peer reviewed~\cite{moylett_quantum_2017} & Routing — TSP on bounded-degree graphs. & Edge-based exact TSP formulation represented through a backtracking tree. (3) & Gate-based quantum algorithm combining quantum backtracking with quantum minimum finding. & Theoretical. & N/A. & N/A. & N/A. \\ \midrule
Nałęcz-Charkiewicz et al. (2025); journal article; peer reviewed~\cite{Na_e_cz_Charkiewicz_2025} & Routing — CVRP. & MILP model with visit, capacity, and subtour-elimination constraints compared with QUBO reformulations. Nearby customers are first merged (graph coarsening) and routes rebuilt afterwards. (2) & Hybrid annealing. & D-Wave hybrid solver compared with classical simulated annealing and Gurobi. & Public benchmarks. 7 Christofides CVRP benchmarks (CMT01–05, CMT11–12). & CMT05 with 200 locations and 16 vehicles. & Embedding from problem variables into physical qubits is not reported. \\ \midrule
Neukart et al. (2017); journal article; peer reviewed~\cite{neukart_traffic_2017} & Routing — Traffic flow optimisation. & QUBO with 1 binary variable per car-route choice. Penalties force 1 route per car and quadratic terms discourage shared road segments. (1) & Hybrid annealing. & D-Wave 2X. & Historic. 1 week of GPS traces from over 10,000 Beijing taxis and OpenStreetMap data. 418 taxis were optimised. & 1 scenario of 418 taxis with up to 3 candidate routes each. & 1,254 logical variables in the full model. Embedding from problem variables into physical qubits is not reported. \\ \midrule
Osaba et al. (2024); journal article; peer reviewed~\cite{osaba_solving_2024} & Routing — Package delivery. & Constrained quadratic models for individual routes, with capacity, priority-deadline, and working-day constraints. Classical logic selects trucks and concatenates the partial routes. (2) & Hybrid annealing. & D-Wave Advantage. Classical comparisons ran on a laptop. & Synthetic. 6 delivery scenarios generated under guidance from the logistics company Ertransit, with 14-29 deliveries each. & 29 deliveries (30 locations including the depot) served by 4 trucks, formulated with approximately 1,600 model variables overall. & Embedding from problem variables into physical qubits is not reported. \\ \midrule
Palackal et al. (2023); conference paper; peer reviewed~\cite{palackal_quantum-assisted_2023} & Routing — CVRP. & QUBOs are formulated for the full CVRP and its decomposition into subproblems for clustering and TSP. The TSP QUBO uses penalties which force each city into exactly one tour position. (2) & QAOA (p=5) and VQE. & Classical simulation using IBM Qiskit and execution on 27-qubit IBM quantum processor. & Synthetic. A stylised waste-collection case example. Performance tests used 3 nested subproblems of 4, 5, and 6 locations. & 6 location TSP subproblem. & 9, 16, and 25 physical qubits for the executed subproblems. \\ \midrule
Papalitsas et al. (2019); journal article; peer reviewed~\cite{papalitsas_qubo_2019} & Routing — TSP with time windows. & QUBO combining route constraints, travel cost, and time-window penalties. Equations omit subtour elimination. (1) & Proposed formulation for quantum annealing. & Analytic. Symbolic derivation and 2 hand-worked examples, with no simulation or hardware execution. & Synthetic. 2 closely related hand-constructed examples with 4 and 5 locations. & No instance was solved. Largest of the 2 worked examples had 5 locations. & N/A. \\ \midrule
Qian et al. (2023); journal article; peer reviewed~\cite{qian_comparative_2023} & Routing — TSP. & QUBO with city-based encoding. Penalties to constrain each city visited only once. (2) & QAOA (p=4,6) comparing X, cyclic-XY, and row-swap mixers. VQE ansatz as an additional comparison. & Classical simulations using IBM Qiskit, including gate noise. & Synthetic. Random weighted networks with 3, 4, and 5 cities. & 5 cities. & 4-16 simulated qubits. \\ \midrule
Ruan et al. (2020); journal article; peer reviewed~\cite{ruan_quantum_2020} & Routing — TSP. & TSP is formulated on a complete weighted graph by assigning 1 binary variable to each edge. A QUBO objective minimises the total weight of the selected edges. (2) & QAOA (p=4) with a custom mixer which restricts evolution to edge sets forming valid Hamiltonian cycles. & Classical simulation on i7 desktop with Nelder-Mead optimisation for angles. & Synthetic. A single 5-city network. & 5 cities and 10 edges. & 11 simulated qubits. \\ \midrule
Salehi et al. (2022); journal article; peer reviewed~\cite{salehi_unconstrained_2022} & Routing — TSP with time windows. & 3 different binary encodings. An edge-based QUBO, a higher-order node-based model, and an integer-programming-derived QUBO. Constraints (visit once, connectivity, time windows) enforced via penalties. (1) & Quantum annealing. & D-Wave Advantage. & Synthetic. 30 examples of 3-5 cities with random costs and time windows. Public benchmarks. 40 cases were analysed for size only. & 5 cities on hardware, with approximately 2,000 physical qubits and 250 logical variables. & Approximately 2,000 physical qubits. \\ \midrule
Sato et al. (2023); conference paper; peer reviewed~\cite{Sato_2023} & Routing — TSP. & Feasibility-only QUBO marking valid city-position assignments. Distance objective is deliberately excluded. (3) & Divide-and-conquer Grover-style quantum search. & Classical simulation using IBM Qiskit. & Synthetic. 3, 4, and 5 cities. & 5 cities. & 9-25 simulated qubits, one per city-position pair. Ancilla qubits were not reported. \\ \midrule
Sharma et al. (2025); journal article; peer reviewed~\cite{Sharma_2025} & Routing — TSP. & Fault-tolerant gate-model construction. Edge weights are unitary eigenphases, permutation registers represent routes, phase estimation for tour costs, and density-matrix exponentiation is proposed for finding minimum tour. (3) & Eight-step theoretical algorithm using controlled swaps, quantum phase estimation, conditional rotations, density-matrix exponentiation, measurement, and classical decoding. & Theoretical. & Synthetic. 4-city network with 2 missing connections used as a worked example, but not executed in simulation or on hardware. & N/A. & N/A. \\ \midrule
Silva et al. (2021); journal article; peer reviewed~\cite{Silva_2021} & Routing — TSP. & QUBO assigning each city to a tour position. Penalties enforce valid tours. (1) & Quantum annealing. & D-Wave Advantage. & Synthetic. Randomly generated instances with 3-8 cities. & 8 cities was attempted but the annealer found no valid optimum. & 9-64 logical variables. Embedding from problem variables into physical qubits is not reported. \\ \midrule
Sinno et al (2025); journal article; peer reviewed~\cite{sinno_analyzing_2025} & Routing — CVRP. & Arc-flow CVRP with binary route variables and MTZ subtour elimination represented in QUBO form. (2) & Hybrid annealing. & D-Wave hybrid solver. & Public benchmarks. 3 standard vehicle-routing instances: Augerat A-series A-n32-k5, A-n60-k9, and A-n80-k10. & 80 locations and 10 vehicles. Formulated with 63,990 logical variables and 62,519 constraints. & Embedding from problem variables into physical qubits is not reported. \\ \midrule
Spyridis et al. (2023); conference paper; peer reviewed~\cite{spyridis_variational_2023} & Routing — Multiple TSP. & Integer linear program converted to QUBO. (1) & QAOA at p=1. & Evaluation method unclear, inferred to be classical simulation. & Synthetic. A single 4-city example with 2 salesmen. & 4 cities and 2 salesmen using 6 logical variables. & 6 qubits. \\ \midrule
Tambunan et al. (2023); conference paper;~\cite{Tambunan_2023} & Routing — VRP with weighted segments. & One-hot QUBO in which each vehicle selects exactly one route, with costs penalising vehicles that share road segments. (1) & Hybrid annealing. & D-Wave hybrid solver. & Synthetic. A road network of 9 junctions and 12 segments with assumed integer weights. & 5 vehicles choosing among 3 routes each, encoded with 15 logical variables. & Embedding from problem variables into physical qubits is not reported. \\ \midrule
Tszyunsi and Beterov (2023); journal article; peer reviewed~\cite{Tszyunsi_2023} & Routing — TSP. & Route lengths are encoded as phases in a unitary operator. (3) & Quantum phase estimation followed by a modified Grover-type minimum search. & Theoretical with classical simulation using IBM Qiskit on small-scale example. & Synthetic. 4-city network with hand-specified edge lengths. Symmetry leaves 3 distinct candidate routes. & 4 cities. & 12 simulated qubits for the phase-estimation stage and 6 for the minimum search. \\ \midrule
Vargas et al. (2024); conference paper;~\cite{Vargas_2024} & Routing — CVRP with time windows. & Time-expanded binary integer program converted to a penalised QUBO, with slack variables encoding the weight and volume capacity limits. (1) & Unclear, but likely hybrid annealing. & D-Wave hybrid solver. & Historic. 4 months of proprietary Siemens logistics requests. & 12 delivery locations. & Embedding from problem variables into physical qubits is not reported. \\ \midrule
Villar-Rodríguez et al. (2022); conference paper; peer reviewed~\cite{villar-rodriguez_analyzing_2022} & Routing — TSP. & Two different QUBO formulations with 1-visit-per-city and 1-city-per-position constraints. A heuristic variant additionally penalises unpromisingly long connections. (1) & Quantum annealing. & D-Wave Advantage with 400-microsecond anneal times. & Public benchmarks. 6 cases derived from TSPLIB Burma14 coordinates. & 14 cities, solved only through classical partitioning and merging. The largest problem annealed whole had 7 cities. & Up to 49 logical variables annealed directly (196 in the full 14-city model). Embedding from problem variables into physical qubits is not reported. \\ \midrule
Weinberg et al. (2023); journal article; peer reviewed~\cite{weinberg_supply_2023} & Routing — Multi-truck pickup and delivery. & Higher-order binary model built for a single truck at a time and converted to QUBO form. Each returned route updates remaining demand before the next truck. (1) & Hybrid annealing within a classical framework. & D-Wave hybrid annealing. & Commercial. Logistics for Aisin Corporation across 23 Japanese facilities using real input data. & A 23 facility network, comprising approximately 100 PUBO/200 QUBO variables in the single-truck subproblems and approximately 350 PUBO/2,500 QUBO variables in the simulated-annealing subproblems. & Embedding from problem variables into physical qubits is not reported. \\ \midrule
Xie et al. (2024); journal article; peer reviewed~\cite{xie_feasibility-preserved_2024} & Routing — CVRP. & Constraint-preserving encoding with customer-visit ordering plus depot-return choices. Capacities are enforced during decoding rather than by penalty terms. (2) & A feasibility-preserving variant of QAOA simulated and compared against conventional penalty-based QAOA. & Classical simulation using IBM Qiskit. & Synthetic. 2 constructed 4-customer cases and 48 randomly generated 3-customer cases. & 4 customers plus a depot, encoded in 19 decision variables. & 27 simulated qubits. \\ \midrule
Xu and Créput (2025); conference paper; peer reviewed~\cite{XU_2025} & Routing — TSP. & A classical self-organising map deforms a neuron ring into a TSP tour. Quantum circuits are used to identify the nearest neuron to each city. & Quantum machine learning. A quantum distance-estimation circuit inside a classical self-organising map. & Classical simulation using IBM Qiskit. & Public benchmarks. 4 Euclidean TSPLIB instances: berlin52, st70, kroA100, and bier127. & bier127 with 127 cities. & Approximately 4-12 simulated qubits per circuit. \\ \midrule
Yarkoni et al. (2020); conference paper; peer reviewed~\cite{yarkoni_quantum_2020} & Routing — Traffic flow optimisation. & QUBO with 1 binary variable per car-route choice. Penalties force 1 route per car and quadratic terms discourage shared road segments. (1) & Hybrid annealing. & D-Wave 2000Q. & Live operational data. 9 buses serving a Lisbon conference with real-time traffic-aware candidate routes. 162 trips and 1,275 optimisation requests over 4 days. & 9 active buses. & Embedding from problem variables into physical qubits is not reported. \\ \midrule
Yarkoni et al. (2021); conference paper; peer reviewed~\cite{yarkoni_solving_2021} & Routing — Shipment re-routing problem. & Mixed-integer program converted to a QUBO with 1-route-per-shipment and truck-capacity constraints. This conversion bins shipment volumes and can tighten the capacity limits. (1) & Hybrid annealing. & D-Wave hybrid annealing with additional classical methods: simulated annealing, tabu search, and Gurobi. & Synthetic. Dataset abstracted from European freight operator DB Schenker; 4 generated shipment sets of 30-100 shipments. & 100 shipments with 925 candidate routes, formulated with approximately 3,300 binary model variables. & Embedding from problem variables into physical qubits is not reported. \\ \midrule
Zawalska and Rycerz (2023); conference paper; peer reviewed~\cite{Zawalska_2023} & Routing — TSP. & QUBO with penalties forcing each city into exactly one tour position. (1) & QAOA with a hybrid feedforward neural-network optimiser. & Classical simulation using TensorFlow Quantum. & Synthetic. 220 randomly generated 4-city instances. & 4 cities. & 16 simulated qubits. \\ \midrule
\multicolumn{8}{c}{\textbf{Scheduling}} \\ \cmidrule(lr){1-8}
Bickert et al. (2021; version 3 revised 2023); preprint; not peer reviewed~\cite{bickert_optimising_2023} & Scheduling — Assign rail trips to trains with maintenance stops. & Penalty-based QUBO with trip-assignment and non-overlap constraints and maintenance as soft terms, compared with a constraint-programming model that enforces maintenance strictly. & Hybrid annealing. & D-Wave annealer, compared with classical constraint-programming and tabu search. & Historic. Deutsche Bahn data: A simplified day of intercity timetable data across 5 German cities, downsampled to 4 problem sizes. & 284 trips and 150 trains over 1 day, formulated with approximately 87,000 binary variables. & Embedding from problem variables into physical qubits is not reported. \\ \midrule
Dalyac et al. (2021); journal article; peer reviewed~\cite{dalyac_qualifying_2021} & Scheduling — EV charging. & Two graph models. (1) Charger assignment is mapped to weighted max-k-cut. (2) interval scheduling is mapped to maximum-independent-set problem prepared by a classical CPLEX placement step. & QAOA including an analogue neutral-atom (Rydberg) proposal. & Classical simulation on a supercomputer using Atos QLM and QuTiP. & Synthetic. Instances derived from 2,250 historical charging records from Paris public charging stations, with randomised priorities and groupings. & 250 charging jobs studied analytically at p=1. Simulated QAOA covered graphs of up to approximately 15 jobs and p=7. & 20 simulated qubits. \\ \midrule
Domino et al. (2022); journal article; peer reviewed~\cite{domino_quadratic_2022} & Scheduling — Railway rescheduling. & Time-indexed QUBO with penalties for departure times, headways, track and platform occupancy. Higher-order terms are reduced using auxiliary variables. & Quantum annealing and hybrid annealing. & D-Wave Advantage and D-Wave hybrid solver. & Synthetic. 4 small railway cases designed by the authors. & 5 trains and 5 stations (341 binary variables). & Approximately 740-1,420 physical qubits. \\ \midrule
Domino et al. (2023); journal article; peer reviewed~\cite{domino_quantum_2023} & Scheduling — Resolve railway conflicts. & QUBO in which each train selects 1 departure delay per station, with penalties for single-track occupation, passing-time, and deadlock rules. An equivalent linear integer program is also presented. & Quantum annealing. & D-Wave Advantage and D-Wave 2000Q. & Historic. A Polish single-track line with a real official timetable and disruption scenario. Synthetic. A line with planned infrastructure and synthetic timetable and delays. & A 3-train conflict was solved successfully. 6-train cases returned no feasible solution, and an 18-train case ran on hardware but produced only infeasible output. Largest instance had up to 594 logical variables and 5,552 QUBO connections. & Embedding from problem variables into physical qubits is not reported. \\ \midrule
Grange (2024); doctoral thesis; peer reviewed~\cite{grange_design_2024} & Scheduling — Railway optimisation. & QUBO and higher-order (PUBO) penalty models of simplified train-selection and passenger-assignment problems, with group-coverage and train-capacity constraints. & QAOA and analytical gate-based algorithms combining dynamic programming with quantum search. & Classical simulation using IBM's 32-qubit simulator. Analytical algorithm proposals. & Mixed synthetic and realistic SCNF instances. & A proprietary timetable slice with 32 candidate trains and approximately 4,000 passengers in 40 groups. & 32 simulated qubits. \\ \midrule
Kea et al. (2023); journal article; peer reviewed~\cite{kea_leveraging_2023} & Scheduling — Electric vehicle charging. & 0-1 knapsack model with 1 binary decision per vehicle. A power-capacity constraint is enforced by penalty operations inside the quantum circuit. & Constraint-extended QAOA. & Classical simulation using IBM Qiskit and hardware execution of IBM quantum processor at p=3. & Synthetic. 5 charging scenarios with 2-4 vehicles each. & 4 vehicles evaluated in simulation, 3 vehicles in hardware. & 5-9 qubits. \\ \midrule
Koniorczyk et al. (2025); journal article; peer reviewed~\cite{koniorczyk_solving_2025} & Scheduling — Railway rescheduling. & Integer linear program with headway, track-occupancy, dwell, and rolling-stock constraints. Converted to constrained QUBO form for hybrid annealing. & Hybrid annealing. & D-Wave hybrid solver compared against CPLEX on a local desktop computer (i7). & Synthetic. Instances on 3 model railway lines. Historic. 10 scenarios on a real Polish regional network with a 2021 timetable and constructed disruptions. & 59 trains over a 3-hour horizon, formulated with approximately 1,650 model variables. Cases on the real-network involve 27 trains. & Embedding from problem variables into physical qubits is not reported. \\ \midrule
Mohammad\-bagher\-poor et al. (2021); preprint; not peer reviewed~\cite{mohammadbagherpoor_exploring_2021} & Scheduling — Airline gate assignment. & QUBO for gate assignment recast as graph colouring, with penalties preventing overlapping flights sharing a gate. A compact binary encoding reduces the qubit count. & VQE. & Classical simulation using IBM Qiskit and direct execution on a 27-qubit IBM quantum processor. & Historic. Airline test schedule assembled by Delta Air Lines. & 20-4 flights and 8 gates in simulation. Hardware handled a reduced case of 8 flights. & Hardware runs used 15 physical qubits with the compact encoding versus 25 with the standard encoding. \\ \midrule
Q-CTRL (2025); corporate case study; not peer reviewed~\cite{q-ctrl_accelerating_2025} & Scheduling — Railway assignment. & Station routing selects compatible train-route options as a maximum-weight independent-set model. Timetabling is translated into a satisfiability (MaxSAT) model. Full details are undisclosed. & Unclear, likely VQA. & Q-CTRL's Fire Opal error-suppressed gate-based optimisation on IBM quantum processors. & Historic. Arrival and departure times for London Bridge station with an indicative constraint set. & 26 trains across 15 platforms over an 18-minute window at London Bridge station, formulated as a 103-qubit problem. & 103 physical qubits reported for station routing and over 100 for timetabling. \\ \midrule
Stollenwerk et al. (2020); journal article; peer reviewed~\cite{stollenwerk_quantum_2020} & Scheduling — Deconflicting airplane trajectories. & QUBO in which each flight selects 1 discretised departure delay with penalties on conflicting flight pairs. Classical preprocessing is used to split the flights into independent conflict clusters. & Quantum annealing. & D-Wave 2X and D-Wave 2000Q. 10,000-1,000,000 anneals per QUBO at 20 microseconds each. & Historic. 1 day of 984 real transatlantic flight trajectories, simplified to departure-delay decisions only. & The largest reported 2000Q embedding was 64 flights with 261 conflicts. & Embeddings reached 200 logical variables mapped to approximately 1,100 physical qubits. \\ \midrule
Svensson et al. (2023); journal article; peer reviewed~\cite{svensson_hybrid_2023} & Scheduling — Airline tail assignment. & Binary set-partitioning model translated into QUBO with penalties. Embedded within branch-and-price formalism. & QAOA. & Classical simulations with up to p=40. & Synthetic. 35 small restricted master problem instances derived from real airline tail-assignment data. & Restricted master problems with up to 20 route variables. & 6-20 simulated qubits. \\ \midrule
Vikstål et al. (2020); journal article; peer reviewed~\cite{vikstal_applying_2020} & Scheduling — Airline tail assignment. & Binary set-partitioning model translated into QUBO with penalties. & QAOA. & Classical simulation including depolarising noise. & Historic. Proprietary airline data. & 25 candidate routes covering 278 flights. & 8-25 simulated qubits. \\ \midrule
Xu et al. (2023); journal article; peer reviewed~\cite{xu_high-speed_2023} & Scheduling — Rail timetable optimisation. & Space-time arc-based MILP adapted to QUBO. Candidate train paths are encoded as binary variables and the formulation penalsies conflicting pairs. & Coherent Ising machine. & Classical simulation using QBoson's Kaiwu-SDK. & Historic. Cases derived from historical timetables on a high-speed corridor. & One hundred candidate train paths (100 binary variables). The simulator's solution at this size was described as infeasible. & Embedding from problem variables into physical qubits is not reported. \\ \midrule
\multicolumn{8}{c}{\textbf{Traffic signalling}} \\ \cmidrule(lr){1-8}
Hussain et al. (2020); journal article; peer reviewed~\cite{hussain_optimal_2020} & Traffic signalling — Optimal control. & Macroscopic traffic model coupled with QUBO. The QUBO encodes traffic-clearing rewards, neighbouring-signal coordination rewards, and a 1-mode-per-intersection constraint. & Hybrid annealing. & D-Wave hybrid solver. & Synthetic. Traffic on a 6 \ensuremath{\times} 6 grid of 36 intersections with 4,320 simulated cars. & A 36-intersection grid producing 30 successive 216-variable problems over a 150-second simulation. Simulation updated every 5 seconds. & Embedding from problem variables into physical qubits is not reported. \\ \midrule
Inoue et al. (2021); journal article; peer reviewed~\cite{inoue_traffic_2021} & Traffic signalling — Optimal control. & QUBO with 1 binary signal variable per intersection and interactions between neighbouring intersections. Embedded within a simple macroscopic traffic flow model. & Hybrid annealing. & D-Wave 2000Q with Metis software for problem decomposition. & Synthetic. Traffic on an idealised 50 \ensuremath{\times} 50 grid of intersections with uniform turning behaviour. & A 2,500-intersection grid, solved as 42 separate subproblems of fewer than 64 intersections each. & Embedding from problem variables into physical qubits is not reported. \\ \midrule
Inoue et al. (2024); journal article; peer reviewed~\cite{inoue_traffic_2024} & Traffic signalling — Optimal control. & A model-predictive control problem converted into a QUBO model. This uses one binary signal variable per intersection and is embedded within a microscopic traffic flow model. & Quantum annealing. & D-Wave Advantage with SUMO for traffic simulations. & Synthetic. Traffic based on a real Sapporo road network, plus 9 idealised square grids of 25-625 intersections. & Quantum annealer solved a 100-intersection grid. Larger cases up to 625 intersections used classical simulated annealing. & Embedding from problem variables into physical qubits is not reported. \\ \midrule
Marchesin et al. (2023); journal article; peer reviewed~\cite{marchesin_improving_2023} & Traffic signalling — Optimal control. & Macroscopic traffic model coupled with QUBO. The QUBO groups route directions that share signal timing and controls downstream congestion. Hard penalties constrain one signal mode to be active at a time whereas soft penalties regulate signal sequencing. & Quantum annealing. & D-Wave Advantage, with simulated annealing used as a classical comparison. & Synthetic. Square grids of intersections with corresponding traffic signals. & 225 intersections with 784 traffic signals solved using simulated annealing and 64 intersections with 196 traffic signals using quantum annealing. & Embedding from problem variables into physical qubits is not reported. \\ \midrule
Shikanai et al. (2025); journal article; peer reviewed~\cite{shikanai_quadratic_2025} & Traffic signalling — Optimal control. & Microscopic traffic model coupled with QUBO. The QUBO rewards traffic clearing and coordinated neighbouring signals. Constraints enforce one signal mode active per intersection. & Quantum annealing. & D-Wave Advantage for QUBO optimisation compared with Gurobi and simulated annealing. Microscopic transport model performed with SUMO. & Historic. Real road-map geometry for a district of Aomori City, Japan. Synthetic. Traffic with 10 random demand instances at 5 different traffic levels. & A network of 48 intersections with 100 binary variables. & Embedding from problem variables into physical qubits is not reported. \\ \midrule
\multicolumn{8}{c}{\textbf{Transport network design}} \\ \cmidrule(lr){1-8}
Chandra et al. (2022); conference paper; peer reviewed~\cite{Chandra_2022} & Transport network design — EV charger placement. & QUBO formulation which encodes candidate charger locations as binary variables. Formulated to balance relative proximity of chargers with points of interest. Solutions are subsequently refined with a classical genetic algorithm. & Hybrid annealing. & D-Wave hybrid annealing. & Synthetic. Chargers positioned on grids between sizes 15 \ensuremath{\times} 20 to 100 \ensuremath{\times} 100, with 5-20 points of interest, 1-4 existing chargers, and 3-4 new chargers. & A 100 \ensuremath{\times} 100 grid with 20 points of interest, 4 existing chargers, and 4 new chargers to place. & Embedding from problem variables into physical qubits is not reported. \\ \midrule
Dixit and Niu (2023); journal article; peer reviewed~\cite{dixit_quantum_2023} & Transport network design — Link capacity expansion. & Bi-level model. A QUBO selects link expansions within a budget while classical traffic assignment computes the resulting flows. & Quantum annealing and hybrid annealing. & D-Wave Advantage compared with tabu search on an i7 desktop computer. & Public benchmarks. 10 road networks from 76 to 6,674 links and 24-4,660 nodes. & A road network with 6,674 links solved using hybrid annealing and tabu search. & Embedding from problem variables into physical qubits is not reported. \\ \midrule
Gabbassov (2022); journal article; peer reviewed~\cite{gabbassov_transit_2022} & Transport network design — Bus-stop retention. & Binary variables encode whether to retain or remove a bus stop along a route. A QUBO is formulated by truncating a nonlinear spatial-interaction objective. This aims to find a given number of stops which maximise population accessibility while minimising competition between them. & Hybrid annealing. & D-Wave quantum annealer within broader classical workflow. & Historic. Vancouver B20 bus route with 49 stops, including geographic, census, population, ridership, and dwell data. Synthetic. 30 instances derived from the B20 route with 50 facilities and 135 demand nodes. & 50 candidate stops with 135 demand points per synthetic instance. & Embedding from problem variables into physical qubits is not reported. \\ \midrule
Nourbakhsh et al. (2024); journal article; peer reviewed~\cite{Nourbakhsh_2024} & Transport network design — Bike share station clustering. & Particle swarm optimisation and load flow used to determine bus loads and station counts. A QUBO then encodes 1 binary variable for every station-cluster assignment. Formulation rewards assigning similar stations to the same cluster and enforces that each station belongs to a single cluster. & Quantum annealing. & D-Wave Advantage embedded within a classical data-preparation and evaluation workflow. & Historic. Toronto bike-share data including 625 stations, approximately 950,000 trips, and census population. & 625 stations. Logical variables not reported, but inferred as 1,875. & Embedding from problem variables into physical qubits is not reported. \\ \midrule
Okada et al. (2026); journal article; peer reviewed~\cite{okada_joint_2026} & Transport network design/Routing — Simultaneous optimisation of electric vehicle routes and charging station placement. & Two nested QUBO formulations. An outer learned model selects charging sites. Each candidate placement triggers an inner time-indexed routing QUBO, with battery feasibility checked classically. & Quantum annealing. & D-Wave quantum annealing solves the small station-selection model inside a classical Bayesian learning loop. & Synthetic. A 20-location case with artificial elevation-based energy costs and 16 candidate charger sites. & The single 20-location case. Only its 16-variable station-selection subproblem was solved on the annealer and required 16 logical variables. & Embedding from problem variables into physical qubits is not reported. \\ \midrule
Rao and Sodhi (2023); journal article; peer reviewed~\cite{Rao_2022} & Transport network design — Electric-vehicle charger placement. & QUBO over candidate sites which weighs ownership cost and demand-capacity mismatch. A penalty term constrains the station counts per bus. & Hybrid annealing and VQA. & D-Wave hybrid solver plus VQA simulation (using TensorFlow Quantum/Cirq) trained by a classical neural network. & Synthetic. Candidate locations and spatial demand regions overlaid on a power-grid system. Each candidate charger was assigned to a load bus and the required charger count per bus determined classically. & 2,000 candidate charging sites as a 2,000-variable QUBO with 100 sites selected. Solved through hybrid annealing. & 200-2,000 logical variables for the hybrid solver and 20 simulated qubits for the variational circuit. \\ \midrule
\multicolumn{8}{c}{\textbf{Computer vision}} \\ \cmidrule(lr){1-8}
Innan et al. (2025); journal article; peer reviewed~\cite{innan_qnn-vrcs_2025} & Computer vision — Traffic light colour classification. & Quantum machine learning. A quantum neural network plus 2 quantum similarity classifiers over heavily downscaled images. & Quantum neural network and VQA using angle encoding. & Classical simulation, including simulated hardware noise and compared against a convolutional neural network. & Public benchmarks. 2 image datasets: 1,909 driving-simulator images and a real-world traffic-light collection. & Classification over the 2 datasets with images reduced to 2 \ensuremath{\times} 2 or 4 \ensuremath{\times} 4 pixels. & 4 simulated qubits for the main classifier and 3-12 for the image-encoding variants. \\ \midrule
Kang et al. (2025); conference paper; peer reviewed~\cite{Kang_2025} & Computer vision — Number plate digit recognition. & Modified pretrained ResNet-18 features are compressed to 2 values, transformed by a 2-qubit quantum circuit, fused with classical features through a learned alpha, and classified into 10 digits. & Hybrid quantum-classical neural network. Quantum circuit uses 2 data-parameterised Ry gates and 1 CNOT. & Classical simulation using IBM Qiskit. & Public benchmarks. Handwritten digit benchmark: a 12,000-image training subset and an evaluation set of approximately 10,000 images. & A 28 \ensuremath{\times} 28-pixel digit image which was compressed to 2 features for the quantum layer. & 2 simulated qubits per image with 1,000 measurement shots. \\ \midrule
Kuros and Kryjak (2023); conference paper; peer reviewed~\cite{kuros_traffic_2023} & Computer vision — Road sign classification. & Preprocessed traffic-sign images are passed through convolutional, max-pooling, and dropout layers. The resulting feature maps are flattened and processed by dense layers to assign 1 of 43 classes of traffic signs. & Hybrid quantum-classical neural network. Input feature maps are first produced through quantum rather than conventional convolution. This is achieved through a shallow quantum circuit. & Classical simulation using PennyLane. Random "quanvolutional" feature maps were produced through RY rotations, Pauli-X gates, and 2 random circuit layers. & Public benchmarks. A 7,000-image subset of the German Traffic Sign Recognition Benchmark covering 43 classes of traffic signs. & The largest training batch had 512 images. However, quantum filter was applied to a 2 \ensuremath{\times} 2-pixel patch, comprising 4 pixels. & 4 simulated qubits. \\ \midrule
Meghanath et al. (2025); journal article; peer reviewed~\cite{meghanath_qdcnn_2025} & Computer vision — Correct image shadows, extract road lanes, and classify a simulated vehicle's intended direction as left, right or straight. & Quantum machine learning classifiers. A 1-qubit overlap test and a 2-qubit variational circuit inside a largely classical image-processing pipeline. & Parameterised quantum circuit. & Classical simulation using IBM Qiskit. & Synthetic. Driving-simulator imagery. & 1 simulated qubit for the overlap classifier and 2 for the variational classifier (4 trained parameters). & 1 and 2 simulated qubits. \\ \midrule
\multicolumn{8}{c}{\textbf{Drive cycle estimation}} \\ \cmidrule(lr){1-8}
Dixit and Jian (2022); journal article; peer reviewed~\cite{dixit_quantum_2022} & Drive cycle estimation — Estimate the dominant frequencies of vehicle acceleration and braking patterns. & An \$N=2\textasciicircum{}n\$ binary drive-cycle signal is loaded into an \$n\$-qubit state, Fourier transformed, and measured as frequency probabilities. & Gate-model quantum Fourier transform, followed by a classical statistical correction that mitigates quantum hardware noise. & Classical simulation plus direct execution on a 15-qubit IBM quantum processor. & Historic. Three real 256-second drive-cycle signals across one day. Synthetic. 22 short generated sequences for the hardware runs. & A 256-sample signal (8 qubits) in simulation, hardware runs were limited to 16-sample sequences (4 qubits). & 8 simulated and 4 physical qubits. \\ \midrule
\multicolumn{8}{c}{\textbf{Traffic forecasting}} \\ \cmidrule(lr){1-8}
Hari\-krishna\-kumar et al. (2023); journal article; peer reviewed~\cite{harikrishnakumar_forecasting_2023} & Traffic forecasting — Bike-share demand. & Separate LSTM and Gaussian-process forecasts are inverse-RMSE weighted, discretised into 4 states, represented as a 3-node Bayesian network, and sampled as a compositional quantum Bayesian network. & Quantum Bayesian network. & Classical simulation using IBM Qiskit. & Historic. Dataset derived from approximately 9.6 million New York City bike-share trips over 6 months. & A 3-node Bayesian network forecasting a single station's hourly demand. & 10 simulated qubits (6 data qubits plus 4 auxiliary qubits). \\ \midrule
Schetakis et al. (2025); journal article; peer reviewed~\cite{schetakis_quantum_2025} & Traffic forecasting — Loop detector counts. & A window of past traffic-flow measurements is compressed by a pretrained LSTM autoencoder. The resulting embedding is passed to a regressor predict future traffic. & Quantum machine learning. In the hybrid model, either a fully connected layer or an LSTM layer is replaced by a variational quantum circuit using angle encoding. & Classical simulation using PennyLane. & Historic. Approximately 40,000 traffic-flow measurements over 40 days from a loop detector on a road in Athens. & One-step forecasting on the 40-day series. & 2-14 simulated qubits across 6 model sizes. \\ \midrule
\end{longtable}
\end{landscape}

\begin{landscape}
\renewcommand{\thetable}{S2B}
\setlength{\LTleft}{0pt}
\setlength{\LTright}{0pt}
\setlength{\tabcolsep}{2.4pt}
\renewcommand{\arraystretch}{1.15}
\footnotesize
\begin{longtable}{>{\RaggedRight\arraybackslash}p{3.30cm}>{\RaggedRight\arraybackslash}p{7.40cm}>{\RaggedRight\arraybackslash}p{7.40cm}>{\RaggedRight\arraybackslash}p{7.90cm}}
\caption{Advantage-evidence assessment (C1--C3)}\label{tab:s2b}\\
\toprule
\textbf{Study} & \textbf{C1: best classical comparator} & \textbf{C2: computational costs} & \textbf{C3: quantum advantage/scaling evidence} \\
\midrule
\endfirsthead
\multicolumn{4}{l}{\textit{Table S2B: Advantage-evidence assessment (C1--C3) (continued)}}\\
\toprule
\textbf{Study} & \textbf{C1: best classical comparator} & \textbf{C2: computational costs} & \textbf{C3: quantum advantage/scaling evidence} \\
\midrule
\endhead
\midrule
\multicolumn{4}{r}{\textit{Continued on next page}}\\
\endfoot
\bottomrule
\endlastfoot
\multicolumn{4}{c}{\textbf{Routing}} \\ \cmidrule(lr){1-4}
Ariño Sales and Palacios Araos (2023); journal article; peer reviewed~\cite{arino_sales_adiabatic_2023} & \cellcolor{EvidenceP}P — Capacity-aware k-medoids clustering and Google OR-Tools routing were run on the same benchmark family. Neither was established as the best available classical method for CVRP. & \cellcolor{EvidenceN}N — No timing or memory costs reported for any quantum or classical component. & \cellcolor{EvidenceN}N — No compelling scaling evidence demonstrated. \\ \midrule
Azad et al. (2023); journal article; peer reviewed~\cite{azad_solving_2023} & \cellcolor{EvidenceP}P — IBM's CPLEX solver was run on the same instances. Relevant baseline for exact solution, but not established state-of-the-art. & \cellcolor{EvidenceN}N — No timing reported for either the quantum simulations or CPLEX. & \cellcolor{EvidenceN}N — Only instances of 4 or 5 locations were tested. No runtime or quality scaling with problem size was reported. \\ \midrule
Bai and Shang (2025); journal article; peer reviewed~\cite{Bai_2025} & \cellcolor{EvidenceN}N — No classical method. Quantitative comparisons were theoretical and against 2 earlier quantum algorithms. & \cellcolor{EvidenceN}N — Theoretical gate, query, and qubit counts were given, but no runtime, compiled circuit sizes, or end-to-end costs were measured. & \cellcolor{EvidenceN}N — Simulations stopped at 8 cities, and the algorithm's search cost still grows with the square root of a factorial. No practical crossover of quantum advantage was demonstrated. \\ \midrule
Bennett et al. (2021); journal article; peer reviewed~\cite{bennett_quantum_2021} & \cellcolor{EvidenceN}N — The quantum method was compared only to random sampling. & \cellcolor{EvidenceN}N — The comparison counted only final circuit evaluations (iterations) and excluded the required classical parameter search. No timing reported. & \cellcolor{EvidenceN}N — A single synthetic 8-location instance was tested with only iteration depth varied. \\ \midrule
Bentley et al. (2022); preprint; not peer reviewed~\cite{bentley_quantum_2022} & \cellcolor{EvidenceN}N — No classical method. The reported 97\% success was compared only against 2 alternative quantum circuits and an ideal-circuit reference. & \cellcolor{EvidenceN}N — No runtimes, shot counts, or other computational costs were reported for any component. & \cellcolor{EvidenceN}N — Only a 2-passenger circuit was executed. Claimed scaling benefits rely on analytical qubit-count formulas and hardware roadmaps, not comparative measurements to classical methods. \\ \midrule
Borowski et al. (2020); conference paper; peer reviewed~\cite{borowski_new_2020} & \cellcolor{EvidenceP}P — Simulated annealing, a bee algorithm, and 2 other established metaheuristics were run on the same cases, but their tuning and best-available status were not established. & \cellcolor{EvidenceN}N — No timing reported. & \cellcolor{EvidenceN}N — 60 cases from 1 to 200 orders gave breadth, but no scaling or performance law was established. Moreover, quantum annealing was only performed on small instances. \\ \midrule
Chence et al. (2025); journal article; peer reviewed~\cite{niu_applying_2025} & \cellcolor{EvidenceP}P — CPLEX solver for MIP is relevant baseline but not established as best classical method. & \cellcolor{EvidenceP}P — Annealing run times reported, unclear if they represent execution time or end-to-end costs. & \cellcolor{EvidenceN}N — Moderate problem size solved through hybrid annealing. Scaling advantage cannot be assessed due to inconsistent target solution quality for each run. \\ \midrule
Ciacco et al. (2025); conference paper; peer reviewed~\cite{Ciacco_2025} & \cellcolor{EvidenceP}P — Gurobi and classical simulated annealing were executed on the same instances. Relevant baselines, though no best-known solver or optimality gap was established. & \cellcolor{EvidenceP}P — Execution run-times present, but unclear if they represent end-to-end time with embedding and classical processing. Instances solved with hybrid all solved to optimality, providing consistent quality target. & \cellcolor{EvidenceN}N — Valid direct-annealer results stopped at 5 vertices and the hybrid solver at 9. No explicit scaling analysis presented between classical and quantum. \\ \midrule
Codognet (2024); conference paper; peer review~\cite{Codognet_2024} & \cellcolor{EvidenceN}N — 2 classical GPU annealing solvers were executed and scored against published optimal tour values, but no established classical TSP method was run. & \cellcolor{EvidenceN}N — Only solver time-to-solution under a 60-second cap was reported. & \cellcolor{EvidenceN}N — No quantum execution at scale and no runtime-scaling analysis. \\ \midrule
Dixit et al. (2024); journal article; peer reviewed~\cite{dixit_quantum_2024} & \cellcolor{EvidenceP}P — CPLEX solver is relevant baseline but not established as best classical method. & \cellcolor{EvidenceP}P — Annealing run times reported, unclear if they represent just execution time or total end-to-end costs. & \cellcolor{EvidenceN}N — Annealing runtimes were chosen to be linear with effective problem size, and therefore must display a linear scaling. Scaling advantage cannot be assessed due to inconsistent target solution quality for each run. \\ \midrule
Dong and Huang (2020); journal article; peer reviewed~\cite{Dong_2020} & \cellcolor{EvidenceN}N — Simulated annealing and simulated quantum annealing were run on the same 2 samples. The authors acknowledged stronger classical methods but did not execute them. & \cellcolor{EvidenceN}N — Only iteration and Monte Carlo step counts were reported. No explicit run-time presented. & \cellcolor{EvidenceN}N — Only 2 point sets of unreported size were tested, with no empirical or theoretical scaling analysis. \\ \midrule
Fagiolo and Vescera (2025); conference paper; peer reviewed~\cite{Fagiolo_2025} & \cellcolor{EvidenceN}N — No classical method. & \cellcolor{EvidenceN}N — No timing reported, and total circuit evaluations and classical optimisation costs were not counted. & \cellcolor{EvidenceN}N — 9 small synthetic instances did not establish scaling, and 7-city behaviour already degraded. \\ \midrule
Feld et al. (2019); journal article; peer reviewed~\cite{feld_hybrid_2019} & \cellcolor{EvidenceN}N — A purely classical version of the same workflow was executed on 1 instance, but no modern classical routing method was run. & \cellcolor{EvidenceP}P — End-to-end and component times (including remote-access overheads) were reported, but only for a single run on a 50-location instance. & \cellcolor{EvidenceN}N — Solution quality worsened as tours grew from 14 to 38 cities, quantum subproblems never exceeded 20 variables, and runtime was reported on 1 instance only. \\ \midrule
Finžgar et al. (2022); conference paper; peer reviewed~\cite{finzgar_quark_2022} & \cellcolor{EvidenceN}N — Simulated annealing on identical QUBO models were run. Uncompetitive formulation for the TSP variant. & \cellcolor{EvidenceM}M — End-to-end times covering mapping, solving (including embedding), and validation were measured. & \cellcolor{EvidenceN}N — Several sizes per problem were tested, but quantum solution quality collapsed as instance size grew. \\ \midrule
Fitzek et al. (2024); journal article; peer reviewed~\cite{fitzek_applying_2024} & \cellcolor{EvidenceN}N — Exact answers came from brute-force enumeration of the tiny instances, but no classic method was performed for comparison. & \cellcolor{EvidenceN}N — Runtimes were reported only as relative comparisons among the classical parameter optimisers. & \cellcolor{EvidenceN}N — Only 3 instances of 3 or 4 cities were simulated, with success probability falling sharply at the larger sizes. Authors call these too small for meaningful comparison. \\ \midrule
Garcia de Andoin et al. (2023); conference paper; peer reviewed~\cite{De_Andoin_2023} & \cellcolor{EvidenceN}N — No classical method. & \cellcolor{EvidenceN}N — No timing reported. & \cellcolor{EvidenceN}N — Only 1 toy instance was tested. \\ \midrule
Haba et al. (2025); journal article; peer reviewed~\cite{haba_routing_2025} & \cellcolor{EvidenceP}P — Gurobi optimiser and greedy algorithm were run on the same subproblems. Gurobi is a relevant baseline, but not demonstrated state-of-the-art. & \cellcolor{EvidenceP}P — Full annealer access time reported, but not broken down into individual components. Classical post-processing not reported. & \cellcolor{EvidenceN}N — All 629 tested problems had between 5 and 50 variables from a single simulated network. No TTS growth trend measured, and problem size too small to establish scaling. \\ \midrule
Harwood et al. (2021); journal article; peer reviewed~\cite{harwood_formulating_2021} & \cellcolor{EvidenceN}N — CPLEX was executed, but only to characterise the formulation. Quantum runs were scored against random sampling rather than a competitive routing solver. & \cellcolor{EvidenceN}N — No timing reported. & \cellcolor{EvidenceN}N — Model sizes were counted across 4 planning horizons, but quantum performance was tested at a single small size only. \\ \midrule
He (2024); conference paper; peer reviewed~\cite{He_2024} & \cellcolor{EvidenceM}M — Strong exact solvers (Concorde, Gurobi) and a leading heuristic were tabulated on the claimed same dataset. These results were imported from literature rather than rerun. & \cellcolor{EvidenceN}N — Reported times mix unlabelled components on unknown hardware, and decomposition, training, and search costs were not aligned across methods. & \cellcolor{EvidenceN}N — 3 benchmark sizes were tabulated, but the quantum simulation itself never exceeded 10-city subproblems, so no genuine quantum scaling was shown. \\ \midrule
Holliday et al. (2024); conference paper; peer reviewed~\cite{Holliday_2024} & \cellcolor{EvidenceN}N — Best-known solution values and published heuristic results give quality context, but no strong modern classical was executed. & \cellcolor{EvidenceN}N — 1 hour time cap and stopping rules reported, no other relevant time frames mentioned. & \cellcolor{EvidenceN}N — 13 instances span 10-200 locations over 2 problem families, but quantum subproblem scale is not reported. \\ \midrule
Hussain et al. (2025); conference paper; peer reviewed~\cite{Hussain_2025} & \cellcolor{EvidenceN}N — No classical method. & \cellcolor{EvidenceN}N — No timing reported. & \cellcolor{EvidenceN}N — No scaling evidence demonstrated. \\ \midrule
Irie et al. (2019); conference paper; peer reviewed~\cite{irie_quantum_2019} & \cellcolor{EvidenceN}N — No classical method. & \cellcolor{EvidenceN}N — The reported 0.2 seconds covered annealing only. & \cellcolor{EvidenceN}N — Only one small instance was studied. \\ \midrule
Jain (2021); journal article; peer reviewed~\cite{jain_solving_2021} & \cellcolor{EvidenceN}N — Brute force enumeration gave exact optima and a simple 2-opt heuristic ran on the same instances. & \cellcolor{EvidenceN}N — The quantum run-time counted only 2 milliseconds of annealing, excluding embedding and all other overheads. & \cellcolor{EvidenceN}N — Only 3 city sizes were trialled, with failure to find solution beyond 8 cities. \\ \midrule
Kieu (2019); journal article; peer reviewed~\cite{Kieu_2019} & \cellcolor{EvidenceN}N — No classical method. & \cellcolor{EvidenceN}N — Space and energy requirements and a necessary time bound were discussed, but the actual running time and success probability were not reported. & \cellcolor{EvidenceN}N — The claimed efficient scaling is conditional. Author notes the rigorous run-time analysis remains open. \\ \midrule
Le et al. (2023); conference paper; peer reviewed~\cite{le_quantum_2023} & \cellcolor{EvidenceN}N — No classical method. & \cellcolor{EvidenceP}P — Quantum-processor access, sampling, and postprocessing times are reported. Omits embedding time. & \cellcolor{EvidenceN}N — Only 5-7 cities were tested, the largest unsuccessfully. \\ \midrule
Leonidas et al. (2024); journal article; peer reviewed~\cite{leonidas_qubit_2024} & \cellcolor{EvidenceN}N — No classical method. & \cellcolor{EvidenceN}N — No timing reported. & \cellcolor{EvidenceN}N — No compelling scaling evidence demonstrated. \\ \midrule
Li and Mazumder (2023); conference paper; peer review not assessable~\cite{Li_2023} & \cellcolor{EvidenceP}P — Exact dynamic programming was run on the same nine graphs and was faster than the hybrid on eight of them, but was not established as the strongest available classical solver. & \cellcolor{EvidenceP}P — The reported runtimes had an undefined boundary, with embedding, queueing, decoding, and reconstruction costs not itemised. & \cellcolor{EvidenceN}N — Only 9 graphs of 4-20 locations were tested. Quantum outputs become unusably noisy above 16 cities. \\ \midrule
Li and Ruan (2024); conference paper; peer reviewed~\cite{Li_2024} & \cellcolor{EvidenceP}P — 4 established learned routing models were trained on the same platform for comparison, but no classical solvers considered. & \cellcolor{EvidenceN}N — No timing reported. & \cellcolor{EvidenceN}N — All results pertain to 5-city random instances. No scaling with problem size was examined. \\ \midrule
Li et al. (2025); journal article; peer reviewed~\cite{li_quantum_2024} & \cellcolor{EvidenceN}N — No classical method. & \cellcolor{EvidenceN}N — No timing reported. & \cellcolor{EvidenceN}N — Executed problems stayed at 17-20 simulated qubits and performance worsened as qubit counts grew. \\ \midrule
Liu et al. (2025); journal article; peer reviewed~\cite{liu_solving_2025} & \cellcolor{EvidenceN}N — No classical method. The claimed quadratic speed-up is measured only against a theoretical count of classical search steps. & \cellcolor{EvidenceN}N — No timing reported. Complexity analysis counts only oracle queries and excluded circuit construction, repeated searches, and post-processing. & \cellcolor{EvidenceN}N — Executed cases stopped at 4 locations, with no runtime, resource, or quality trend measured against problem size. \\ \midrule
Maciejunes et al. (2025); conference paper; peer reviewed~\cite{Maciejunes_2025} & \cellcolor{EvidenceN}N — Google OR-Tools was executed but unable to find minima on small problem sizes. & \cellcolor{EvidenceN}N — No timing reported. & \cellcolor{EvidenceN}N — Only 7-to-13-location cases on undisclosed graphs were tested. No runtime scaling or larger executions were reported. \\ \midrule
Makhanov et al. (2024); conference paper; peer reviewed~\cite{makhanov_quantum_2024} & \cellcolor{EvidenceP}P — The classical comparison is a simple Dijkstra implementation with standard minimum-search routine. This is a strong baseline. However, the authors note that Fibonacci-heap Dijkstra has better asymptotic scaling, but did not execute it. & \cellcolor{EvidenceP}P — Simulator times, projected gate times, and circuit-construction time were reported. Hardware execution and end-to-end route computation costs were excluded. & \cellcolor{EvidenceN}N — The simulated search covers 4-256 candidates, but no scaling of the complete route workflow on quantum hardware was demonstrated. \\ \midrule
Markevich and Trushechkin (2019); journal article; peer reviewed~\cite{Markevich_2019} & \cellcolor{EvidenceP}P — Classical branch-and-bound was executed, but was not established as state-of-the-art. & \cellcolor{EvidenceN}N — Only counts of bounding-function calls were compared. Preprocessing for the quantum method and run-time costs not reported. & \cellcolor{EvidenceN}N — Only 1 problem size (20 cities) was evaluated. \\ \midrule
Mohanty et al. (2023); journal article; peer reviewed~\cite{mohanty_analysis_2023} & \cellcolor{EvidenceN}N — No classical method. & \cellcolor{EvidenceN}N — Only an aggregate 219 processor-hours of simulation time was reported, with no per-run timing or comparable classical costs. & \cellcolor{EvidenceN}N — Only 2 toy cases at 6 and 12 simulated qubits were tested. \\ \midrule
Mohanty et al. (2024); journal article; peer reviewed~\cite{mohanty_solving_2024} & \cellcolor{EvidenceN}N — Results were scored against a classical minimum value. No comparison to classical VRP solver. & \cellcolor{EvidenceN}N — Only a study-wide total of 150 processor-hours was reported. & \cellcolor{EvidenceN}N — Only 3- and 4-city cases were studied. \\ \midrule
Moosavi and Farooq (2025); conference paper; peer reviewed~\cite{Moosavi_2025} & \cellcolor{EvidenceN}N — Executed baselines were 2 classical reinforcement-learning methods. No exact solver or strong routing heuristic was trialled. & \cellcolor{EvidenceN}N — No timing reported. & \cellcolor{EvidenceN}N — All results use 1 synthetic problem size. \\ \midrule
Moylett et al. (2017); journal article; peer reviewed~\cite{moylett_quantum_2017} & \cellcolor{EvidenceM}M — The proved running times were compared against the best classical algorithms for each setting. & \cellcolor{EvidenceM}M — The major algorithmic costs are counted in the asymptotic analysis. & \cellcolor{EvidenceM}M — Quadratic scale speedup for graphs below degree 7 shown through asymptotic complexity analysis. The scale at which quantum advantage occurs is not demonstrated. \\ \midrule
Nałęcz-Charkiewicz et al. (2025); journal article; peer reviewed~\cite{Na_e_cz_Charkiewicz_2025} & \cellcolor{EvidenceP}P — Gurobi solver and simulated annealing control were executed. Relevant baselines but not established as the best available routing method. & \cellcolor{EvidenceP}P — Solve times were reported, but under different time limits and qualities for each solver. Preprocessing and partitioning overheads were not listed. & \cellcolor{EvidenceN}N — No compelling scaling evidence demonstrated. \\ \midrule
Neukart et al. (2017); journal article; peer reviewed~\cite{neukart_traffic_2017} & \cellcolor{EvidenceN}N — No classical method. & \cellcolor{EvidenceN}N — Only a 22-second minimum solver time was reported. & \cellcolor{EvidenceN}N — A single 418-taxi scenario was tested. No compelling scaling evidence demonstrated. \\ \midrule
Osaba et al. (2024); journal article; peer reviewed~\cite{osaba_solving_2024} & \cellcolor{EvidenceN}N — Google OR-Tools was executed, but only to re-optimise individual routes already chosen by the quantum workflow. No classical method solved the complete delivery problem. & \cellcolor{EvidenceN}N — No timing reported for either the hybrid workflow or OR-Tools. & \cellcolor{EvidenceN}N — 6 cases of increasing size were each solved once with no runtimes reported. \\ \midrule
Palackal et al. (2023); conference paper; peer reviewed~\cite{palackal_quantum-assisted_2023} & \cellcolor{EvidenceN}N — No classical method. & \cellcolor{EvidenceN}N — Circuit-evaluation counts were reported, but shots, hardware access, compilation, and end-to-end times were not. & \cellcolor{EvidenceN}N — Performance was tested on just 1 subproblem per size at 4-6 locations, and hardware feasibility collapsed beyond 4. \\ \midrule
Papalitsas et al. (2019); journal article; peer reviewed~\cite{papalitsas_qubo_2019} & \cellcolor{EvidenceN}N — No classical method. & \cellcolor{EvidenceN}N — No timing reported. & \cellcolor{EvidenceN}N — No instances solved. \\ \midrule
Qian et al. (2023); journal article; peer reviewed~\cite{qian_comparative_2023} & \cellcolor{EvidenceN}N — No classical method. & \cellcolor{EvidenceN}N — No timing reported. Circuit depth and gate counts per layer were documented. & \cellcolor{EvidenceN}N — Only 3- to 5-city simulations were performed. \\ \midrule
Ruan et al. (2020); journal article; peer reviewed~\cite{ruan_quantum_2020} & \cellcolor{EvidenceN}N — No classical method. & \cellcolor{EvidenceN}N — No timing reported. & \cellcolor{EvidenceN}N — Only a 5-city instance was simulated. \\ \midrule
Salehi et al. (2022); journal article; peer reviewed~\cite{salehi_unconstrained_2022} & \cellcolor{EvidenceN}N — No classical method. Simulated annealing runs served only to tune penalty terms. & \cellcolor{EvidenceN}N — Only a programmed 50-microsecond anneal time was reported. Tuning, embedding, access, readout, and comparable classical costs were excluded. & \cellcolor{EvidenceN}N — Hardware results stopped at 5 cities with collapsing performance. \\ \midrule
Sato et al. (2023); conference paper; peer reviewed~\cite{Sato_2023} & \cellcolor{EvidenceN}N — No classical method. & \cellcolor{EvidenceN}N — Only iteration-count proxies were reported. Gate counts, simulation time, and cost of eventual tour search were omitted. & \cellcolor{EvidenceN}N — A single configuration of 3, 4, and 5 cities was simulated. Success probability fell sharply as size grew. \\ \midrule
Sharma et al. (2025); journal article; peer reviewed~\cite{Sharma_2025} & \cellcolor{EvidenceN}N — The exponential speed-up was compared against brute-force enumeration only. & \cellcolor{EvidenceN}N — Cost analysis omits major components. This includes loading the network data into the circuit and repeated runs of threshold procedure. & \cellcolor{EvidenceN}N — No compelling scaling evidence demonstrated. \\ \midrule
Silva et al. (2021); journal article; peer reviewed~\cite{Silva_2021} & \cellcolor{EvidenceN}N — Classical simulated annealing was performed, but no established TSP solvers. & \cellcolor{EvidenceP}P — The quantum timing counts on-device sampling only, excluding access, embedding, and preparation costs. Classical timings were not benchmarked against comparable solution qualities. & \cellcolor{EvidenceN}N — Only graphs up to 8 vertices were considered. \\ \midrule
Sinno et al (2025); journal article; peer reviewed~\cite{sinno_analyzing_2025} & \cellcolor{EvidenceN}N — Solutions were scored against published best-known route costs, but no classical solver was executed for comparison. & \cellcolor{EvidenceN}N — No timing reported. & \cellcolor{EvidenceN}N — Solution quality worsened with instance size. No runtime scaling was reported. \\ \midrule
Spyridis et al. (2023); conference paper; peer reviewed~\cite{spyridis_variational_2023} & \cellcolor{EvidenceN}N — No classical method. & \cellcolor{EvidenceN}N — No timing reported. & \cellcolor{EvidenceN}N — A single 4-city example was presented. \\ \midrule
Tambunan et al. (2023); conference paper; peer-review status not assessable~\cite{Tambunan_2023} & \cellcolor{EvidenceN}N — No classical method. & \cellcolor{EvidenceN}N — No timing reported. & \cellcolor{EvidenceN}N — Only 3-5 vehicles on 1 small synthetic network were tested. No runtime or quality trend with size was reported. \\ \midrule
Tszyunsi and Beterov (2023); journal article; peer reviewed~\cite{Tszyunsi_2023} & \cellcolor{EvidenceN}N — No classical method. & \cellcolor{EvidenceN}N — No timing reported. & \cellcolor{EvidenceN}N — Only a 4-city example was simulated. \\ \midrule
Vargas et al. (2024); conference paper;~\cite{Vargas_2024} & \cellcolor{EvidenceP}P — The open-source CBC branch-and-cut solver was executed and gave slightly better routes overall. This is likely a strong baseline, but not established as the strongest. & \cellcolor{EvidenceP}P — Model-construction, embedding and solver times were each reported, but the comparative plot excluded the dominant preprocessing and no common solution quality with the classical solver was defined. & \cellcolor{EvidenceN}N — Timings span 8 sizes between 2 and 12 nodes from a single data source. While QA solve time scales faster than CBC in this regime (refer to Figure 1), advantage disappears when embedding and classical pre-processing were included. \\ \midrule
Villar-Rodríguez et al. (2022); conference paper; peer reviewed~\cite{villar-rodriguez_analyzing_2022} & \cellcolor{EvidenceN}N — No classical method. & \cellcolor{EvidenceP}P — Only the annealing time per read and read counts were reported. Access overheads, classical processing, and end-to-end times were excluded. & \cellcolor{EvidenceN}N — Only up to 7 cities considered with direct quantum annealing. \\ \midrule
Weinberg et al. (2023); journal article; peer reviewed~\cite{weinberg_supply_2023} & \cellcolor{EvidenceN}N — A simulated-annealing version of the workflow was performed largely on subproblems, but was not directly comparable to the quantum version. & \cellcolor{EvidenceN}N — No timing reported. & \cellcolor{EvidenceN}N — The application is genuinely commercial in scale, but only a single instance was solved, with no reported scaling trend. \\ \midrule
Xie et al. (2024); journal article; peer reviewed~\cite{xie_feasibility-preserved_2024} & \cellcolor{EvidenceN}N — No classical method. & \cellcolor{EvidenceN}N — No timing reported. Theoretical gate-count formulas were provided. & \cellcolor{EvidenceN}N — Problem size never exceeded 4 customers. \\ \midrule
Xu and Créput (2025); conference paper; peer reviewed~\cite{XU_2025} & \cellcolor{EvidenceN}N — While a task-equivalent classical self-organised map was used as a baseline, it is not the best known method for the TSP. & \cellcolor{EvidenceP}P — Values were reported for the presumed end-to-end quantum runtime, but it is unclear if this includes costs such as state preparation and postprocessing. & \cellcolor{EvidenceN}N — No compelling scaling evidence demonstrated. \\ \midrule
Yarkoni et al. (2020); conference paper; peer reviewed~\cite{yarkoni_quantum_2020} & \cellcolor{EvidenceN}N — No classical method. & \cellcolor{EvidenceP}P — Service response times were reported for all 1,275 live requests, but component-level costs and the solver's internal time split were not. & \cellcolor{EvidenceN}N — Only 9 buses in 1 city and no scaling data. \\ \midrule
Yarkoni et al. (2021); conference paper; peer reviewed~\cite{yarkoni_solving_2021} & \cellcolor{EvidenceP}P — Gurobi, simulated annealing, Tabu search, and a direct-shipment baseline were all executed. Gurobi proved small-case optima but solved a slightly different exact model, and no specialised state-of-the-art method was tried. & \cellcolor{EvidenceP}P — Time budgets and a provider-reported quantum-processing total were provided. However, budgets differed across methods and end-to-end times, hardware, and preprocessing costs were absent. & \cellcolor{EvidenceN}N — The study shows quantum performance decreases with problem size, and authors explicitly caution against current QUBO conversion. \\ \midrule
Zawalska and Rycerz (2023); conference paper; peer reviewed~\cite{Zawalska_2023} & \cellcolor{EvidenceN}N — No classical method. & \cellcolor{EvidenceN}N — No timing reported. & \cellcolor{EvidenceN}N — Every instance had 4 cities. Only circuit depth was varied, which is not problem-size scaling. \\ \midrule
\multicolumn{4}{c}{\textbf{Scheduling}} \\ \cmidrule(lr){1-4}
Bickert et al. (2021; version 3 revised 2023); preprint; not peer reviewed~\cite{bickert_optimising_2023} & \cellcolor{EvidenceP}P — A task-specific constraint-programming solver and tabu search on the same model were executed. Potentially the strongest baseline, but not explicitly demonstrated as such. & \cellcolor{EvidenceP}P — Model-building and solver times were reported, but postprocessing times are unreported. & \cellcolor{EvidenceN}N — No compelling scaling evidence demonstrated. \\ \midrule
Dalyac et al. (2021); journal article; peer reviewed~\cite{dalyac_qualifying_2021} & \cellcolor{EvidenceN}N — No classical method. & \cellcolor{EvidenceN}N — No timing reported. & \cellcolor{EvidenceN}N — No compelling scaling evidence demonstrated. \\ \midrule
Domino et al. (2022); journal article; peer reviewed~\cite{domino_quadratic_2022} & \cellcolor{EvidenceP}P — Exact integer-programming (CBC) optima and same-problem simulated annealing were executed. These are relevant baselines, though neither was established as a state-of-the-art railway rescheduler. & \cellcolor{EvidenceN}N — Only the 250-microsecond anneal time per read was reported. & \cellcolor{EvidenceN}N — Direct annealing found no feasible schedule beyond 176 variables. \\ \midrule
Domino et al. (2023); journal article; peer reviewed~\cite{domino_quantum_2023} & \cellcolor{EvidenceP}P — Exact enumeration, CPLEX, an equivalent linear integer program solved optimally, and dispatching heuristics were all run. Relevant baselines, but none were established as the best modern method. & \cellcolor{EvidenceP}P — Quantum run times reported only the anneals themselves. Embedding, access, and decoding were excluded, and most classical runtimes were not given. & \cellcolor{EvidenceN}N — Performance worsened with size. Larger cases were infeasible or could not be embedded, while a classical solver handled every instance in at most 0.03 seconds. \\ \midrule
Grange (2024); doctoral thesis; peer reviewed~\cite{grange_design_2024} & \cellcolor{EvidenceM}M — The central exact quantum scheduling results are compared at the same solution standard with cited best-known classical exact worst-case complexities; one weighted/unweighted comparator row requires qualification, and the empirical railway QAOA experiments retain only weak random-sampling baselines. & \cellcolor{EvidenceM}M — The asymptotic analysis includes classical preprocessing and bounded-error quantum search, but assumes constant-time coherent access to exponentially large QRAM. & \cellcolor{EvidenceM}M — Theoretical scaling rates are proved for the exact algorithms, but assume access to QRAM. Empirical evidence stopped at 32 simulated qubits, with success probability collapsing as instances grew. \\ \midrule
Kea et al. (2023); journal article; peer reviewed~\cite{kea_leveraging_2023} & \cellcolor{EvidenceN}N — No classical method. & \cellcolor{EvidenceP}P — The study reports run-time between 5-7 seconds for the quantum hardware, but it is unclear if this includes classical optimisation of angles. & \cellcolor{EvidenceN}N — Only 5 synthetic cases of 2-4 vehicles were tested. No trend in runtime with problem size was reported. \\ \midrule
Koniorczyk et al. (2025); journal-author manuscript/preprint; peer review not assessable~\cite{koniorczyk_solving_2025} & \cellcolor{EvidenceP}P — CPLEX solver was run to proven optimality on the same model as a relevant baseline, though not demonstrated as state of the art for the given problem. & \cellcolor{EvidenceM}M — Solver times and quantum-processor access times were reported, but the D-Wave cloud service and local CPLEX had unmatched hardware and timing boundaries. & \cellcolor{EvidenceN}N — No compelling scaling evidence demonstrated. \\ \midrule
Mohammad\-bagher\-poor et al. (2021); preprint; not peer reviewed~\cite{mohammadbagherpoor_exploring_2021} & \cellcolor{EvidenceN}N — No classical method. & \cellcolor{EvidenceN}N — Two hardware runtimes were reported but had unclear definitions. & \cellcolor{EvidenceN}N — Only a single simulated case and reduced hardware case were tested. \\ \midrule
Q-CTRL (2025); corporate case study; not peer reviewed~\cite{q-ctrl_accelerating_2025} & \cellcolor{EvidenceN}N — Brief comparison made to a local greedy search, but otherwise no established rail-scheduling or optimisation solver was reported. & \cellcolor{EvidenceN}N — No timing reported. & \cellcolor{EvidenceN}N — No compelling scaling evidence demonstrated. \\ \midrule
Stollenwerk et al. (2020); journal article; peer reviewed~\cite{stollenwerk_quantum_2020} & \cellcolor{EvidenceN}N — Exact solvers and a classical heuristic supplied reference optima, but no state-of-the-art solver was executed. & \cellcolor{EvidenceP}P — The annealing time was reported, but embedding, tuning, readout, and all classical processing were excluded, and no classical runtimes were given. & \cellcolor{EvidenceN}N — No compelling scaling evidence demonstrated. \\ \midrule
Svensson et al. (2023); journal article; peer reviewed~\cite{svensson_hybrid_2023} & \cellcolor{EvidenceN}N — No classical method. & \cellcolor{EvidenceN}N — No timing reported. & \cellcolor{EvidenceN}N — The largest instance had 20 variables against a projected practical need of thousands. \\ \midrule
Vikstål et al. (2020); journal article; peer reviewed~\cite{vikstal_applying_2020} & \cellcolor{EvidenceN}N — No classical method. & \cellcolor{EvidenceP}P — A dimensionless time-to-solution measure was reported. & \cellcolor{EvidenceN}N — Only 8-25 variables were simulated with no scaling analysis. \\ \midrule
Xu et al. (2023); journal article; peer reviewed~\cite{xu_high-speed_2023} & \cellcolor{EvidenceP}P — Gurobi supplied solutions labelled optimal for every case and simulated annealing ran on the same model and computer. Relevant baselines but not proven state-of-the-art. & \cellcolor{EvidenceP}P — Matched solver times were reported for the 2 heuristics. No timing reported for Gurobi or tuning and data-preparation costs. & \cellcolor{EvidenceN}N — No compelling scaling evidence demonstrated. \\ \midrule
\multicolumn{4}{c}{\textbf{Traffic signalling}} \\ \cmidrule(lr){1-4}
Hussain et al. (2020); journal article; peer reviewed~\cite{hussain_optimal_2020} & \cellcolor{EvidenceP}P — Classical tabu search was run on the same models and performed slightly better and faster than hybrid annealing. Tabu search is a relevant baseline but not shown to be the strongest classical method. & \cellcolor{EvidenceP}P — Solver times were compared, but their quality benchmarks were unclear and excluded the surrounding classical workflow. & \cellcolor{EvidenceN}N — Only one grid size tested and no scaling behaviour demonstrated. \\ \midrule
Inoue et al. (2021); journal article; peer reviewed~\cite{inoue_traffic_2021} & \cellcolor{EvidenceP}P — Simulated annealing and a conventional local signalling rule were run on the same model, but no modern traffic-control method or strongest classical solver was tested. & \cellcolor{EvidenceN}N — No timing reported. & \cellcolor{EvidenceN}N — No compelling scaling evidence demonstrated. \\ \midrule
Inoue et al. (2024); journal article; peer reviewed~\cite{inoue_traffic_2024} & \cellcolor{EvidenceP}P — Simulated annealing and other simple control methods were run on the same model, but no modern traffic-control method or strongest classical solver was tested. & \cellcolor{EvidenceP}P — End-to-end quantum annealing and simulated annealing run-times were present, but unclear whether both used the same benchmark quality. & \cellcolor{EvidenceN}N — The quantum annealer was tested at a single size of 100 intersections. Larger networks were solved classically and no hardware scaling was reported. \\ \midrule
Marchesin et al. (2023); journal article; peer reviewed~\cite{marchesin_improving_2023} & \cellcolor{EvidenceP}P — Classical comparisons included simulated annealing on the same QUBO and a simple local signal controller. No modern traffic-control method was tested. & \cellcolor{EvidenceP}P — Solve times were reported for the annealer and simulated annealing, but excluded embedding and model-building costs. Classical run times not reported. & \cellcolor{EvidenceN}N — Quantum hardware was tested on a 64-intersection network only. Larger grids were solved classically, and no quantum performance scaling was reported. \\ \midrule
Shikanai et al. (2025); journal article; peer reviewed~\cite{shikanai_quadratic_2025} & \cellcolor{EvidenceP}P — Gurobi and simulated annealing were relevant classical baselines, but not established as the most efficient solver. Results were not compared to a best-known traffic optimiser. & \cellcolor{EvidenceP}P — Run-times reported, but each solver reports a different timing metric. No common end-to-end runtime or overhead costs were reported. & \cellcolor{EvidenceN}N — No scaling reported, all instances comprised 100 binary variables. \\ \midrule
\multicolumn{4}{c}{\textbf{Transport network design}} \\ \cmidrule(lr){1-4}
Chandra et al. (2022); conference paper; peer reviewed~\cite{Chandra_2022} & \cellcolor{EvidenceN}N — A randomly seeded genetic algorithm was run on the same cases, but its settings were largely unreported. & \cellcolor{EvidenceN}N — No timing reported. & \cellcolor{EvidenceN}N — No compelling scaling evidence demonstrated. \\ \midrule
Dixit and Niu (2023); journal article; peer reviewed~\cite{dixit_quantum_2023} & \cellcolor{EvidenceP}P — Tabu search is a relevant comparator, but is not established as the strongest method for the complete bilevel problem. & \cellcolor{EvidenceP}P — Solver timings are compared across ten network sizes, but QUBO construction, embedding and classical lower-level traffic assignment costs are excluded. Solver budgets use the D-Wave Ocean QPU-access-time estimator; time to a common solution-quality target is not reported. & \cellcolor{EvidenceN}N — Reported linear hybrid and exponential tabu-search runtime trends are encouraging solver-level results. Without complete workflow costs and matched convergence or solution-quality targets, they do not establish practical end-to-end quantum advantage. \\ \midrule
Gabbassov (2022); journal article; peer reviewed~\cite{gabbassov_transit_2022} & \cellcolor{EvidenceN}N — Classical solvers were executed only to establish optimality. No best classical comparison reported. & \cellcolor{EvidenceN}N — No timing reported. & \cellcolor{EvidenceN}N — All instances considered either 49 or 50 stops. No compelling scaling evidence demonstrated. \\ \midrule
Nourbakhsh et al. (2024); journal article; peer reviewed~\cite{Nourbakhsh_2024} & \cellcolor{EvidenceP}P — Results compared with K-means, but timed under a different formulation and objective function. & \cellcolor{EvidenceP}P — End-to-end runtimes were reported, but omit critical processing and embedding times. & \cellcolor{EvidenceN}N — No compelling scaling evidence demonstrated. Favourable annealing run-times disappear once embedding and processing times included. \\ \midrule
Okada et al. (2026); journal article; peer reviewed~\cite{okada_joint_2026} & \cellcolor{EvidenceN}N — Simulated annealing was used for comparison. It was not demonstrated that the nested formulation is competitive. & \cellcolor{EvidenceN}N — No timing reported. & \cellcolor{EvidenceN}N — Only a single problem size was tested. No scaling with the number of locations or charger sites was reported. \\ \midrule
Rao and Sodhi (2023); journal article; peer reviewed~\cite{Rao_2022} & \cellcolor{EvidenceP}P — Simulated annealing solves the same QUBO and a binary lightning-search algorithm provides another relevant baseline. Neither is established as state-of-the-art or evaluated under clearly matched stopping conditions. & \cellcolor{EvidenceP}P — QUBO construction takes approximately 5,000 seconds for 2,000 variables, compared with 5 seconds for hybrid annealing and 30 seconds for simulated annealing. This informative breakdown omits embedding time. & \cellcolor{EvidenceN}N — The common classical QUBO-construction cost dominates the workflow, making hybrid and simulated-annealing total times nearly identical once included. The reported results do not establish end-to-end quantum advantage. \\ \midrule
\multicolumn{4}{c}{\textbf{Computer vision}} \\ \cmidrule(lr){1-4}
Innan et al. (2025); journal article; peer reviewed~\cite{innan_qnn-vrcs_2025} & \cellcolor{EvidenceP}P — A CNN was trained on the same datasets and resolutions using the same stated split, epochs, and batch size. It was under-specified and not established as the best available classifier. & \cellcolor{EvidenceN}N — No timing was reported for preprocessing, circuit simulation, CNN or QNN training, or inference. & \cellcolor{EvidenceN}N — Only fixed small models on tiny images were tested. No scaling of accuracy or cost with image size or qubit count was reported. \\ \midrule
Kang et al. (2025); conference paper; peer reviewed~\cite{Kang_2025} & \cellcolor{EvidenceP}P — The network without its quantum layer was run on the same dataset. This is a useful ablation, but no tuned standalone classical classifier was tested. & \cellcolor{EvidenceP}P — The simulated quantum layer was reported as 8–10 ms per batch with 1,000 shots per sample. Complete training, inference, data-transfer, ResNet, and comparator costs were not reported. & \cellcolor{EvidenceN}N — A single fixed two-qubit circuit was tested on one dataset. No scaling with qubits, data size, or task difficulty was reported. \\ \midrule
Kuros and Kryjak (2023); conference paper; peer reviewed~\cite{kuros_traffic_2023} & \cellcolor{EvidenceP}P — A classical CNN was trained on the same traffic-sign data and consistently attained high accuracy. It was not established as the strongest available classifier. & \cellcolor{EvidenceN}N — No timing was reported for preprocessing, quantum convolution, classical or hybrid training, or inference. & \cellcolor{EvidenceN}N — Only batch size was varied on a fixed dataset and circuit. No scaling with data volume, image size, or qubit count was reported. \\ \midrule
Meghanath et al. (2025); journal article; peer reviewed~\cite{meghanath_qdcnn_2025} & \cellcolor{EvidenceP}P — A ResNet-50 DNN was run on the two Udacity tracks. It was task relevant but under-specified, and the CARLA runs had no classical comparator. & \cellcolor{EvidenceN}N — No measured quantum runtime was reported. The quoted value is a projection from photonic gate times that excludes preprocessing, encoding, sampling, optimisation, and postprocessing. & \cellcolor{EvidenceN}N — Only fixed one- and two-qubit circuits were simulated. Image size, feature count, qubits, and end-to-end runtime were not scaled. \\ \midrule
\multicolumn{4}{c}{\textbf{Drive cycle estimation}} \\ \cmidrule(lr){1-4}
Dixit and Jian (2022); journal article; peer reviewed~\cite{dixit_quantum_2022} & \cellcolor{EvidenceM}M — The classical fast Fourier transform was executed on the same signals and is the most appropriate task-aligned baseline. & \cellcolor{EvidenceN}N — No measured timing was reported for the classical transform, QFT, state preparation, measurement, or correction workflow. & \cellcolor{EvidenceN}N — Only two input sizes were tested: 256 samples in simulation and 16 on hardware. No task-equivalent runtime scaling was demonstrated. \\ \midrule
\multicolumn{4}{c}{\textbf{Traffic forecasting}} \\ \cmidrule(lr){1-4}
Hari\-krishna\-kumar et al. (2023); journal article; peer reviewed~\cite{harikrishnakumar_forecasting_2023} & \cellcolor{EvidenceN}N — LSTM, GPR, and Netica were executed, but served as pipeline components or probability references rather than an end-to-end classical alternative to the QBN ensemble. & \cellcolor{EvidenceN}N — No timing was reported for LSTM/GPR training, QBN simulation, Netica inference, or the complete forecasting pipeline. & \cellcolor{EvidenceN}N — One fixed three-node network for one station was simulated. No scaling with network size, stations, forecasting scope, or data volume was tested. \\ \midrule
Schetakis et al. (2025); journal article; peer reviewed~\cite{schetakis_quantum_2025} & \cellcolor{EvidenceP}P — Paired FC and LSTM baselines were trained on the same series and folds across several model sizes. Stronger graph-based forecasting methods identified by the authors were not executed. & \cellcolor{EvidenceP}P — Epoch and batch counts were reported, and simulated quantum training took approximately twice as long for each data-re-uploading block. No matched wall-clock or end-to-end timing was provided. & \cellcolor{EvidenceN}N — Models were simulated at fixed sizes on one traffic-flow series. No hardware execution or scaling with sites, horizons, data complexity, or complete computational cost was demonstrated. \\ \midrule
\end{longtable}
\end{landscape}

\clearpage
\section*{Database search query}

The following query was executed in the Scopus Advanced Search and IEEE Xplore interfaces on 22 July 2026 yielding 444 total reports. Given the prevalence of transport-related terms in physics, robotics, and logistics, many keywords were excluded from the title field.

\begin{Verbatim}[breaklines, breakanywhere]
( TITLE-ABS-KEY ( "quantum computing" OR "quantum computer" OR "quantum computers" OR "quantum annealing" OR "quantum annealer" OR "quantum algorithm" OR "quantum approximate optimization" OR "QAOA" OR "quantum walk" OR "quantum neural network" OR "QUBO" OR "quadratic unconstrained binary" OR "quantum machine learning" OR "variational quantum" OR "grover" OR "adiabatic quantum" OR "quantum heuristic" OR "quantum optimization" OR "quantum optimisation" ) AND TITLE ( "vehicle routing" OR "vrp" OR "traveling salesman" OR "travelling salesman" OR "tsp" OR "traffic signal" OR "traffic flow" OR "traffic control" OR "traffic optim*" OR "transportation" OR "timetabl*" OR "fleet management" OR "air traffic" OR "railway" OR "rail network" OR "rail scheduling" OR "rail traffic" OR "train schedul*" OR "urban mobility" OR "route planning" OR "route optim*" OR "ride sharing" OR "ride-sharing" OR "bike sharing" OR "shortest path" OR "last mile" OR "package delivery" OR "parcel delivery" OR "delivery rout*" OR "freight" OR "aviation" OR "flight trajectory" OR "flight gate" OR "flight rout*" OR "cargo" OR "public transport" OR "transport network" OR "transport planning" OR "transport system" OR "intelligent transport" OR "vehicle schedul*" OR "crew schedul*" OR "train timetab*" OR "shipment" OR "shipping" OR "electric vehicle" OR "ev charging" OR "ev rout*" OR "road network" OR "urban traffic" OR "intersection" OR "commuter" ) ) AND PUBYEAR > 2016 AND PUBYEAR < 2026 AND ( LIMIT-TO ( LANGUAGE , "English" ) ) AND NOT TITLE ( "quantum dot" OR "quantum transport" OR   "spin" OR "photon*" OR "nanowire" OR "graphene" OR "topological" OR "superconducti*" OR "key distribution" OR "quantum cryptograph*" OR "quantum communication" OR "post-quantum" OR "post quantum" OR "drug delivery" OR "molecular docking" OR "protein" OR "quantum-inspired" OR "quantum inspired" OR "robot" OR "robotic" OR "uav"  OR "unmanned aerial" OR "auv" OR "underwater vehicle" OR "warehouse" OR "supply chain" OR "inventory" OR "logistics" )
\end{Verbatim}

\begingroup
\renewcommand{\familydefault}{\rmdefault}
\normalfont
\urlstyle{same}
\printbibliography[title={References}]
\endgroup
\clearpage
\includepdf[pages=-,fitpaper=true,pagecommand={}]{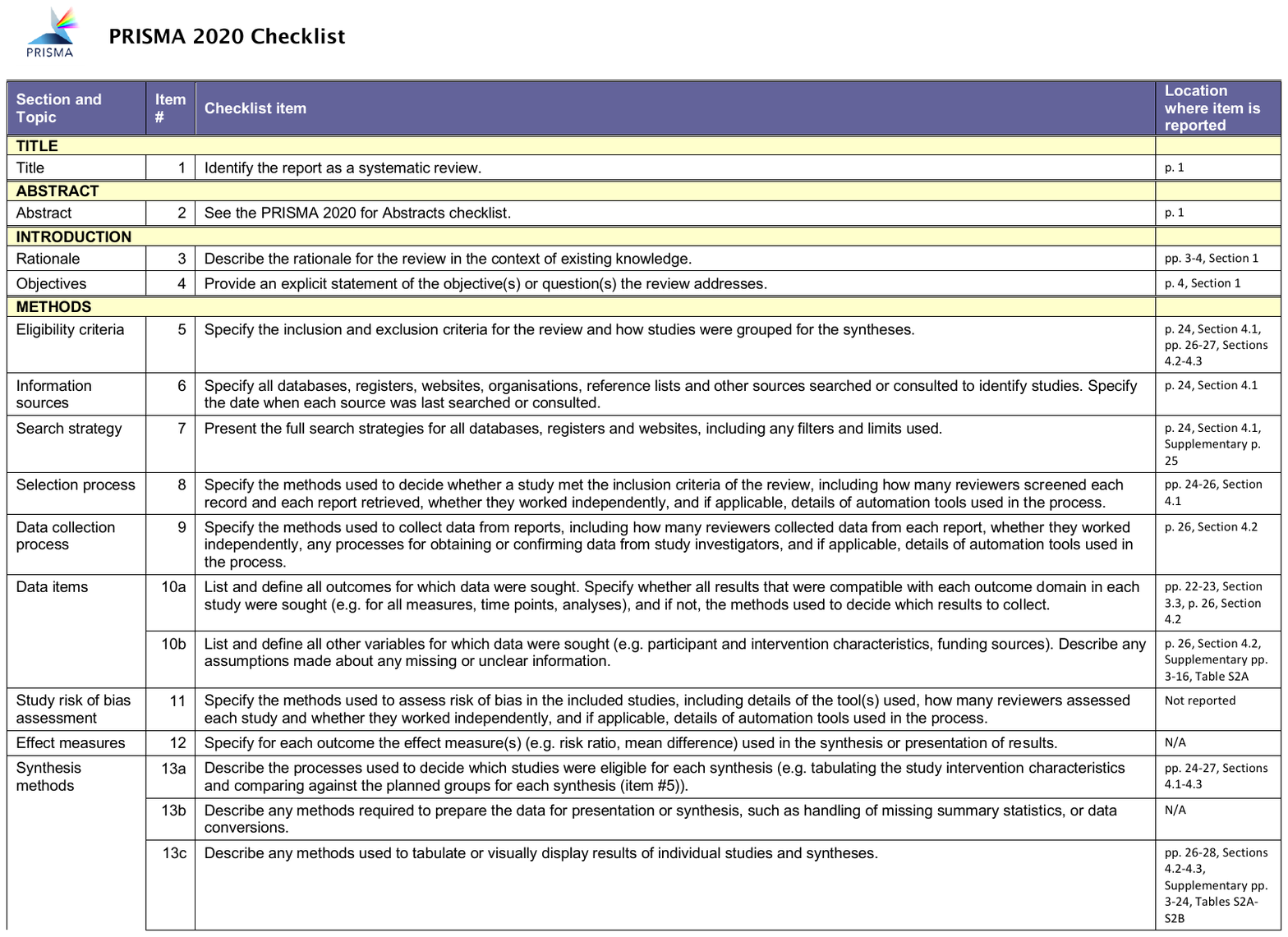}